\documentclass[a4paper,11pt]{article}
\pdfoutput=1 

\usepackage{jheppub} 

\usepackage[T1]{fontenc} 
\usepackage[all]{xy}
\usepackage{rotating}

\def\cA{\mathcal{A}}

\def\cT{\mathcal{T}}
\def\cD{\mathcal{D}}
\def\cI{\mathcal{I}}
\def\cH{\mathcal{H}}
\def\cN{\mathcal{N}}
\def\bR{\mathbb{R}}

\def\diag{\mathop{\rm diag}\nolimits}

\DeclareMathOperator{\tr}{tr}
\def\Re{\mathop{\mathrm{Re}}}
\def\Im{\mathop{\mathrm{Im}}}
\usepackage{slashed}

\def\a{\alpha}

\def\CA{{\cal A}}
\def\CB{{\cal B}}

\def\CG{{\cal G}}

\def\CI{{\cal I}}

\def\CK{{\cal K}}

\def\CM{{\cal M}}
\def\CN{{\cal N}}

\def\CT{{\cal T}}

\def\BC{{\mathbb C}}

\def\BR{{\mathbb R}}

\def\BZ{{\mathbb Z}}

\def\U{\mathrm{U}}

\def\beq#1\eeq{\begin{align}#1\end{align}}

\title{\boldmath Symmetry enhancement and closing of knots in 3d/3d correspondence}

\preprint{IPMU-18-0045}

\author[a]{Dongmin Gang}
\author[b]{and Kazuya Yonekura}

\affiliation[a]{Center for Theorectical Physics, Seoul National University, Seoul 08826, Korea}
\affiliation[b]{Kavli Institute for the Physics and Mathematics of the Universe, \\ University of Tokyo, Kashiwa, Chiba 277-8583, Japan}

\abstract{We revisit Dimofte-Gaiotto-Gukov's construction of 3d gauge theories  associated to 3-manifolds with a torus boundary. After clarifying their construction from a viewpoint of compactification of a 6d $\CN=(2,0)$ theory of $A_1$-type on a 3-manifold, 
we propose a topological criterion for $SU(2)/SO(3)$ flavor symmetry enhancement for the $u(1)$ symmetry in the theory associated to a torus boundary, which is expected from the 6d viewpoint. 
Base on the understanding of symmetry enhancement, we generalize the construction to closed 3-manifolds by identifying the gauge theory counterpart of  Dehn filling operation. The generalized construction predicts infinitely many 3d dualities from surgery calculus in knot theory.
Moreover, by using the symmetry enhancement criterion, we show that theories associated to all hyperboilc twist knots have surprising $SU(3)$ symmetry enhancement which is unexpected from the 6d viewpoint.  }

\begin{document} 
\maketitle
\flushbottom

\section{Introduction and Summary}
\label{sec:intro}

3-dimensional (3d) quantum field theory exhibits several interesting aspects. Unlike higher dimensional case, Abelian gauge interaction in 3d is strongly coupled at infrared (IR)  and   gives non-trivial IR physics. Different gauge theories at  ultraviolet (UV) could end  at the same IR fixed point along renormalization group (RG) and such phenomena  is called ``duality''. Refer to \cite{Intriligator:1996ex,deBoer:1996mp,Aharony:1997bx} for  examples of dualities among 3d gauge theories. There could be enhanced symmetries in the IR fixed point which is invisible in the UV gauge theory. From purely field theoretic viewpoint, these phenomena are not easy to understand or predict. 

In this paper, we consider a certain subclass of 3d  quantum field theories  with $\mathcal{N}=2$ (4 supercharges) supersymmetry which can be engineered by a twisted compactification of the 6d  $(2,0)$-superconformal field theory (SCFT) of $A_1$ type.  The  6d theory is the simplest  maximally supersymmetric conformal field theory  and describes the low energy world volume theory of two coincident M5-branes in M-theory. The 6d theory has $SO(5)$ R-symmetry and allows a 1/2 BPS regular co-dimension  two defect. The concrete set-up of this paper  is as follows
\begin{align}
\begin{split}
&\textrm{6d $A_1$ (2,0)-SCFT on  $\mathbb{R}^{1,2}\times M$ with a partial topological twisting along $M$ }
\\
&\textrm{with a regular co-dimension two defect along $\mathbb{R}^{1,2}\times K$} 
\\
&\xrightarrow[\text{}]{\quad \text{compacitification along $M$}\quad} T^{6d}[M,K]\textrm{  on $\mathbb{R}^{1,2}$}\;. \label{6d set-up}
\end{split}
\end{align}
Here $M$ is a compact (closed) 3-manifold and $K$ is a knot inside $M$.\footnote{The system can be generalized to the case when a knot $K$ is replaced by a link $L$ with several components. We  use the letter $K$ for knot and $L$ for link.} Using the vector $SO(3)$ subgroup of $SO(5)$ R-symmetry, we perform a topological twisting along $M$ which preserves $1/4$ supersymmetries.  After the  compactification, we obtain a 3d $\mathcal{N}=2$ quantum field theory, say $T^{6d}[M,K]$, determined  by  the topological choice of $M$ and  $K$. 
These theories are 3d analogy of 4d $\mathcal{N}=2$ theories of class S \cite{Gaiotto:2009we,Gaiotto:2009hg}.  In the analogy, closed Riemann surface corresponds to $M$ and a regular puncture on the surface corresponds to $K$. The 6d picture predicts the existence of  $su(2)$ flavor symmetry associated to the knot in the resulting 3d gauge theory. 

One non-trivial task is finding field theoretical description of the  3d  theory $T^{6d}[M,K]$.  A hint comes from so called 3d/3d relations  \cite{Yagi:2013fda,Cordova:2013cea,Lee:2013ida,Dimofte:2014zga,Gang:2015wya} which says that the partition functions of the  $T^{6d}[M,K]$ theory on supersymmetric curved backgrounds are equal to the partition functions of purely bosonic $SL(2,\mathbb{C})$ Chern-Simons (CS) theories on $M$ with a  monodromy defect along $K$.   
State-integral models \cite{2007JGP,Dimofte:2011gm,Andersen:2011bt,Dimofte:2012qj,Dimofte:2014zga} give integral expressions for complex CS  partition functions while
localization techniques \cite{Kim:2009wb,Kapustin:2009kz,Hama:2010av,Imamura:2011su} give similar integral expressions for the supersymmetric partition functions of 
3d field theories. 

Base on  the technical developments,   field theoretic algorithm of constructing  3d gauge theory $T^{DGG}[M,K]$ labelled by  the choice of $(M,K)$ is proposed by Dimoft-Gaiotto-Gukov \cite{Dimofte:2011ju}.  Their construction guarantees that the localization integrals of the  $T^{DGG}[M,K]$ theory are identical to the corresponding state-integral models.  In the original paper, the 3d gauge theory $T^{DGG}[N,X_A]$ is actually labelled by a choice of a knot complement $N$ and  a primitive boundary cycle $A \in H_1 (\partial N , \mathbb{Z})$. But there is a one-to-one map between  the two topological choices, $(M,K)$ and $(N,A)$, and we can labell them by the choice of $(M,K)$ which has more clear meaning in  the 6d compactification \eqref{6d set-up}. The explicit map between two topological choices is explained around Figure~\ref{fig:(M,K)and(N,A)}. 
From the non-trivial match of supersymmetric partition functions,  it is tempting to conclude that the $T^{DGG}[M,K]$ is actually  $T^{6d}[M,K]$. However, there are two manifest differences between two theories. Firstly, only some subset of irreducible flat $SL(2,\mathbb{C})$ connections on the knot complement $N:=M\backslash K$ appears as vacua on $\mathbb{R}^2\times S^1$ of $T^{DGG}[M,K]$ theory while all flat connections are expected to appear as the vacua of $T^{6d}[M,K]$ theory.  This point was already emphasized in \cite{Chung:2014qpa}. Secondly, the $T^{DGG}[M,K]$ theory generically has $U(1)$ flavor symmetry, denoted as $U(1)_{X_A}$,  associated to the knot $K$ while $T^{6d}[M,K]$ has a $su(2)$ flavor symmetry. Motivated from the similarity and differences of two theories, we propose the precise relation \eqref{Relation-T6d-TDGG} between them,
which we reproduce here:
\begin{align}
T^{6d} \xrightarrow[\text{}]{\textrm{\quad on a vacuum $P_{\rm SCFT}$} \quad} T^{6d}_{\rm irred} \xrightarrow[\text{}]{\textrm{\quad deformed by $\delta W = \mu^3$} \quad} T^{DGG} \label{Relation-T6d-TDGG:intro}
\end{align}
Here $\mu=(\mu^1, \mu^2,\mu^3)$ is
a chiral operator in the triplet representation of $su(2)$, and this operator is associated the co-dimension two defect along $K$. 
Each of the arrows in the above equation are nontrivial RG flows which are explained below.

The proposed relation explains why the $T^{DGG}[M,K]$ theory generically has only $U(1)$ symmetry associated to the knot while $T^{6d}[M,K]$ has  $su(2)$ flavor symmetry. The $su(2)$ symmetry of $T^{6d}[M,K]$ is broken by the superpotential deformation $\delta W = \mu^3$ in \eqref{Relation-T6d-TDGG:intro} which is typically a relevant deformation in the RG sense.   After $S^1$-reduction, the 6d theory becomes 5d maximally supersymmetric $su(2)$ Yang-mills theory (SYM) and the co-dimension two defect in 6d theory is realized by coupling a copy of the 3d $\mathcal{N}=4$ $T[SU(2)]$ theory~\cite{Gaiotto:2008ak} to the 5d theory.  
Then $\mu$ is the $su(2)$ moment map operator of the 3d $\CN=4$ $T[SU(2)]$ theory. 

As an intermediate step, we introduce a 3d SCFT $T^{6d}_{\rm irred}[M,K]$ appearing in \eqref{Relation-T6d-TDGG:intro}
which is the IR fixed point of $T^{6d}[M,K]$ on a particular point $P_{SCFT}$ of the vacuum moduli space.  
Unlike $T^{6d}[M,K]$, $T^{6d}_{\rm irred}[M,K]$ might not contain the $su(2)$ moment map operator $\mu$ after taking the IR limit.  In that case,   
the superpotential deformation is not possible (or more precisely, it is irrelevant) and  thus the $T^{DGG}[M,K]$ still has the $su(2)$ symmetry. By carefully analyzing the coupled system, 5d SYM+3d $T[SU(2)]$, we find a topological condition on $(N,A)$ which guarantees  the absence of moment map operator and thus the $su(2)$ symmetry in $T^{DGG}[N,X_A]$ theory.  The topological condition is summarized in Table~\ref{Sym enhancement in DGG}. For example, we expect $su(2)$ symmetry enhancement when $M$ is a Lens-space and do not expect the enhancement when $M$ is hyperbolic.   

As an application of the symmetry enhancement criterion, we show that the $T^{DGG}[M=S^3,K]$ theory for all hyperbolic twist knots $K$ has a surprising $SU(3)$ symmetry.  As a simplest example, we claim that  the following 3d $\mathcal{N}=2$ theory has $SU(3)$ symmetry. 
\begin{align}
\begin{split}
&T^{DGG}[M=S^3,K=\textrm{figure-eight knot}] 
\\
&= \textrm{A $U(1)_0$ vector multiplet coupled to two chiral multiplets of charge +1}\;.
\end{split}
\end{align}
The theory only has manifest $SU(2)\times U(1)$ symmetry where the $SU(2)$ rotates the two chirals and the $U(1)$ comes from the topological symmetry of the dynamical abelian gauge field. The $U(1)_{X_A}$ symmetry associated to the knot is a linear combination of two Cartans of the $SU(2)\times U(1)$ which is  expected to be enhanced to $SO(3)$ according to the criterion in Table~\ref{Sym enhancement in DGG}. From a group theoretical analysis, the enhancement   implies that the $SU(2)\times U(1)$ should be enhanced to $SU(3)$. We  checked the symmetry enhancement by explicitly constructing the corresponding conserved current multiplet. 

Base on the proposed relation between $T^{6d}[M,K]$ and $T^{DGG}[M,K]$, we identify the field theoretical operation on $T^{DGG}[M,K]$  corresponding to Dehn filling operation on the knot complement $N=M\backslash K$. The operation is only possible when the $T^{DGG}[M,K]$ has $su(2)$ flavor symmetry.  The Dehn filling operation is analogous to closing of punctures on Riemann surface in 4d/2d correspondence~\cite{Tachikawa:2013kta}. By applying the Dehn filling operation, we can extend the DGG's construction to 3d gauge theories labelled by a closed 3-manifold $M$. The theory is denoted as $T^{6d}_{\rm irred}[M]$ and has similar 6d interpretation as $T^{6d}_{\rm irred}[M,K]$.   As concrete examples, field theoretic descriptions of $T^{6d}_{\rm irred}[M]$ for three smallest hyperbolic 3-manifolds are given in \cite{Gang:2017lsr}.  One interesting aspect of our construction of $T^{6d}_{\rm irred}[M]$ is that we can relate surgery calculus in knot theory to 3d $\mathcal{N}=2$ dualities.  One way of representing closed 3-manifold is using so called Dehn surgery representation. A closed 3-manifold $M$ has infinitely many different surgery descriptions and surgery calculus tell when two surgery descriptions give the same 3-manifold. Different surgery representations  of a closed 3-manifold give different field theoretical descriptions of $T^{6d}_{\rm irred}[M]$ which are related by 3d dualities. One illustrative example   is given around eq.~\eqref{duality-surgery-example}. Since the 3d theory depends on only the topology of the 3-manifold,  every physical quantities of the theory are topological invariants of the 3-manifold. As an example, we introduce a new 3-manifold  invariant called ``3d index'' which is nothing but the superconformal index of the $T^{6d}_{\rm irred}[M]$.



The paper is organized as follows. In section \ref{sec:T[M]}, we introduce two ways of associating the choice of 3-manifold $M$ and a knot $K$ inside it with a 3d gauge theory $T[M,K]$. One is  through the construction by Dimofte-Gaiotto-Gukov \cite{Dimofte:2011ju} (DGG) and the corresponding gauge theory is denoted as $T^{DGG}[M,K]$. The other is through a twisted compactification of 6d $A_1$ (2,0) theory on $M$ with a regular co-dimension two defect along $K$. 
The resulting 3d gauge theory is denoted as $T^{6d}[M,K]$. After explaining the two constructions in detail, we propose a precise relation \eqref{Relation-T6d-TDGG} between two constructions. Base on the proposed relation, in section \ref{sec:symm},  we give a topological criterion on $(M,K)$ which determines when the $U(1)_{X_A}$ symmetry  $T^{DGG}[M,K]$ theory is enhanced to $SU(2)$ or $SO(3)$. The criterion is summarized in Table~\ref{Sym enhancement in DGG}. In section \ref{sec : Dehn-filling}, we identify field theoretic operation corresponding to Dehn filling operation in 3-manifold side in 3d/3d correspondence. It allows us to extend the DGG's construction to the case when the knot is absent. In section \ref{sec: Duality/Surgery}, we discuss how the surgery calculus in knot theory predicts infinitely many 3d $\mathcal{N}=2$ dualities.

\section{3d $\mathcal{N}=2$ Superconformal field theories labelled by 3-manifolds}
\label{sec:T[M]}

In this section, we introduce two ways of associating a 3-manifold $M$ with a knot $K$ in it to a 3d $\mathcal{N}=2$ gauge theory $T[M,K]$.
\begin{align}
\begin{split}
&(M,K) \; \rightsquigarrow \; (\textrm{a 3d $\mathcal{N}=2$ gauge theory $T[M,K]$})\;, \textrm{ where}
\\
&\textrm{$M$ is a closed 3-manifold}\;, \quad\textrm{$K\subset M$ is a knot in $M$}\;.
\end{split}
\end{align}
One way is through a twisted compactification of a 6d $\mathcal{N}=(2,0)$ theory of $A_1$ type on a closed 3-manifold $M$ with a regular co-dimension two defect along a knot $K$ on $M$. The other way is using the construction  by Dimofte-Gaiotto-Gukov \cite{Dimofte:2011ju} (DGG) based on an ideal triangulation of the knot complement $M\backslash K$. 
These two theories are argued to be related \cite{Dimofte:2011ju}, and 
we will propose the more precise relation between them with supporting evidences.
We describe the relation after reviewing basic aspects of two approaches.

Before going to detailed analysis, let us first introduce an alternative labelling for the topological choice, ($M$ and $K$), which will be used throughout the paper.
\begin{figure}[h]
\begin{center}   
\includegraphics[width=.50\textwidth]{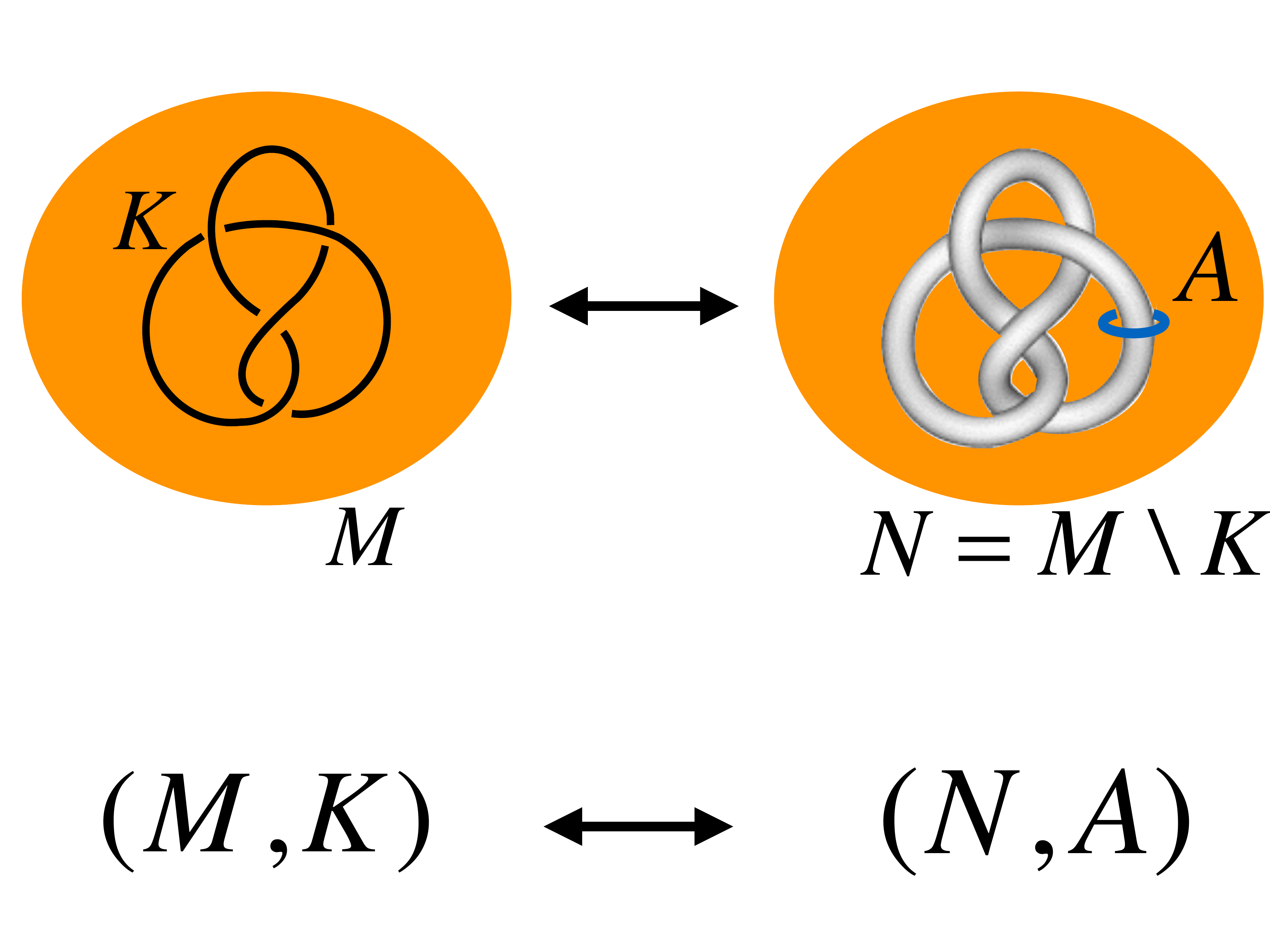}
\end{center}
\caption{ The choice of a knot $K$ inside a closed 3-manifold $M$ can be alternatively described by a choice of knot complement $N$ and a boundary cycle $A \in H_1 (\partial N, \mathbb{Z})$.  } 
\label{fig:(M,K)and(N,A)}
\end{figure}
The choice  can be replaced by  
\begin{align}
\begin{split}
&(M,K) \leftrightarrow (N,A)\;, \textrm{ where}
\\
&\textrm{$N$ is a knot complement and $A\in H_1 (\partial N,\mathbb{Z})$ is a primitive boundary cycle .}
\end{split}
\end{align}
For a given $(M,K)$, the corresponding $(N,A)$ is given by 
\begin{align} 
\begin{split}
&N =M\backslash  K:=M-N_K\;, \quad (N_K \textrm{ : Tubular neighborhood of a knot $K$}) 
\\
&A \; : \;\textrm{A primitive boundary cycle  around the knot $K$. } 
\\
& \qquad  \textrm{(i.e. the contractible cycle in the removed tubular neighborhood $N_K$)}
\end{split} 
\end{align}
For given $(N,A)$, on the other hand, $(M,K)$ is determined by 
\begin{align}
\begin{split}
M & = N_{A} : =(\textrm{A closed 3-manifold obtained from $N$ by closing a cycle  $A$ in $\partial N$}) 
\\
&:=(N\cup (D_2\times S^1))/\sim\; \;\textrm{with } A \sim \textrm{(contractible boundary cycle of $D_2\times S^1$)}
\\
K &:=\{p\} \times S^1 \subset D_2 \times S^1, \text{ where $p \in D_2$ is the origin of $D_2$.}  
\end{split}
\end{align}
Using the  map, we can use two choices interchangeably. For example,
\begin{align}
T[N ,A]  = T[M, K]\;.
\end{align}
In most part of this paper, we assume that $N$ is a knot complement with one torus boundary but our discussion can be easily generalized to the case when $N$ is a link complement with several torus boundaries.

\subsection{6d $A_1$ (2,0) theory on 3-manifolds : $T^{6d}$ and $T^{6d}_{\rm irred}$}\label{sec:6d}
We define
\begin{align}
\begin{split}
&T^{6d}[M,K] = (\textrm{Twisted compactification of 6d (2,0)  $A_1$ theory on $M$} 
\\ 
& \qquad \qquad \quad \qquad   \textrm{with a regular co-dimension two defect along $K \subset M$})\;.
\end{split} \\
\begin{split}
&T^{6d}_{\rm irred}[M,K] = (\textrm{The low energy limit of $T^{6d}[M,K]$ }
\\
& \qquad \qquad \quad \qquad   \textrm{on a particular point $P_{\rm SCFT}$ on the moduli space of vacua} )\;.
\end{split} \label{eq:defof6dtheory}
\end{align}
As a simpler set-up, we can also consider the case when the defect is absent. In that case, the resulting 3d theory is denoted as $T^{6d}[M]$ and $T^{6d}_{\rm irred}[M]$, respectively. For the $T^{6d}_{\rm irred}[M,K]$ to be defined, we assume that $N=M\backslash K$ is a {\it hyperbolic} knot complement.

The reason that we consider $T^{6d}_{\rm irred}[M,K] $ is as follows. 
The moduli space of vacua of $T^{6d}[M,K]$ in general contains several different connected components.
Then, we have to decide which point of the moduli space we consider before taking the low energy limit.
The typical distances between different components of the moduli space are of the order of the compactification scale on $M$, which set the cutoff scale
of the low energy effective 3d theory. Therefore, we cannot expect that there is a single effective 3d theory which describes the entire
moduli space of vacua. Only after specifying a point on the moduli space, we can obtain a low energy effective field theory which describes
the physics near that point.\footnote{
A simple example which illustrates the point is the $T^2$ compactification of the 6d $\CN=(2,0)~A_1$ theory on $T^2$.
The moduli space of this theory is $[\BR^5 \times S^1]/\BZ_2$, where $S^1$ comes from the integral of the 2-form field on $T^2$.
On the other hand, the moduli space of 4d $\CN=4$ SYM is $\BR^6/\BZ_2$.
Only after picking a point on $[\BR^5 \times S^1]/\BZ_2$ and taking the low energy limit, the 6d $\CN=(2,0)$ theory on $T^2$ becomes
the 4d $\CN=4$ SYM. In this case the moduli space is connected, but still there is no single 4d effective theory describing the whole moduli space of vacua.}

In other words, $T^{6d}[M,K]$ is not a genuine 3d theory, but should be considered more appropriately as
the 6d theory compactified on $M$. However, we will be sometimes sloppy and call it a 3d theory in this paper.

In $T^{6d}_{\rm irred}[M,K]$, we pick up a point and take the low energy limit. 
The low energy limit may be described by a 3d SCFT (which can be empty or a topological theory). 
Below we will specify which point on the moduli space of vacua we take, by using reduction to 5d SYM.

\paragraph{$T^{6d}$ on $\mathbb{R}^2\times S^1$ via 5d SYM} 
The structure of the moduli space of vacua becomes simpler if we compactify the 3d spacetime to $\BR^2 \times S^1$.
This is because we can use the 5 dimensional maximally supersymmetric Yang-Mills (5d SYM)  theory description.
The set-up is\footnote{On general grounds, one may only expect that 5d SYM describes the moduli space only in the limit of very small radius of $S^1$.
However, somewhat miraculously, it is believed that 5d SYM describes even a finite radius of $S^1$.}
\begin{equation}
\xymatrix{
\text{6d }\cN=(2,0)\text{ on }\mathbb{R}^2 \times S^1 \times M   \ar[d] \ar[dr]  &   \\
\text{5d }\cN=2\text{ SYM on } \mathbb{R}^2 \times M \ar[r] & \text{theory }T^{6d}[M] \text{ on }\mathbb{R}^2 \times S^1
}\label{eq:comm}
\end{equation}
The bosonic components of the 5d SYM theory are gauge fields $A_{I}~(I=0,\cdots,4)$ and scalar fields $\phi_k~(k=0,\cdots,4)$.
After compactification on $M$, the supersymmetry is defined on $\bR^2$, and we split these fields as
\begin{align}
(A_{I=0,1,2,3,4}, \phi_{k=1,2,3,4,5}) \to (A_{\mu=0,1}, \phi_{k=0,1}) \oplus ( \cA_i :=A_{i} + i \phi_{i} )_{i=2,3,4}
\end{align}
From the point of view of the super-algebra on $\BR^2$,
the ${\cal V}= (A_{\mu=0,1}, \phi_{k=0,1})$ is the vector multiplet and $\cA_i=A_{i} + i \phi_{i}~(i=2,3,4)$ are twisted chiral fields.
The reason that we regard $\cA$ as twisted chiral fields rather than chiral fields is that the relation between 5d SYM and $T^{6d}[M]$ is a kind of mirror symmetry
analogous to the case of 4d class S theories.

The twisted superpotential is given by complex Chern-Simons action as
\begin{align}
\widetilde{W}_\text{YM} = \frac{1}{g_{\rm YM}^2} CS[\cA] :=\frac{1}{g_{\rm YM}^2} \int_M  \frac{1}{2} \mathrm{Tr} \left(  \cA d \cA +\frac{2}{3} \cA^3 \right). \label{W_YM}
\end{align}
where $g_{\rm YM}^2$ is the gauge coupling of 5d SYM which is related to the radius $R$ of $S^1$ as
\beq
\frac{1}{g_{\rm YM}^2 } =\frac{1}{8\pi^2R}.
\eeq
This is the results in \cite{Yagi:2013fda,Lee:2013ida,Cordova:2013cea} in the limit $S^2 \to \bR^2$.
This twisted superpotential corresponds to the twisted superpotential obtained in DGG's construction discussed in Sec.~\ref{sec:DGG}

The regular co-dimension two defect along a knot  $K \subset M$ can be realized as coupling the 3d $T[SU(2)]$ theory~\cite{Gaiotto:2008ak} to the fields of
5d SYM~\cite{Benini:2009gi,Benini:2010uu,Gaiotto:2011xs,Chacaltana:2012zy,Yonekura:2013mya}.  
The  theory $T[SU(2)]$ is reviewed in Appendix~\ref{app:brief}. This
is a 3d $\cN=4$ SCFT given by $U(1)$ vector multiplet coupled two fundamental hypermultiplets $(E^a,\tilde{E}_a)_{a=1,2}$. 
The theory has $su(2)_H\times su(2)_C$  flavor symmetry and let 
\begin{align}
\begin{split}
&\widetilde{\mu} := \textrm{holomorphic moment map operator of $su(2)_C$}\;,
\\
&\widetilde{\nu}:= \textrm{holomorphic moment map operator of $su(2)_H$} \;.
\end{split}
\end{align}
Then the twisted superpotential coupling of the $T[SU(2)]$ and the 5d SYM is given by
\begin{align}
\widetilde{W}_\text{YM-defect} = \int_K \textrm{tr} (\widetilde{\nu} \cA). \label{W_YM-defect}
\end{align}
This means that we integrate the one-form $\textrm{tr}(\widetilde{\nu} \cA_i)dy^i$ over $K$.\footnote{
The gauge invariance is preserved as follows. The supersymmetry is considered in the two dimensional space $\BR^2$,
and hence the direction along the knot $K$ is considered as a kind of ``internal manifold''. Let $t$ be the coordinate along $K$.
Then, the kinetic term along this direction comes not from the Kahler potential, 
but from the twisted superpotential as $\widetilde{W} \supset \tilde{E} \partial_t E$.
This term combines with \eqref{W_YM-defect} to form a covariant derivative $\tilde{E}(\partial_t +\CA)E$,
where we have used $\widetilde{\mu} \sim E\tilde{E}$ (see Appendix~\ref{app:brief}).}

We can also include (complexified) mass terms to the defect as 
\begin{align}
\widetilde{W}_{\rm mass} = \int_K ds \;\textrm{tr} (m \widetilde{\mu} ) \label{W_mass}
\end{align}
where $ds$ is the line element on $K$, and $m$ is the mass. The mass of defect is related to the eigenvalues of $\widetilde{\nu}$:
\begin{align}
\textrm{Eigenvalues of $\widetilde{\nu}$} = \{ m, -m \}\;.
\end{align}
See \cite{Yonekura:2013mya} for detailed explanations of the coupling of 5d SYM to $T[SU(2)]$ in the context of 4d class S theories. 
The analysis there may be extended to the 3d/3d case,
but we do not perform a detailed analysis.

By solving F-term equations for the twisted superpotenal in  \eqref{W_YM} and \eqref{W_YM-defect}, 
a part of the moduli space of vacua\footnote{When the connection $\CA$ is reducible, we can turn on the expectation values of 
the vector multiplets ${\cal V}= (A_{\mu=0,1}, \phi_{k=0,1})$. These branches are very important in 4d class S theories~\cite{Yonekura:2013mya,Xie:2014pua}.
However, in the 3d theories considered in this paper, we only consider points on the moduli space on which $\CA$ is irreducible.
Therefore, we can neglect those branches.
} on $\mathbb{R}^2\times S^1$ with mass parameter $m$ is given by 
\begin{align}
\begin{split}
&\mathcal{M}_{\rm vacua} (\textrm{$T^{6d}[M,K]$ on $\mathbb{R}^2\times S^1$}) 
\\
&= \{ \cA : d\cA+\cA\wedge \cA =\widetilde{ \nu} \delta (K) ,~\text{eigenvalues of } \widetilde{\nu}=\{m,-m \}   \}/ \CG 
\label{flat connections}\;.
\end{split}
\end{align}
where $\delta(K)$ is  the  delta  function  localized on $K$, and $\CG$ is the group of $PSL(2,\mathbb{C}) $ gauge transformations on $M$.   
This  is  the  space  of  flat  connections  of  the complexifield gauge group $PSL(2,\mathbb{C})$ with the holonomy $e^{\widetilde{\nu} }$ around $K$.
\begin{align}
 \rho_{\rm hol}(A):=(\textrm{$PSL(2,\mathbb{C})$ holonomy matrix along $A$-cycle})= e^{\widetilde{\nu } }\;. \label{HolA and nu}
\end{align}
Notice that the eigenvalues of $e^{\widetilde{\nu}}$ are determined by the mass parameter $m$.

Now we can specify the point $P_{\rm SCFT}$ on the moduli space of vacua which is taken in the definition \eqref{eq:defof6dtheory}.
First, let us consider more generally. For simplicity we assume that the moduli space of vacua on $\BR^3$,
$\mathcal{M}_{\rm vacua} (\textrm{$T^{6d}[M,K]$ on $\mathbb{R}^3$}) $, is a discrete set. 
Let us take an arbitary point $P \in \mathcal{M}_{\rm vacua} (\textrm{$T^{6d}[M,K]$ on $\mathbb{R}^3$ }) $.
Then, if we compactify the theory on $S^1$ with a radius which is large enough compared to potential barriers between different points on 
$\mathcal{M}_{\rm vacua} (\textrm{$T^{6d}[M,K]$ on $\mathbb{R}^3$}) $, then the point $P$ goes to a subset $\CM(P)$ of 
the moduli space of vacua on $\BR^2 \times S^1$ denoted as $\mathcal{M}_{\rm vacua} (\textrm{$T^{6d}[M,K]$ on $\mathbb{R}^2\times S^1$})$, 
\beq
P \to \CM(P) \subset \mathcal{M}_{\rm vacua} (\textrm{$T^{6d}[M,K]$ on $\mathbb{R}^2\times S^1$}).
\eeq
This $\CM(P)$ need not be a single point, but may have several points whose number is related to the Witten index of the 3d effective theory on $P$.
Because of the supersymmetry, the condition that the radius of $S^1$ is large may be dropped since there is no phase transition under  change of the radius.

The explicit forms of $\mathcal{M}_{\rm vacua} (\textrm{$T^{6d}[M,K]$ on $\mathbb{R}^3$}) $ and $\CM(P)$ are not known
and they are defined just by the abstract field theoretical considerations as above.
However, later we will propose how $\CM(P_{\rm SCFT})$ may be given concretely in terms of flat $PSL(2,\BC)$ connections.

To consider a superconformal point, we set the mass $m$ to be zero. 
Then the point $P_{\rm SCFT}$ is defined as follows.
After compactification on $S^1$, the moduli space of vacua of the theory on $P_{\rm SCFT}$ becomes
a subset $\CM(P_{\rm SCFT})$ of the moduli space of vacua on $\BR^2 \times S^1$.
Then, the point $P_{\rm SCFT}$ is defined by the condition that $\CM(P_{\rm SCFT})$ contains the connection 
$\CA^{\rm \overline{hyp}} \in \mathcal{M}_{\rm vacua} (\textrm{$T^{6d}[M,K]$ on $\mathbb{R}^2\times S^1$})$ 
which is determined by the unique complete hyperbolic metric on $N=M \backslash K$. 
More explicitly, using the spin-connection $\omega$ and dreibein $e$ of the complete hyperbolic metric, the flat connection can be expressed as
\begin{align}
\CA^{\rm \overline{hyp}} =  \omega - i e\;.
\end{align}
This flat connection has the greatest value of $\textrm{Im}(CS[\CA^\a])$ among all flat connections $\CA^\a$ with parabolic boundary holonomy and is conjectured to be the only vacua contributing to a squashed 3-sphere partition function \cite{Hama:2011ea} of  $T^{6d}_{\rm irred}[M,K]$.  Refer to  \cite{Andersen:2011bt,Gang:2014ema,Bae:2016jpi,Mikhaylov:2017ngi,Gang:2017hbs} for discussions on the  conjecture from various respects, state-integral model of the complex CS theory, holographic principal and resurgent analysis.   To other physical quantities of $T^{6d}_{\rm irred}[M,K]$ such as superconformal index,  on the other hand, other flat connections in  $\CM(P_{\rm SCFT})$ may contributes. 

Now we give a conjecture about how $\CM(P_{\rm SCFT})$ is given concretely in terms of flat connections. 
First we consider the case where a knot $K$ exists.
For this purpose, we define $\CM(P_{\rm SCFT})$ even for nonzero mass $m$ by continuity from $m=0$.
Namely, $\CM(P_{\rm SCFT})$ is just the set of vacua of the 3d effective theory near $P_{\rm SCFT}$ with mass $m$.
We make the dependence on $m$ explicit by writing it as $\CM(P_{\rm SCFT},m)$.
We also define $\chi(N)$ as
\beq
\chi(N) =  \bigcup_m \{ \cA : d\cA+\cA\wedge \cA =\widetilde{ \nu} \delta (K),~\text{eigenvalues of } \widetilde{\nu}=\{m,-m \}   \}/ \CG \label{eq:chidef1}
\eeq
This means that we consider all flat connections with varying holonomy around the knot.
Then we propose
\begin{align}
\bigcup_m \CM(P_{\rm SCFT},m) &=\; \{\textrm{the connected component of $\chi(N)$ containing $\CA^{\rm \overline{hyp}}$} \} \nonumber \\
& := \chi_0(N)
 \;. \label{Dehn-surgery com}
\end{align}
In \cite{tillmann2012degenerations}, the component is called {\it Dehn surgery component}. 
Another way of representing the above equation is $\CM(P_{\rm SCFT},m)  = \chi_0(N) \cap \{ \text{eigenvalues of }\nu=\{ m,-m \} \}$.

Next, consider the case where there is no knot on $M$.
If the closed 3-manifold is represented by a Dehn filling operation on a hyperbolic knot complement $N=M \backslash K$ for some $K$ along a boundary cycle $A$,
\begin{align}
M = N_{A}
\end{align}
we propose that the $ \CM(P_{\rm SCFT}) $ is given by
\beq
& \CM(P_{\rm SCFT}) \textrm{ of the closed manifold } M  \nonumber \\
&= \chi_0(N) \cap \{ \rho_{\rm hol}(A)=1 \}
\eeq
The definition of the right hand side contains a knot $K$,
but we assume that it is independent of the choice $K \subset M$.
Notice that $\rho_{\rm hol}(A)=1$ is stronger than $m=0$, since we could have a nonzero upper-right component of $\nu$ even if its eigenvalues are zero.

\subsection{Dimofte-Gaiotto-Gukov's construction : $T^{DGG}$}\label{sec:DGG}
In \cite{Dimofte:2011ju}, a combinatorial way of constructing a 3d SCFT, which we denote $T^{DGG}[N,X_A]$, for given choice of $(N,A)$ is proposed. Empirically, the theory associated to non-hyperbolic $N$ is a trivial theory only with topological degrees of freedom.  In this subsection we focus on the case when $N$ is hyperbolic. Here we give a  summary of the DGG's construction with a modification on superpotential deformation associated to  `hard' internal edges (see \eqref{DGG : final}) which play a crucial role in the symmetry enhancement of the theory. 

\paragraph{Mechanics of ideal triangualtion} The construction is based on a choice of an ideal triangulation $\cT$ of $N$.
\begin{align}
\cT \;: \; N = \bigg{(}\bigcup_{i=1}^k \Delta_i \bigg{)}/\sim\;.
\end{align}
Here $\Delta_i$ denote the $i$-th tetrahedron in the triangulation. Ideal tetrahedron can be embedded into  a hyperbolic upper half plane $\mathbb{H}^3$ in a way that all vertices are located on the boundary of $\mathbb{H}^3$ and both of edges and faces are geodesics. Hyperbolic structures on an ideal tetrahedron can be parameterized by a complex parameter $z$ (with $0<\textrm{Im}[Z]<\pi$, $Z:=\log z$), which is the cross-ratio of the positions of its vertices on $\partial \mathbb{H}^3$. We assign edge parameters $(z,z',z'')$ to each pair of edges of ideal tetrahedron as in the figure below.
\begin{figure}[h]
\begin{center}   
\includegraphics[width=.30\textwidth]{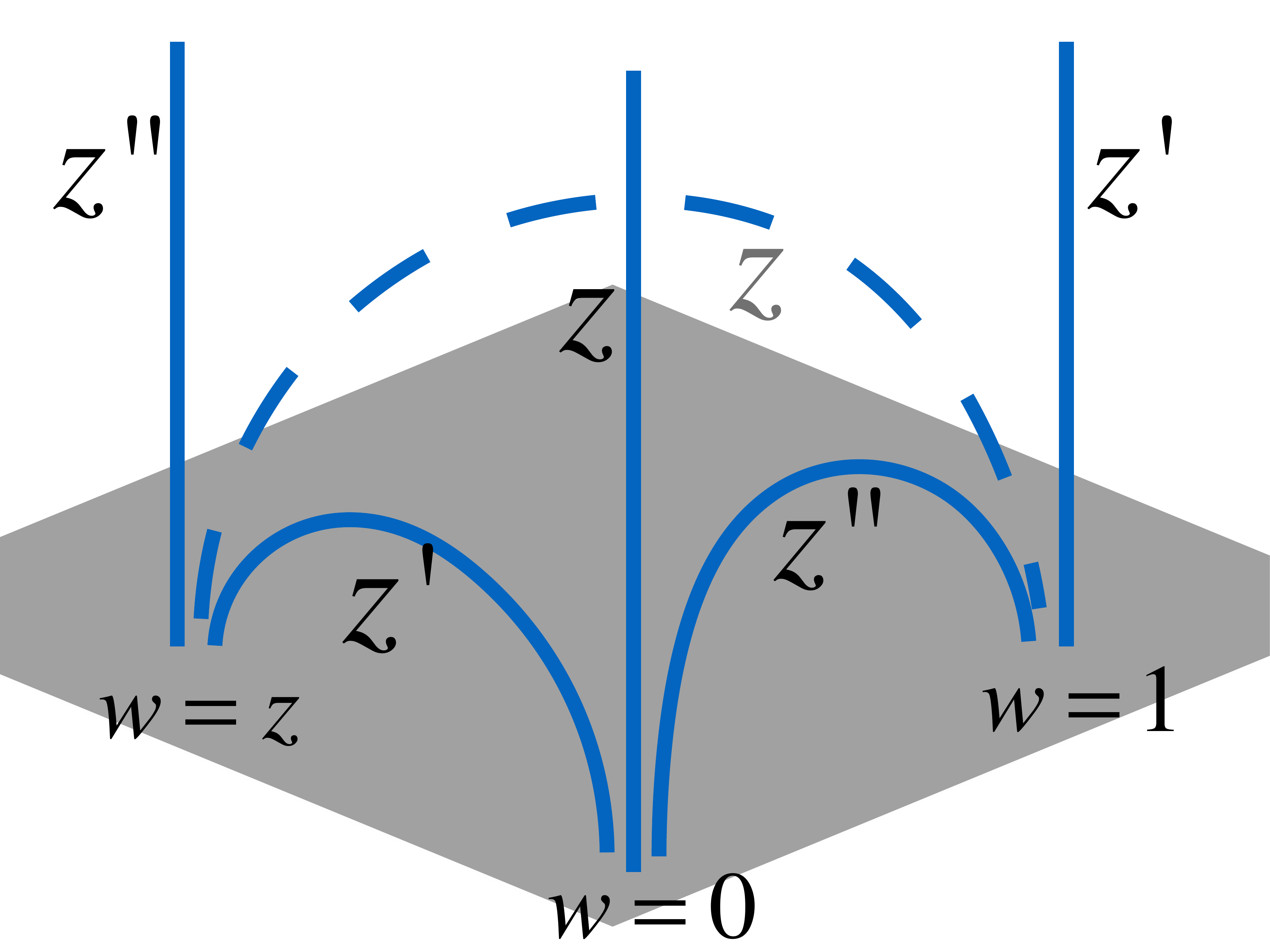}
\quad 
\includegraphics[width=.25\textwidth]{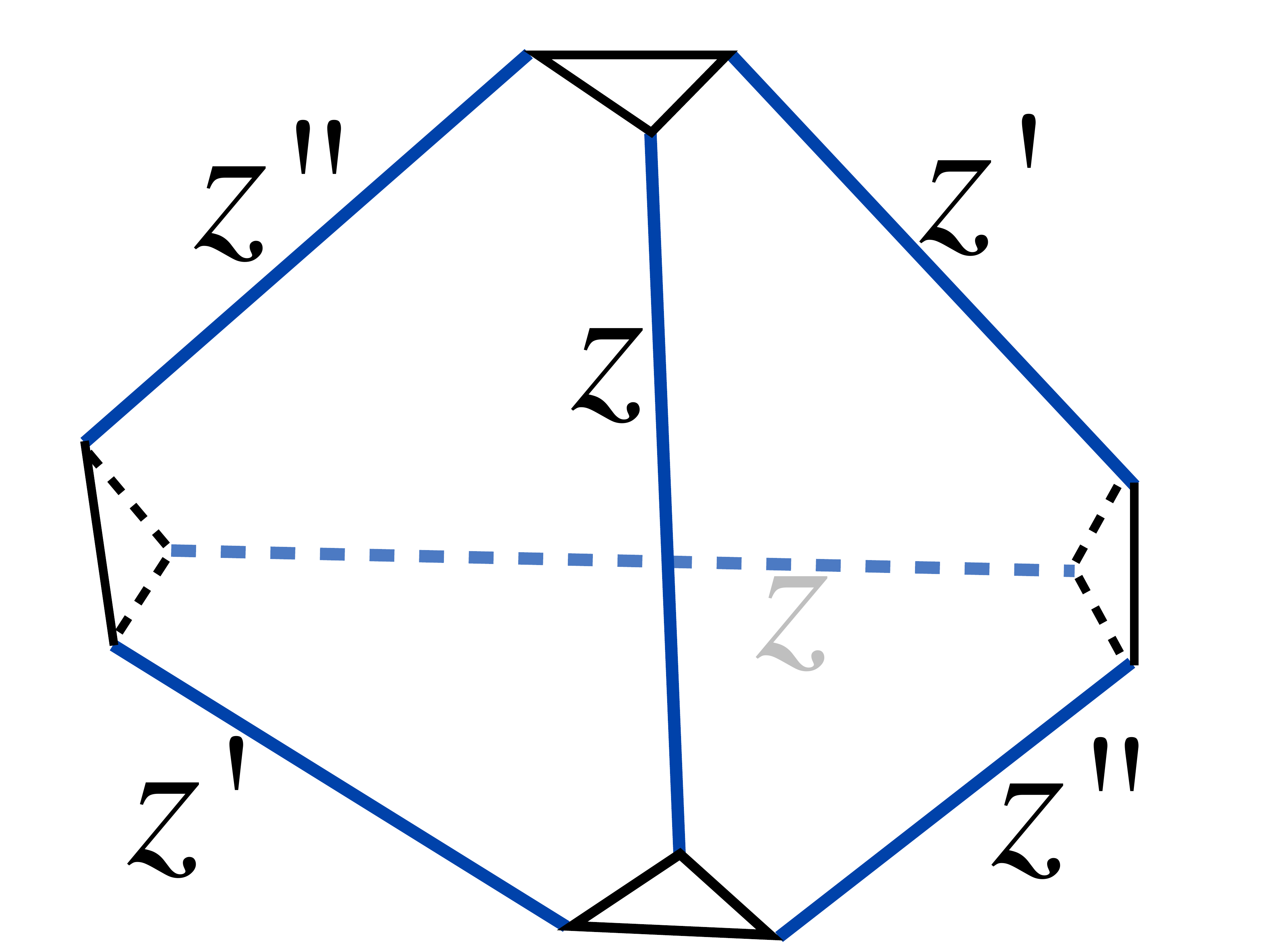}
\end{center}
\caption{Edge parameters $(z,z',z'')$ of an ideal tetrahedron. Left: ideal tetrahedron in $\mathbb{H}^3=\{(y,w):y\in \mathbb{R}_+, w \in \mathbb{C}\}$ with metric $ds^2 (\mathbb{H}^3) = \frac{dy^2 +d w d \bar{w}}{y^2}$. Using the isometry of $\mathbb{H}^3$, $PSL(2,\mathbb{C})$, four asymptotic vertices can be placed at  $(y,w) = (0,0),(0,1),(0,z)$ and $(\infty, \cdot)$. Right: topologically, ideal tetrahedron is a tetrahedron with truncated vertices. } 
\label{fig:ideal-tetrahedron}
\end{figure}
\\
Geometrically, these edge parameters correspond to 
\begin{align}
&\{ Z,Z',Z''\}:=\{ \log z, \log z',\log z'' \}  \nonumber
\\
&=i(\textrm{dihedral angle between two faces meeting on the edge})+(\textrm{torsion}) \;.
\end{align}
Here ``torsion'' is a quantity which measures the twisting of hyperbolic metric around the edge. 
For an ideal tetrahedron in $\mathbb{H}^3$, these parameters satisfy
\begin{align}
\begin{split}
&Z_i+Z'_i+Z''_i=i \pi\; (\textrm{sum of angles in small boundary triangle equals to $\pi$})\;, 
\\
&e^{-Z_i}+e^{Z''_i}=1\;. \label{hyperbolic on Delta}
\end{split}
\end{align}
The second equation follows directly from the geometric definition of $(z,z',z'')$ as equivalent cross-ratios.
These constraints are compatible with the following cyclic symmetry of ideal tetrahedron:
\begin{align}
\mathbb{Z}_3 \;: \; (Z,Z',Z'')\rightarrow  (Z',Z'',Z)\rightarrow (Z'',Z,Z')\;. \label{Cyclc-in-Z-Zp-Zpp}
\end{align}

An hyperbolic structure on a knot complement $N$ can be obtained  by gluing the hyperbolic structure on each tetrahedron in a smooth way. 
For the smooth gluing, we need to impose the following conditions
\begin{align}
\begin{split}
C_I  &:= (\textrm{sum of all logarithmic edge variables associated to }
\\
& \quad \quad  \textrm{edges meeting at the $I$-th internal edge in the gluing})  
\\
& =\sum_{i=1}^{k}(G_{Ii}Z_i +G'_{Ii}Z'_i +G''_{Ii}Z''_i)\;, \quad G_{Ii},G'_{Ii},G''_{Ii}\in \{ 0,1,2\}\;.
\\
& =2\pi i \;. \label{internal edges}
\end{split}
\end{align}
There are $k$-internal edges in an ideal triangulation with $k$ ideal tetrahedra. A solution to these gluing equations  \eqref{hyperbolic on Delta} and \eqref{internal edges} with conditions $0<\textrm{Im}[Z_i]<\pi$ for all $i$ gives a  hyperbolic  (generally  incomplete) structure on $N$. 

\paragraph{$\CM(P_{SCFT})$ from ideal triangulation}  More generally, a solution to the exponentiated gluing equations  gives an irreducible flat $PSL(2,\mathbb{C})=SL(2,\mathbb{C})/\langle \pm 1\rangle$ connections on $N$. 
Consider the algebraic variety $\cD[N,\cT]$ determined by gluing equations of an ideal triangulation $\cT$,
\begin{align}
\cD[N,\cT]:=\{ z_i, z'_i, z''_i \in \mathbb{C}\backslash \{ 0,1\}: z_i z_i' z_i''=-1, z_i^{-1}+z''_i-1=0, \prod_{i=1}^k z_i^{G_{Ii}}(z'_i)^{G'_{Ii}} (z''_i)^{G''_{Ii}}=1\}\;.
\label{deformation variety}
 \end{align}
The variety  is called a {\it deformation variety}.  A point in $\cD[N,\cT]$ gives an irreducible flat-connection via a map $\chi_T$%
 \begin{align}
 \chi_{\cT} : \; \cD[N,\cT] \rightarrow \chi (N):=\{  \rho_{\rm hol} \in \textrm{Hom}\big{[}\pi_1 (N) \rightarrow PSL(2,\mathbb{C})\big{]}/(\textrm{conj}) \}\;. 
 \end{align}
where the definition of $\chi(N)$ here is equivalent to that in \eqref{eq:chidef1}.
The map $ \chi_{\cT}$ is injective but not surjective.  
Using the map, holonomy matrix along a primitive boundary  cycle $A\in H_1 (\partial N,\mathbb{Z}) = \pi_1 (\partial N) \subset \pi_1 (N)$ can be written as linear combinations of logarithmic edge parameters
\begin{align}
\begin{split}
& \rho_{\rm hol}(A) = \left(\begin{array}{cc}e^{a/2} & 0 \\ * & e^{-a/2}\end{array}\right) \quad \textrm{where }a=\sum_{i=1}^k (\alpha_i Z_i +\alpha''_i Z''_i)+ i \pi \epsilon 
\\
&\textrm{with integer coefficients }(\alpha_i, \alpha''_i ,  \epsilon)\;. \label{Hol(A) in terms of edges}
\end{split}
\end{align}
Dependence on $\{Z'_i\}$ was eliminated using the linear relations in  \eqref{hyperbolic on Delta}.
The algebraic variety depends on the choice of an ideal triangulation $\cT$ of $N$.  
But it is known that  the Dehn surgery component $\chi_0(N)=\bigcup_m \CM(P_{SCFT},m)$ 
in \eqref{Dehn-surgery com} is always contained in $\cD[N,\cT]$ for any $\cT$  except exotic cases when  $\cD[N,\cT]$ is empty \cite{tillmann2012degenerations},
\begin{align}
\begin{split}
\chi_0(N)=\; &\textrm{the connected component of $\cD[N,\cT]$ for non-exotic $\cT$  }
\\
&\textrm{containing a  solution  of gluing eqns corresponding to $\CA^{\rm \overline{hyp}}$} \;. \label{Dehn-surgery com-2}
\end{split}
\end{align}
We currently do not have the field theoretic understanding of the exotic case and will always work with non-exotic triangulations. 

\paragraph{$SU(2)/SO(3)$-type  of boundary cycle $A$} For later use, we classify a primitive boundary cycle $A$ into two types, $SU(2)$ or $SO(3)$, depending on evenness/oddness of the linear coefficients $(\alpha_i, \alpha''_i)$. 
\begin{align}
A \textrm{ is of }
 \begin{cases}
 \textrm{$SU(2)$-type}\;,\;\textrm{if all $(\alpha_i, \alpha''_i)$ can be chosen as even-integers}
 \\
\textrm{$SO(3)$-type}\;,\;\textrm{otherwise} 
\end{cases} \label{Def : SO/SU type}
\end{align}
Note that the linear coefficients are defined modulo the following shifts due to the last gluing equations in \eqref{deformation variety}
\begin{align}
 \bigg{(}\alpha_i, \alpha''_i \bigg{)}\rightarrow \bigg{(} \alpha_i + \sum_{i=1}^k c_{Ii}(G_{Ii}-G'_{Ii}), \alpha''_i + \sum_{i=1}^k c_{Ii} (G''_{Ii} - G'_{Ii})\bigg{)}\quad \textrm{with some integers $c_{Ii}$}\;,\nonumber
\end{align}
and $A$ is $SU(2)$-type if there is a choice of $c_{Ii}$ which makes all $(\alpha_i, \alpha''_i)$ even-integers. An  alternative definition of $SO(3)/SU(2)$ type without relying an ideal triangulation is 
\begin{align}
A \textrm{ is of }
 \begin{cases}
 \textrm{$SU(2)$-type}\;,\; A \in \textrm{Ker} \big{(} i_* : H_1 (\partial N, \mathbb{Z})\rightarrow H_1 (N, \mathbb{Z}_2) \big{)}
 \\
\textrm{$SO(3)$-type}\;,\;\textrm{otherwise} 
\end{cases} \label{Def : SO/SU type 2}
\end{align}
Two definitions, \eqref{Def : SO/SU type} and  \eqref{Def : SO/SU type 2}, are equivalent \cite{neumann1992combinatorics}.
An explanation of $SU(2)/SO(3)$ types from the 6d $\CN=(2,0)$ theory point of view is discussed in Appendix~\ref{app:susotypes}.

\paragraph{$T^{DGG}[N,X_A]$ from symplectic gluing} The gluing equations are known to have the following symplectic structure \cite{neumann1985volumes,neumann1992combinatorics} which play a crucial role in the DGG's construction. Upon a skew-symmetric bilinear $\{ \;, \;\}$ \textrm{defined by } $\{Z_i, Z'_j  \} =\{Z'_i, Z''_j  \} = \{ Z''_i, Z_j \} = \delta_{ij}$, internal edge variables $\{C_I\}$ and the boundary holonomy variable $a$ around $A$ satisfy the followings:  
\begin{align}
\{ C_I, C_J\} = \{ a , C_I\} =0\;, \quad \textrm{for all $I,J=1,\ldots, k$}
\end{align}
Further we can choose a linearly independent primitive cycle $B\in H_1 (\partial N, \mathbb{Z})$ such that 
\begin{align}
\{a, b\} =- 2\;, 
\end{align}
where $b$ is related to the holonomy along $B$ as in eq.~\eqref{Hol(A) in terms of edges}. The choice of $B$ is not unique but have the following freedom of choice
\begin{align}
B\rightarrow B+ k A\;, \quad k \in \mathbb{Z}\;.
\end{align}
Using the freedom, we will always choose $B$ to have the properties that
\begin{align}
B \textrm{ is of }
 \begin{cases}
 \textrm{$SU(2)$ type}\;,\;\textrm{when $A$ is of  $SO(3)$ type\;,}
 \\
\textrm{$SO(3)$ type}\;,\;\textrm{when $A$ is of $SU(2)$ type\;,} \label{types of (A,B)}
\end{cases}
\end{align}
where the $SU(2)/SO(3)$ types of $B$-cycle is defined in the same way as $A$.

Among $k$-internal edge variables in eq.~\eqref{internal edges},  only $k-1$ of them\footnote{More generally, for an ideal triangulation of  a knot/link complement $N$  with $\sharp_T$ torus boundaries the number of linearly independent internal edge variables are $k-\sharp_T$. } are linearly independent modulo linear relations in \eqref{hyperbolic on Delta}. Let the linearly independent set as $\{C_I\}_{I=1}^{k-1}$. Then, we introduce their conjugate variables $\{\Gamma_I\}_{I=1}^{k-1}$ satisfying
\begin{align}
\{ C_I, \Gamma_J\}= \delta_{IJ}\;, \quad \{ a, \Gamma_I\} = \{ b, \Gamma_I\}=0\;, \quad I,J=1,\ldots, k-1\;.
\end{align}
From the choice of $(A,B,\{ C_I\}, \{\Gamma_I\})$, we associate a $Sp(2k,\mathbb{Z})$ matrix $g_{N}$ and integer-valued $2k$-vector $\nu$ as follows
\begin{align}
&\left(\begin{array}{c}X_A \\C_1  \\ \ldots \\ C_{k-1}\\ P_B \\ \Gamma_1 \\ \ldots \\ \Gamma_{k-1}\end{array}\right) = g_{N} \cdot  \left(\begin{array}{c}Z_1 \\ Z_2  \\ \ldots \\ Z_{k}\\ Z_1'' \\ Z_2'' \\ \ldots \\ Z_{k}'' \end{array}\right) + i \pi \nu_{N}\;, \label{g_N}
\end{align}
where
\begin{align}
&(X_A, P_B) = 
\begin{cases}
 (\frac{a} 2,b)\;,\;\textrm{when $(A,B)$ is of  $\big{(}SU(2),SO(3)\big{)}$ type}
 \\
(a,\frac{b}2 )\;,\;\textrm{when $(A,B)$ is of $\big{(}SO(3),SU(2)\big{)}$ type} \end{cases} \label{(XA,PB)}
\end{align}
Notice that $(X_A, P_B)$ are always linear combinations of $Z_i, Z''_i$ with integer coefficients because of the even-ness condition \eqref{Def : SO/SU type}.

Using the gluing data summarized in $(g_N, \nu_N)$, we can construct the corresponding $T^{DGG}$ theory. 
As a first step, we prepare $k$-copies of a free chiral theory 
\begin{align}
\begin{split}
&T_{\textrm{step I}} = T_\Delta^{\otimes k} = \overbrace{T_\Delta \otimes \ldots \otimes T_{\Delta}}^{\textrm{$k$ times}}\;,
\\
& T_{\Delta}  := (\textrm{a free theory of single chiral multiplet $\Phi$ with  CS level $-1/2$}
\\
&  \quad \quad \quad \textrm{ for non-dynamical background gauge field coupled to $U(1)$ flavor symmetry})\;,
\\
&\mathcal{L}_{T_{\rm step 1}} (V_1,\ldots , V_k) = \sum_{i=1}^k \frac{1}{4\pi}\int d^4\theta (- \frac{1}2 \Sigma_{i} V_{i})   +\int d^4\theta \Phi_{i}^\dagger e^{V_{i}} \Phi_i\;,
\label{DGG : step 1}
\end{split}
\end{align}
where $\Sigma_i$ is the field strength of the vector multiplet $V_i$.
The theory $T_{\textrm{step I}}$ has $u(1)^k$ flavor symmetry and  $\{V_i\}$ are background vector-multiplets  coupled to the flavor symmetries. 

Using the symmetry, one can consider $Sp(2k,\mathbb{Z})$ action on the theory which is a generalization of Witten's $SL(2,\mathbb{Z})$ action \cite{Witten:2003ya} which corresponds to  $k=1$ case. To be more explicit, one needs to decompose a $Sp(2k,\mathbb{Z})$ into products of ``T-type ($g^t_K$),'' ``S-type ($g^s_{J}$),'' and ``GL-type($g^{gl}_U$)'': 
\begin{align}
g^t_K:= \left(\begin{array}{cc}I & 0 \\ K & I\end{array}\right)\ \;,  \quad g^s_J:= \left(\begin{array}{cc}I-J & -J \\ J & I-J\end{array}\right)\ \;,  \quad g^{gl}_U:= \left(\begin{array}{cc}U & 0 \\ 0 & (U^{-1})^t\end{array}\right)\ \;. \label{T,S and GL type}
\end{align}
Here $J$ is a diagonal matrix whose diagonal entries are either 0 or 1.
Let $\mathcal{L}_{T} (\vec{V}:=(V_1, \ldots, V_k))$ be a Lagrangian for a theory $T$ with $U(1)^k$ flavor symmetry. Field theoretic actions of the basic types  are
\begin{align}
\begin{split}
&\mathcal{L}_{g^t_K \cdot T} (\vec{V}) :=   \mathcal{L}_{T} (\vec{V}) + \frac{1}{4\pi } \int d^4\theta \vec{\Sigma} \cdot K \vec{V}\;,
\\
&\mathcal{L}_{g^s_J \cdot T} (\vec{V}) :=   \mathcal{L}_{T} ((I-J)\vec{V}+J\vec{V}') + \frac{1}{2\pi } \int d^4\theta \vec{\Sigma}\cdot J \vec{V}'\;,
\\
&\mathcal{L}_{g^{gl}_U \cdot T} (\vec{V}) :=   \mathcal{L}_{T} (U^{-1}\cdot\vec{V})\;,
\end{split}
\end{align}
where $\vec{V}'$ only has components such that $J\vec{V}'=\vec{V}'$, and they are now dynamical fields.
As for the $SL(2,\mathbb{Z})$ case, the final theory does not depend on the decomposition and depends only on the $Sp(2k,\mathbb{Z})$ element. 

Now the second step of the construction is 
\begin{align}
T_{\textrm{step II}} = g_N \cdot T_{\textrm{step I}}\;, \label{DGG : step 2}
\end{align}
where $g_N$ is the symplectic matrix in \eqref{g_N} obtained from an ideal triangulation of $N$. The $g_N$-transformed theory still has $u(1)^k$ flavor symmetry 
\begin{align}
U(1)_{X_A} \times U(1)_{C_1} \times \ldots \times U(1)_{C_{k-1}}\;,
\end{align}
whose background gauge fields are $V_{X_A}:=V_1,\ldots, V_{C_{k-1}}:=V_{k}$ in $T_{\textrm{step II}}$.

As a final step, we break the $U(1)^k$ to its subgroup by adding chiral operators 
to the superpotential 
\begin{align}
\mathcal{L}_{T^{DGG}[N,X_A]} = \mathcal{L}_{T_{\textrm{step II}}}+\bigg{(} \sum_{\textrm{`easy' $C_I$}}\int d^2 \theta  O_{C_I}+c.c \bigg{)}\;. \label{DGG : final}
\end{align}
An internal edge $C_I = \sum_{i=1}^k (G_{Ii}Z_i +G'_{Ii}Z'_i+G''_{Ii}Z''_i)$ in  \eqref{internal edges} is called `easy' \cite{Dimofte:2011ju} 
if at most one of $G_{Ii}$, $G'_{Ii}$ and $G''_{Ii}$ is nonzero for each $i$,
\begin{align}
\sum_{i=1}^k (G_{Ii}G'_{Ii}+G'_{Ii} G''_{Ii}+G''_{Ii}G_{Ii}) = 0\;
\end{align}
and `hard' otherwise. 
This condition simply means that only one of edge parameters ($Z_i,Z'_i$ and $Z''_i$) of $i$-th tetrahedron appears in $C_I$ for all $i=1\ldots k$. Upon a proper choice of cyclic relabeling \eqref{Cyclc-in-Z-Zp-Zpp} of edge parameters, we can  make such an internal edge $C_I$ as a linear combination of only $Z_i$s: 
\begin{align}
C_I = \sum_{i=1}^k \widetilde{G}_{Ii}Z_i\;, \quad G_{Ii}\in \{ 0, 1, 2\}\;.
\end{align}
Then, the gauge-invariant chiral primary operator $O_{C_I}$ in $T_{\textrm{step II}}$ is given by
\begin{align}
O_{C_I} = \prod_{i=1}^k \Phi_i^{\widetilde{G}_{Ii}}\;.
\end{align}
As will be explained below, different cyclic labelings  give different descriptions of $T_{\textrm{step II}}$ which are related by a sequence of basic dualities in \eqref{T=ST*(T)}. Therefore for each easy internal edge $C_I$, there is a chiral primary operator  $O_{C_I}$ which can be written as the above form in a  duality frame.  The operator is  charged only under $U(1)_{C_I}$. For each hard internal edge, on the other hand, there may only be a corresponding gauge invariant dyonic 1/4 BPS operator with non-zero spin.  There is no way to write down a supersymmetric deformation using the dyonic local operators. 
%
\paragraph{Hard internal edges and accidental symmetries } In the original DGG's construction \cite{Dimofte:2011ju}, they proposed to use ideal triangulations with only easy internal edges. From superficial counting, we expect the resulting $T^{DGG}[N]$ has   flavor symmetry of rank $1$ whose Cartan corresponds to the $U(1)_{X_A}$.
\begin{align}
\begin{split}
&\textrm{If all $C_I$ are easy, we superficially expect that}
\\
&U(1)_{X_A} \times U(1)_{C_1}\times \ldots \times U(1)_{C_{k-1}} \; \textrm{in $T_{\textrm{step II}}$}
\\
&  \xrightarrow[]{\textrm{\quad Superpotential deformation in \eqref{DGG : final} \quad } }   U(1)_{X_A} \; \textrm{in $T^{DGG}[N,X_A]$}
\end{split}
\end{align}
The counting sounds compatible with the 6d construction since the knot gives a flavor symmetry $(su(2))$ of rank 1. But the counting  could be wrong as we will see below  for the case with an ideal triangulation of $N=(\textrm{figure-eight knot complement})$ with $6$-tetrahedra. The correct  rank is always equal or greater than the superficial counting.  
In our modified proposal \eqref{DGG : final}, we can use any ideal triangulation  and  will argue that the resulting  theory is independent of the choice of ideal triangulation regardless of existence of hard edges. One of the consequences is that  rank of the flavor symmetry could be larger than $1$ because the number of independent easy edges  could be less than $(k-1)$.  From the counting of linearly independent easy internal edges, we checked that $T^{DGG}$ theories for most of knot complements in SnapPy's census have additional symmetries. 

For example, we show  the $SU(3)$ symmetry for all hyperbolic twist knots in section  \ref{sec:SU(3)}.
The  additional symmetries are accidental and unexpected from 6d viewpoint. 
The above DGG's construction can be  generalized to higher $\mathcal{K}$ (number of M5-branes) cases \cite{Dimofte:2014zga} and  there is no such an additional symmetry when  $\mathcal{K}$ is sufficiently large. 
For higher $\CK$ one need to use a so-called $\mathcal{K}$-decomposition which replace a single tetrahedron in an ideal triangulation into  $\frac{1}6 \mathcal{K}(\mathcal{K}^2-1)$ copies of finer building blocks, octahedra. The construction of the 3d theory for higher $\mathcal{K}$ is parallel to the construction for $\mathcal{K}=2$ case reviewed above except tetrahedra in an ideal triangulation are replaced by  octahedra in a $\mathcal{K}$-decomposition. 
We  assign 3 complex parameters ($z,z',z''$) to each pair of two vertices of an octahedron and their gluing equations  in a $\CK$-decomposition also possess a symplectic structure. One  difference in higher $\mathcal{K}$ is that there are enough number of easy internal edges (better to call internal vertices for $\mathcal{K}$-decomposition case) to break all $u(1)$ symmetries except the ones expected from 6d viewpoint. 6d viewpoint expect that the 3d theory has a flavor symmetry of rank $(\mathcal{K}-1)$.  A hard internal edge  appears when two edges of a single tetrahedron are glued to the internal edges simultaneously. In  $\mathcal{K}$-decomposition, two different vertices of a single octahedron can not meet at an internal vertex possibly except when the octahedron is located nearest  to one of  vertices of tetrahedrons.  So the number of hard internal vertices will be at most order of $k$ (the number of tetrahedrons in a triangulation) while there are $k \frac{\mathcal{K}(\mathcal{K}^2-1)}6$ internal vertices among which  $(\mathcal{K}-1)$ are linearly dependent. So the number of easy internal vertices are  $k \big{(} \frac{\mathcal{K}(\mathcal{K}^2-1)}6- o(1) \big{)}$ which is large enough to span the  $\big{(} k \frac{\mathcal{K}(\mathcal{K}^2-1)}6 - (\mathcal{K}-1)  \big{)}$-linearly independent internal vertices for sufficiently large $\mathcal{K}$.

\paragraph{Topological invariance of $T^{DGG}[N,X_A]$}  At first glance, the above construction seems to depend on the various choices other than $(N,A)$. For the construction, we  choose  an ideal triangulation of $N$.   
All different ideal triangulations of a given 3-manifold are known to be related by sequence of a basic local move called 2-3 Pachner move. In the DGG's construction, the geometric move corresponds to a mirror symmetry  between a 3d $\mathcal{N}=2$ SQED with two chirals $(\Phi_A, \Phi_B)$ of charge $(+1, -1)$ and a free theory with 3 chirals $(M, T_p, T_m)$:
\begin{align}
\begin{split}
&\int d^4 \theta  \bigg{(}\Phi_A^\dagger e^{V} \Phi_A +\Phi_B^\dagger e^{-V+U} \Phi_B +\frac{1}{4\pi} (U+ 2W) (\Sigma_{V}-\frac{1}2 \Sigma_U)\bigg{)} \quad (\textrm{$ V$ : dynamical})
\\
&\simeq \int d^4 \theta \bigg{(}   M^\dagger e^{U} M + (T_p)^\dagger e^{W} T_p + (T_m)^\dagger e^{-U-W} T_m \bigg{)} +\big{(}  \int d^2 \theta M T_p T_m +c.c \big{)}
\label{3d mirror-1}
\end{split}
\end{align}
Under the duality, gauge-invariant chiral operators  are mapped as follows
\begin{align} 
\Phi_A \Phi_B  \qquad \qquad \qquad \qquad  \qquad \quad &\leftrightarrow&  \qquad \qquad M  \nonumber
\\
V_+ \; (\textrm{BPS monopole operator of magnetic flux $+1$}) \qquad &\leftrightarrow &  T_p \label{3d mirror-2}
\\
V_- \; (\textrm{BPS monopole operator of magnetic flux $-1$}) \qquad &\leftrightarrow &  T_m \nonumber
\end{align}
So the $T^{DGG}$ theory is invariant under the local 2-3 move and thus independent on the choice of $\cT$.
For a given choice of $\cT$, we still have freedoms of choosing cyclic labeling \eqref{Cyclc-in-Z-Zp-Zpp} of edge parameters for each tetrahedron. 
\begin{align}
\begin{split}
 &\left( {\begin{array}{c}
   Z  \\
   Z''  \\
  \end{array} } \right) \; \rightarrow \;  \left( {\begin{array}{c}
   Z'  \\
   Z  \\
  \end{array} } \right)  = ST\cdot  \left( {\begin{array}{c}
   Z  \\
   Z''  \\
  \end{array} } \right) +  i \pi \left( {\begin{array}{c}
   1 \\
   0  \\
  \end{array} } \right)\;,
 \\
 & S := \left( {\begin{array}{cc}
   0 & -1 \\
   1 & 0 \\
  \end{array} } \right)\;, \quad 
  T := \left( {\begin{array}{cc}
   1 & 0 \\
   1 & 1 \\
  \end{array} } \right)\;.
  \end{split}
\end{align}
The invariance $T^{DGG}$ theory under choice is guaranteed from a duality 
\begin{align}
T_\Delta \simeq (ST) \cdot T_\Delta\;. \label{3d mirror-2}
\end{align}
More explicitly, the duality is 
\begin{align}
\begin{split}
&\int d^4 \theta \big{(} \Phi^\dagger e^{U} \Phi -\frac{1}{8\pi} U \Sigma_U \big{)}
\\
&\simeq \int d^4 \theta  \big{(}\Phi^\dagger e^{V} \Phi + \frac{1}{8\pi} V \Sigma_V + \frac{1}{2\pi} U \Sigma_V \big{)} \;, \quad (\textrm{$ V$ : dynamical})\;. \label{T=ST*(T)}
\end{split}
\end{align}
In the construction of $T^{DGG}$ theory, we also need to choose conjugate variables $\{ P_B, \Gamma_I\}$. But these choices only affect the background Chern-Simons coupling coupled to flavor symmetries. So modulo the background CS couplings, the theory only depends on the topological choice $(N,A)$.  To specify the background Chern-Simons coupling of the $U(1)_{X_A}$ flavor symmetry associated to the knot, we sometimes specify the choice of boundary cycle $B$ and denote the theory by 
\begin{align}
T^{DGG}[N,X_A ;P_B]\;.
\end{align}

\begin{figure}[h]
\begin{center}   \includegraphics[width=.35\textwidth]{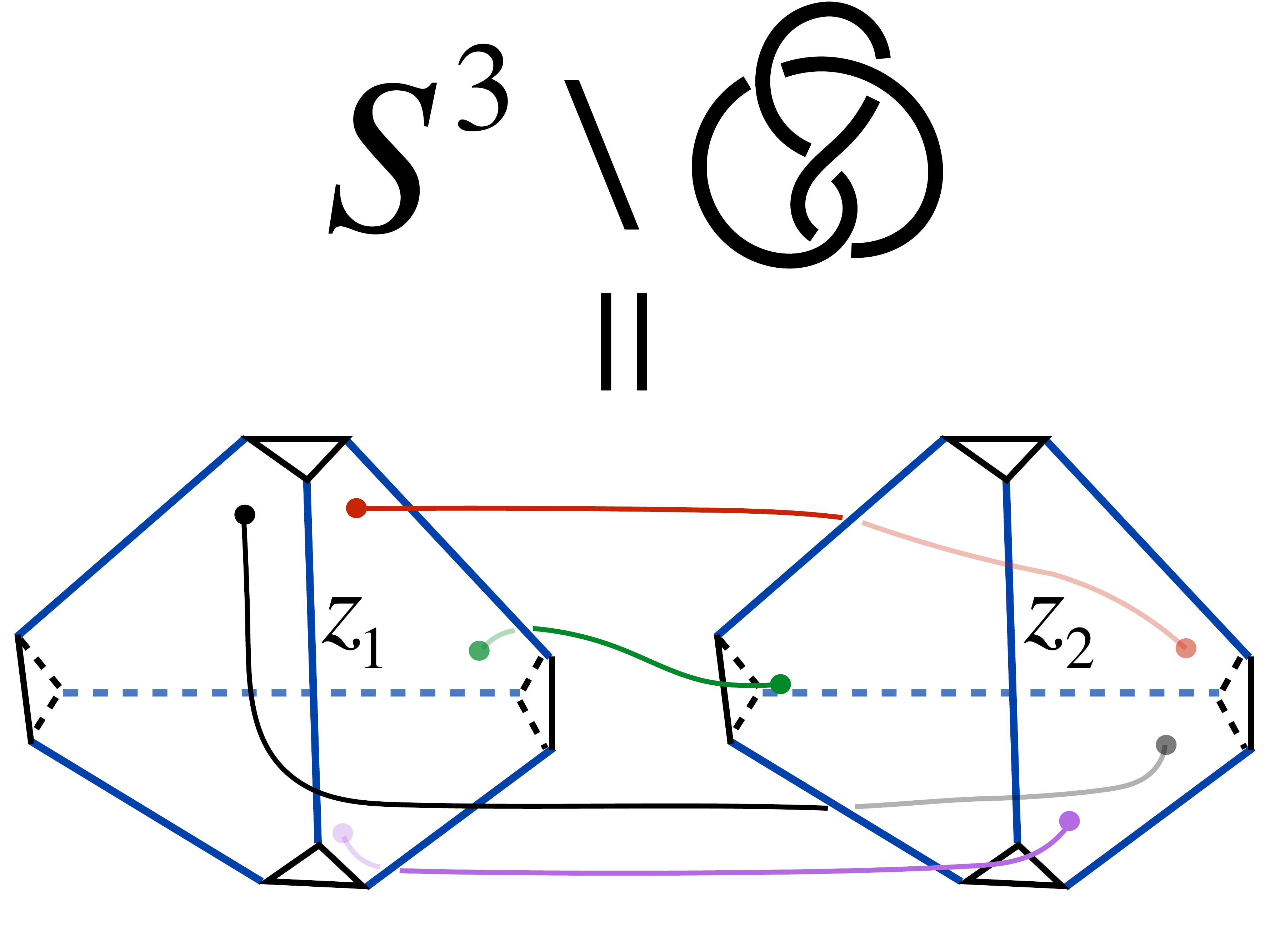}
   \end{center}
   \caption{The simplest ideal triangulation of $m004=S^3\backslash \mathbf{4_1}$.} 
   \label{fig:ideal-triangulation-41}
\end{figure}

\paragraph{Example : $N=S^3 \backslash \mathbf{4}_1 =m004$ with an ideal triangulation with 2 tetrahedra} Here $\mathbf{4}_1$ is a simplified notation, called Alexander-Briggs notation, for  figure-eight knot which is depicted in fig \ref{fig:ideal-triangulation-41}.  The notation simply means that the figure-eight knot is the 1st (simplest) knot with 4 crossings.  The fundamental group of the knot complement is 
\begin{align}
\pi_1 (S^3\backslash \mathbf{4_1}) = & \langle \alpha, \beta,\gamma : \alpha \gamma^{-1} \beta \alpha^{-1}\gamma = \beta \gamma^{-1}\beta^{-1}\alpha= 1 \rangle\;.
\end{align}
 The group contains a  peripheral subgroup $\mathbb{Z}\times \mathbb{Z}$  which can be identified as fundamental group of boundary torus
\begin{align}
\begin{split}
&\pi_1 \big{(}\partial (S^3 \backslash \mathbf{4}_1)\big{)} = \pi_1 (\mathbb{T}^2) = \mathbb{Z} \times \mathbb{Z}=  \langle \mu , \nu \rangle \subset \pi_1 (S^3\backslash \mathbf{4}_1)
\end{split}
\end{align}
Canonical choice of the basis $(\mu, \lambda)$ is   (meridian, longitude). Upon the basis choice, the embedding $i : \pi_1 \big{(}\partial (S^3 \backslash \mathbf{4}_1)\big{)}  \rightarrow \pi_1 (S^3 \backslash \mathbf{4}_1)$ is given by 
\begin{align}
i (\mu) = \alpha \;, \quad i (\nu) = \alpha \gamma^{-1}\beta \gamma \alpha^{-1}\beta^{-1}
\end{align}
The knot complement can be  ideally triangulated by two tetrahedrons. 
\begin{align}
\begin{split}
\cT \; : \;  S^3 \backslash \mathbf{4_1} =(\Delta_1 \cup \Delta_2)/\sim\;.
\end{split}
\end{align}
See fig.~\ref{fig:ideal-triangulation-41} below for the gluing rule $\sim$.
There are two internal edges in the triangulation which are linearly dependent modulo the linear equations in \eqref{hyperbolic on Delta}.
\begin{align}
C_1 =Z_1''+Z_2'+2Z_1'+2Z_2\;, \quad C_2 = Z_1''+Z_2'+2Z_1+2Z_2''\;. \label{internal edges of 4_1}
\end{align}
The deformation variety in this example is
\begin{align}
\begin{split}
\mathcal{D}[S^3\backslash \mathbf{4_1},\cT] = & \{ z_1,z_1', z_1'',z_2,z_2',z_2'' : z_i^{-1}+z_i''-1=0,\; z_i z_i'z_i''=-1,\; z_1'' (z_1')^2  z_2' z_2^2 =1\}_{i=1,2}\;. 
\label{deformation variety-m004}
\end{split}
\end{align}
Each point in the variety gives a $PSL(2,\mathbb{C}) = SL(2,\mathbb{C})/\langle \pm 1 \rangle$ flat connection on the knot complement. 
The holonomy matrices along the basis ($\alpha, \beta,\gamma$) of $\pi_1 (S^3\backslash \mathbf{4_1})$ for the flat connections is 
\begin{align}
\begin{split}
& \rho_{\rm hol}(\alpha)= \left(
\begin{array}{cc}
 \sqrt{\frac{z_1}{z_2}} & 0 \\
 \frac{-1+z_1}{\sqrt{z_1 z_2}} & \sqrt{\frac{z_2}{z_1}} \\
\end{array}
\right) \;,
\\
& \rho_{\rm hol}(\beta)= \left(
\begin{array}{cc}
 \sqrt{\frac{z_2'}{z_1''}} &- \sqrt{\frac{z_2'}{z_1''}} \\
 \sqrt{z_1'' z_2'} & \, -\sqrt{\frac{z_1''}{z_2'}}\left(z_2'-1\right) \\
\end{array}
\right) \;,
\\
& \rho_{\rm hol}(\gamma)= \left(
\begin{array}{cc}
 \frac{z_1'+z_2''-1}{\sqrt{z_1' z_2''}} & \frac{1-z_1'}{\sqrt{z_1' z_2''}} \\
 \frac{z_2''-1}{\sqrt{z_1' z_2''}} & \frac{1}{\sqrt{z_1' z_2''}} \\
\end{array}
\right) \;.
\end{split}
\label{Hol-m004-1}
\end{align}
Boundary (meridian, longitudinal) holonomies  are
\begin{align}
\begin{split}
& \rho_{\rm hol}(\mu)=  {\rm Hol}(\alpha) = \left(\begin{array}{cc}e^{a_\mu /2}& 0 \\ * & e^{-a_\mu /2}\end{array}\right)\;, \quad a(\mu) =  Z_1-Z_2\;,
\\
& \rho_{\rm hol} (\nu) = {\rm Hol} (\alpha \gamma^{-1 }\beta \gamma \alpha^{-1}\beta^{-1})=  \left(\begin{array}{cc}e^{b_\lambda/2}& 0 \\ * & e^{-b_\lambda/2}\end{array}\right)\;, \quad b_\lambda = 2(Z_1-Z_1')\;.
\label{Hol-m004-2}
\end{split}
\end{align}
So, $(\mu,\lambda)$ is of $(SO(3),SU(2))$ type and we choose
\begin{align}
X_{\mu} = a_\mu = Z_1-Z_2\;, \quad P_{\lambda} = b_\lambda/2 = Z_1-Z_1'\;.
\end{align}
With the choices, the $Sp(4,\mathbb{Z})$ matrix $g_{m004}$ in \eqref{g_N} is given by
\begin{align}
g_{m004} = \left(
\begin{array}{cccc}
 1 & -1 & 0 & 0  \\
 -2 & 1 & -1 & -1 \\
 2 & 0 & 1 & 0 \\
 0 & 1 & 0 & 0 \\
\end{array}
\right)\;,
\end{align}
which can be decomposed into $g_{m004} = g^s_{J_{m004}} g^t_{K_{m004}} g^{gl}_{U_{m004}}$  with \eqref{T,S and GL type}
\begin{align}
U_{m004} = \left(
\begin{array}{cc}
 1 & -1  \\
 0 & 1  \\
\end{array}
\right)\;, \quad K_{m004} = \left(
\begin{array}{cc}
 2 & 2 \\
 2 & 1  \\
\end{array}
\right)\;, \quad J_{m004} = \left(
\begin{array}{cc}
 0 & 0  \\
 0 & 1  \\
\end{array}
\right)\;.
\end{align}
Following each steps in eq.~\eqref{DGG : step 1},\eqref{DGG : step 2} and \eqref{DGG : final}, $T^{DGG}[m004,X_\mu;P_\lambda]$ is given by 
\begin{align}
\begin{split}
&\mathcal{L}_{T_{\rm step 1}} (V_1, V_2) = \mathcal{L}_{T^{\otimes 2}_\Delta}
\\
&=\frac{1}{4\pi}\int d^4\theta (- \frac{1}2 \Sigma_{1} V_{1} - \frac{1}2 \Sigma_{2} V_{2} ) +\int d^4\theta (\Phi_{1}^\dagger e^{V_{1}} \Phi_1+\Phi_{2}^\dagger e^{V_{2}} \Phi_2)\;,
\\
&\mathcal{L}_{T^{DGG}[m004, X_\mu ;P_\lambda]} (V_{X}, V_C) = \mathcal{L}_{T_{\rm step 2}} (V_{X}, V_C) 
\\
&=\frac{1}{4\pi} \int d^4 \theta \left(- \frac{1}2 \Sigma_C V_X + \Sigma (2V_C + 3 V_X) \right) + \int d^4 \theta \left( \Phi_1^\dagger e^{V+\frac{V_X}2}\Phi_1 +\Phi_2^\dagger e^{V- \frac{V_X}2}\Phi_2 \right)\;.\label{TDGG[m004]}
\end{split}
\end{align}
Here $V(\textrm{with }\Sigma:=\bar{D}D V)$ is a dynamical $U(1)$ vector multiplet while $V_X(\Sigma_X)$ and $V_C(\Sigma_C)$ are background multiplets coupled to flavor symmetries, $U(1)_{X_A}$ and say $u(1)_C$ respectively. Note that both of $C_1$ and $C_2$ are hard internal edges and we can not break the $u(1)_C$ associated to them. 

\paragraph{Example : $N=S^3 \backslash \mathbf{4}_1 =m004$ with an ideal triangulation with 6 tetrahedra}
The absence of chiral primary operators corresponding to hard edges in the above construction of $T^{DGG}[m004, X_\mu ;P_\lambda]$
using 2 tetrahedra were already noticed in \cite{Dimofte:2011ju}.  
The interpretation there was that this is due to the ``bad" choice of triangulation, which contains hard internal edges, 
and can be cured by choosing a proper ideal triangulation which does not have a hard internal edge. 
As a ``good" ideal triangulation for $m004$, they propose the one using six tetrahedra, $\Delta_{R,S,X,Y,Z,W}$ . 
The internal edges in the triangulation are \cite{Dimofte:2011ju}
\beq
\begin{array}{ll}
C_1 =X+W +2(R' +S' +Z''), \qquad & C_2 =R+Y +2(Z' +W' +S''),  \\
C_3=S+W+2(R''+X''+Y'), \qquad & C_4=R+Z+2(Y''+W''+X'),  \\
C_5=X+Y, \qquad &C_6=S+Z. 
\end{array}
\eeq
Note that there is no hard internal edges in the triangulation and 5 internal edges are linearly independent. 
Superficial counting suggests that the resulting theory have a flavor symmetry of rank $6 -5 = 1$, where five $u(1)$s are broken by superpotential operators.

Our interpretation on this problem is different from \cite{Dimofte:2011ju}. We claim that the theory realized by six tetrahedra is 
actually completely the same as the one realized by two tetrahedra in the low energy limit. Therefore, the theory constructed by six tetrahedra 
has a hidden additional $u(1)$ symmetry in the low energy limit which corresponds to the hard edge in the triangulation with two tetrahedra.

To see it, let us focus on the two tetrahedra $\Delta_X$ and $\Delta_Y$, which are glued in such a way that the system has the internal edge $C_5=X+Y$.
Then, this theory is described by two chiral fields $\Phi_X$ and $\Phi_Y$ with the Lagrangian
\beq
\int d^4\theta (\Phi_X^\dagger \Phi_X+\Phi_Y^\dagger \Phi_Y) + \int d^2\theta \Phi_X \Phi_Y + {\rm h.c.},
\eeq
where we have neglected background fields. The superpotential is due to the presence of the internal edge $C_5=X+Y$.
Then it is clear that these fields $\Phi_X$ and $\Phi_Y$ can be integrated out and the theory becomes empty in the low energy limit.
This means that two tetrahedra $\Delta_X$ and $\Delta_Y$ are eliminated. 
Mathematically this corresponds to the 0-2 move. The invariance of a topological quantity called  3d index (see appendix~\ref{sec : 3d index}) under the 0-2 move is proven in  \cite{2013arXiv1303.5278G}. The definition  of the topological quantity is based on ideal triangulation and is   equivalent to the localization expression for the  superconformal index of  $T^{DGG}$ theory. 
Intuitively, the constraints $0 < \Im[X], \Im[Y] < \pi$ and $C_5=2\pi i $ mean that $\Im[X], \Im[Y] \to \pi$,
and hence these tetrahedra are squashed to be flat. 
The same comment also applies to $\Delta_S$, $\Delta_Z$ and $C_6=S+Z$.

At the level of edge variables, the process of integrating out the massive fields may be done by 
eliminating the variables corresponding to the massive fields. More explicitly, we define
\beq
&C'_1:=C_3+C_4+2C_5-C_6 - 4\pi i=W+R+2W''+2R'' \\
&C'_2:=C_1+C_2 -C_5+2C_6 - 4\pi i=W+R+2W'+2R', 
\eeq
After renaming $W \to Z_1'',~R \to Z_2'$ and so on, these variables $C'_1$ and $C'_2$ become the same
as the ones in the triangulation with two tetrahedra.

\subsection{Relation between the two constructions}\label{sec:rel}
One basic characteristic of the $T^{DGG}[N,X_A]$ theory is that \cite{Dimofte:2011ju}
\begin{align}
\begin{split}
&\mathcal{M}_{\rm parameter}(\textrm{$T^{DGG}[N,X_A]$ on $\mathbb{R}^2\times S^1$}) =\cD[N,\cT] \supseteq  \chi_0(N) \;.
\end{split} \label{eq:moduliID}
\end{align}
Recall the definition of each term of this equation. For simplicity, we only discuss the case where our 3-manifold $N$
only has a torus boundary and hence of the form $N=M \backslash K$.
The deformation variety $\cD[N,\cT] $ defined in \eqref{deformation variety} is a set of flat $PSL(2,\mathbb{C})$ connections on $N$ 
which can be obtained from an ideal triangulation $\cT$. The $\chi_0(N)$  is  a subset of the algebraic variety  defined in   \eqref{Dehn-surgery com} (or \eqref{Dehn-surgery com-2})  which can be seen for any non-exotic ideal triangulation. 
The difference between the two sets are mild, higher codimension, and may be ignorable in our discussion as we discuss later.
Finally the left-hand side $\mathcal{M}_{\rm parameter}(\textrm{$T^{DGG}[N,X_A]$ on $\mathbb{R}^2\times S^1$}) $ is given as follows.
A DGG theory in general consists of chiral fields, dynamical vector fields, and background vector fields.
Let $\Sigma_i~(i=1,\ldots, N_V)$ be twisted chiral fields constructed from dynamical vector multiplets 
whose lowest real component is the real scalar of the vector multiplet
and the imaginary part is the gauge field in the $S^1$ direction. 
The $N_V$ is the number of dynamical vector multiplets, i.e., the gauge group is $u(1)^{N_V}$, and it depends on the details of $g_N$ in \eqref{DGG : step 2} and its decomposition into basic types.
Also, let $X_A$ be the twisted chiral field of the background $u(1) $ field whose real part corresponds to the real mass parameter $m$ and the imaginary part corresponds
to the background flavor gauge field around $S^1$. 
Then, by integrating out the matter chiral fields of the theory on $\BR^2 \times S^1$, we
get a twisted superpotential of $\Sigma_i$ and $X_A$ (in some appropriate normalization),
\beq
\widetilde{\mathcal{W}}( \{\Sigma_i\}_{i=1}^{N_V} ; X_A). 
\eeq
Then we define
\begin{align}
\begin{split}
&\mathcal{M}_{\rm parameter}(\textrm{$T^{DGG}[N,X_A]$ on $\mathbb{R}^2\times S^1$})
\\
&=  \big{\{} 
(e^{X_A},e^{P_B}) : \exp \big{(}  \partial_{\Sigma_k}\widetilde{\mathcal{W}}( \{\Sigma_i\}_{i=1}^{N_V} ; X_A) \big{)}=1, 
\partial_{X_A}\widetilde{\mathcal{W}}( \{\Sigma_i\}_{i=1}^{N_V} ; X_A) \big{)}=P_B
\big{\}} 
\backslash \{ \textrm{singular loci}\}.
\end{split}
\end{align}
The conditions $\exp \big{(}  \partial_{\Sigma_k}\widetilde{\mathcal{W}}( \{\Sigma_i\}_{i=1}^{N_V} ; X_A) \big{)}=1$ are just the 
condition for the vacua on $S^1 \times \BR^2$. The $P_B$ has a definite value (modulo $2\pi i$) at each of the vacua for a given parameter $X_A$.
In other words, the equation $\partial_{X_A}\widetilde{\mathcal{W}}( \{\Sigma_i\}_{i=1}^{N_V} ; X_A) \big{)}=P_B$ gives a polynomial equation
of $(x_A, p_B):=(e^{X_A},e^{P_B})$, and solutions of that equation in terms of $p_B$ for a given $x_A$ correspond to the vacua of the theory with mass parameter $x_A$.

%
%



The relation to localization computation is as follows. The partition function of the $T^{DGG}$ theory on a curved background called squashed 3-sphere 
$(S^3_b)$ can be written in  following form \cite{Hama:2011ea,Dimofte:2011ju}
\begin{align}
\int d\sigma_1 \ldots d\sigma_{N_V} \;\mathcal{I}_b( \{\sigma_i\}_{i=1}^{N_V}  ;X_A) 
\end{align}
where $\sigma_k = \Re[\Sigma_k]$.
In a degenerate limit when $b\rightarrow 0$, which corresponds to the limit where $S^3_b$ become $\mathbb{R}^2 \times S^1$, the leading asymptotic behavior of the integrand is determined by the twisted superpotential 
\begin{align}
\mathcal{I}_b (  \{\sigma_i\}_{i=1}^{N_V} ;X_A)  \xrightarrow[]{\quad b\rightarrow 0 \quad }  
e^{\frac{1}{2\pi i b^2} \widetilde{\mathcal{W}}(  \{\sigma_i\}_{i=1}^{N_V} ;X_A) }\;.
\end{align}
The  equations $\{ \exp (\partial_{\Sigma_i}\widetilde{\mathcal{W}})=1\}_{i=1}^{N_V}$ are equivalent 
to the gluing equations in \eqref{deformation variety} 
with an additional relation $e^{a}=(-1)^{\epsilon}\prod z_i^{\alpha_i} (1-z_i^{-1})^{\alpha''_i}$ where $a=X_A~\text{or}~2X_A$ depending on $SO(3)/SU(2)$ types of boundary cycle $A$,
and the integers $(\alpha_i, \alpha''_i,\epsilon)$ are given in \eqref{Hol(A) in terms of edges} \cite{Dimofte:2012qj}. 

Now, we have
\beq
&\mathcal{M}_{\rm vacua}(\textrm{$T^{DGG}[N,X_A]$ on $\mathbb{R}^2\times S^1$})  \nonumber \\
=& \mathcal{M}_{\rm parameter}(\textrm{$T^{DGG}[N,X_A]$ on $\mathbb{R}^2\times S^1$}) |_{a=2m}. \label{eq:DGGvacua}
\eeq
This means that by taking the parameter $a$ to be a constant fixed value $2m$, we get the vacua of the theory with the mass parameter $m$.

The above equations may have solutions like $X_A=0$. Field theoretically, when the mass parameter is zero,
there could appear some continuous moduli space of vacua spanned by matter chiral fields. Those massless flat directions are subtle, especially when they are generated by
monopole operators because in that case those directions appear by very strong coupling effects which may not be captured by the one-loop computation of 
the twisted superpotential $\widetilde{\mathcal{W}}$. See Sec.~5.2 of \cite{Dimofte:2011ju} for an example.
We may expect that those subtle flat directions might be the reason of the mismatch between $\cD[N,\cT] $ and $\chi_0(N)$
This problem may be avoided if we only consider generic mass parameters. We assume that this is the case.

\paragraph{Comparison of the two constructions}
Now let us compare the constructions in Sec.~\ref{sec:6d} and Sec.~\ref{sec:DGG}.
Comparing the moduli space of $T^{6d}$ in \eqref{flat connections}, and $T^{DGG}$ in \eqref{eq:DGGvacua} we see that 
\begin{align}
\mathcal{M}_{\rm vacua}(\textrm{$T^{DGG}[N,X_A]$ on $\mathbb{R}^2\times S^1$}) \subset \mathcal{M}_{\rm vacua}(\textrm{$T^{6d}[N,A]$ on $\mathbb{R}^2\times S^1$})\;.
\end{align}
This is because that an ideal triangulation captures only a subset of irreducible flat connections on $N$ as emphasized in \cite{Chung:2014qpa}. So we see that $T^{DGG}$ can not be identical to  $T^{6d}$ but can only capture a subsector of $T^{6d}$. This point has already been seen
from the effective field theory point of view in Sec.~\ref{sec:6d}. In general, there is no reason to expect that there exists a genuine 3d theory
which describes all components of moduli space of vacua of $T^{6d}$. So $T^{DGG}$ can, at best, describe the low energy limit of 
some point of the moduli space of vacua of $T^{6d}$.

Then a possibility is that $T^{DGG}$ might be identified with $T^{6d}_{\rm irred}$ in \eqref{eq:defof6dtheory}.
Both of them are genuine 3d theories and they are associated to the hyperbolic connection $\CA^{\rm \overline{hyp}}$ which can be realized in
ideal triangulation. However, it turns out that these two theories are still different as we now explain.

One crucial difference between the two theories is that  $T^{DGG}$ generically has $U(1)_{X_A}$ flavor symmetry associated to the knot while 
$T^{6d}$ and hence $T^{6d}_{\rm irred}$ have $su(2)_A$. Furthermore, the $SL(2,\mathbb{Z})$ action on the canonical variables $(X_A, P_B)$ is realized in field theory
as the $SL(2,\mathbb{Z})$ action of Witten~\cite{Witten:2003ya} using $u(1)$ group on $T^{DGG}$,
while the $SL(2,\mathbb{Z})$ action on the boundary cycle $(A,B)$ is realized in field theory as the $SL(2,\mathbb{Z})$ of
Gaiotto-Witten~\cite{Gaiotto:2008ak} using the $su(2)$ symmetry and $T[SU(2)]$ theory on $T^{6d}$. 

The $su(2)$ $SL(2,\mathbb{Z})$ on $T^{6d}$ is defined as follows. The transformed theory $\varphi \cdot T^{6d}[N,A,B]$ with 
\begin{align}
\varphi =  \left(\begin{array}{cc}r & s \\ p & q\end{array}\right)\; \in SL(2,\mathbb{Z})
\end{align}
can be obtained by 
\begin{align}
\begin{split}
&T^{6d}[N,rA+sB;pA+qB] =  \varphi \cdot T^{6d}[N,A;B]
\\
&:=  \textrm{coupling the duality wall theory $T[SU(2),\varphi]$ to $T^{6d}[N,A;B]$ }\;.
\end{split}
\end{align}
$T[SU(2),\varphi]$ is a 3d $\cN=4$ SCFT which describe the 3d theory  living on a duality domain wall  in 4d $su(2)$ $\cN=4$ SYM associated to  $\varphi \in SL(2,\mathbb{Z})$. The theory has $su(2)_1 \times su(2)_2$ as flavor symmetry. For example, $T[SU(2), \varphi=S] = T[SU(2)]$. In the coupling between $T[SU(2),\varphi]$ and $T^{6d}[N,A;B]$, we introduce a $\cN=2$ vector multiplet to gauge the diagonal $su(2)_{\rm diag}\subset su(2)_{1} \times su(2)_A$ of the two theories with the following superpotential coupling
\begin{align}
\textrm{Tr}(\mu \mu')\;. \label{Superpotential coupling between T[SU(2)] and T[N]}
\end{align}
where $\mu$ is the holomorphic moment map operator associated to the $su(2)_A$ of $T^{6d}[N,A;B]$
which is a chiral operator in the adjoint representation of $su(2)_A$,\footnote{In general, the existence of this operator is guaranteed only for theories with 8
supercharges. However, this operator often exists due to the remnant of 8 supercharges preserved by codimension-2 defects of the 6d $\CN=(2,0)$ theory.
Indeed, the operator $\mu$ is absent only for some special cases. These points will be important and discussed in more detail below.}
and $\mu'$ is the holomorphic moment map operator of $su(2)_{1}$ of $T[SU(2),\varphi]$.

%
%
%

Motivated by the similarities and differences between $T_{\rm irred}^{6d}$ and $T^{DGG}$, we propose the following relation between them:
\begin{align}
T^{6d} \xrightarrow[\text{}]{\textrm{\quad on a vacuum $P_{\rm SCFT}$} \quad} T^{6d}_{\rm irred} \xrightarrow[\text{}]{\textrm{\quad deformed by $\delta W = \mu^3$} \quad} T^{DGG} \label{Relation-T6d-TDGG}
\end{align}
Here, $\mu^3$ is the Cartan component of the moment map operator ${\mu}=\{\mu^1,\mu^2,\mu^3\}$ in the adjoint representation of $su(2)_A$.
The $\delta W = \mu^3$ means the superpotential deformation by the chiral operator $\mu^3$. This deformation breaks $su(2)$ to $U(1)$.

Thus we need two steps from $T^{6d}$ to  $T^{DGG}$. 
First we put the theory on a specific vacuum, $P_{\rm SCFT}$, of the $T^{6d}$ theory on $\mathbb{R}^3$ as explained in Sec.~\ref{sec:6d}. 
%
%
%
As a second step, we  deform the intermediate theory, $T^{6d}_{\rm irred}$, by adding the Cartan component ($\mu^3$) of the $su(2)$ moment map operator 
$\mu$ associated to the knot to the superpotential. We give more evidence for this proposal below.

In the case of 3d $\CN=4$ supersymmetry, the presence of the holomorphic moment map operator associated to a symmetry is guaranteed.
However, when there are only $\CN=2$, it is not guaranteed. What we call the holomorphic moment map operator is a kind of remnant of higher supersymmetry
of the codimension-2 defect of the 6d $\CN=(2,0)$ theory. The $\mu$ may be empty depending on the theory.
However, generically (but not always), the $T^{6d}_{\rm irred}$ theory contains the moment map as chiral operators.
This can be seen as follows. Suppose we are given a theory $\CT$ with $su(2)$ symmetry, and 
then let us perform a transformation 
\beq
ST^k \in SL(2,\BZ) \label{eq:STk}
\eeq acting on this $su(2)$. This is done
by coupling $\CT$ and $T[SU(2)]$ to the $su(2)$ gauge field with Chern-Simons level $k (+\text{ the original value before }T^k)$.
By taking $k$ large enough, the $su(2)$ gauge field is weakly coupled and hence the two theories $\CT$ and $T[SU(2)]$ almost decouple from each other.
The $T[SU(2)]$ theory contains the holomorphic moment map operator $\mu$ associated to the new (ungauged) $su(2)$ global symmetry.
Therefore, we conclude that the total theory has $\mu$ for generic $k$. Moreover, this argument shows that
the scaling dimension of $\mu$ is close to 1, at least if $k$ is large enough, because the scaling dimension of $\mu$ in $T[SU(2)]$ alone is 1 by $\CN=4$
supersymmetry. Therefore, the deformation by $\mu^3$ is a relevant deformation
and it triggers RG flows. The case that $\mu$ is absent happens only in rather exceptional situations, 
and this will be very important in Sec.~\ref{sec:symm}

In particular, the deformation breaks the $su(2)_A$ in $T_{\rm irred}^{6d}[N,A]$ to $U(1)$ which can be identified with $U(1)_{X_A}$ in $T^{DGG}[N,X_A]$. 
\begin{align}
\big{(} su(2)_A   \textrm{ of $T^{6d}_{\rm irred}$}\big{)}\;\xrightarrow[\text{}]{\quad \delta W = \mu^3 \quad }\; \big{(} U(1)_{X_A} \textrm{ of $T^{DGG}$}\big{)}\;.
\end{align}

\paragraph{More details on $SL(2,\BZ)$ transformations of $U(1)$ and $SU(2)/SO(3)$ types.}
The deformation explains not only why $T^{DGG}$ theory generically has only $U(1)_{X_A}$ associated to the knot, but also why the $SL(2,\mathbb{Z})$ action on the boundary $\mathbb{T}^2$ of the knot complement corresponds to the $U(1)$ $SL(2,\mathbb{Z})$ action on $U(1)_{X_A}$. 
Roughly, the relation is given by the following diagram;
\begin{align}
\xymatrix{
T^{6d}_{\rm irred}[N,A;B] \ar[d]^{su(2) \; SL(2,\mathbb{Z})} \ar[r]^{ \delta W = \mu^3 \;\; } & T^{DGG}[N,X_A;P_B]\ar[d]^{U(1)\; SL(2,\mathbb{Z})}\\
T^{6d}_{\rm irred}[N,A';B'] \ar[r]^{ \delta W = \mu^3 \;\;} &T^{DGG}[N,X_{A'}; P_{B'}]}
\end{align}
However, there are more subtle details.

We denote the $U(1)$-type and $su(2)$-type $SL(2,\BZ)$ transformations as $SL(2,\BZ)_1$ and $SL(2,\BZ)_2$, respectively.
The $SL(2,\BZ)_1$ acts on the canonical variables $(X_A, P_B)$, while $SL(2,\BZ)_2$ acts on $(A,B)$ and hence on the variables $(a,b)$.
Recall that they are related as
\beq
(X_A, P_B) = 
\begin{cases}
 (\frac{a} 2,b)\;,\;\textrm{when $(A,B)$ is of  $\big{(}SU(2),SO(3)\big{)}$ type}
 \\
(a,\frac{b}2 )\;,\;\textrm{when $(A,B)$ is of $\big{(}SO(3),SU(2)\big{)}$ type} \end{cases}
\eeq
Therefore, generic elements of $SL(2,\BZ)_1$ and $SL(2,\BZ)_2$ do not exactly correspond to each other.

The $S$-transformation $S_1 \in SL(2,\BZ)_1$ and $S_2 \in SL(2,\BZ)_2$ correspond with each other;
\beq
S_1 &: (X_A,P_B) \to (-P_B, X_A) \nonumber \\
S_2 &: (a,b) \to (-b,a) 
\eeq
and hence
\beq
S_1  \longleftrightarrow S_2. \label{eq:Srelation}
\eeq
Notice that the $S_2$ exchanges the $SU(2)/SO(3)$ types of the cycles.
This is natural, because in 4d $\CN=4$ theory, the gauge groups $SU(2)$ and $SO(3)$ are exchanged under the S-duality.
This exchange can also be shown in purely 3d language and is explained in Appendix~\ref{app:susotypes}.

On the other hand, the $T$-transformation $T_1 \in SL(2,\BZ)_1$ and $T_2 \in SL(2,\BZ)_2$ act as
\beq
T_1 &: (X_A,P_B) \to (X_A, P_B+X_A) \nonumber \\
T_2 &: (a,b) \to (a, b+a) 
\eeq
and hence they are related as
\beq
(T_1)^2  \leftrightarrow T_2 \qquad \textrm{when $(A,B)$ is of  $\big{(}SU(2),SO(3)\big{)}$ type} \label{eq:Trelation1} \\
T_1 \leftrightarrow (T_2)^2 \qquad \textrm{when $(A,B)$ is of  $\big{(}SO(3),SU(2)\big{)}$ type} \label{eq:Trelation2}
\eeq
This also has a natural field theory interpretation. 
Let $A$ be an $su(2)$ gauge field. The global structure may be either $SU(2)$ or $SO(3)$.
Its Chern-Simons 3-form is defined as
\beq
{\rm CS}(A) =\frac{1}{4\pi} \tr (AdA +\frac{2}{3}A^3),
\eeq
where the trace is taken in the doublet representation of $su(2)$.
When the symmetry is $SU(2)$ and is broken down to $U(1)$, we embed a $U(1)$ gauge field $a$ inside $A$ as 
$A = \diag(a, -a)$. Then, under this embedding, we get
\beq
{\rm CS}(A) \to \frac{2}{4\pi} ada = 2 {\rm CS}(a) \label{eq:U1SU2}
\eeq
where 
\beq
 {\rm CS}(a)=\frac{1}{4\pi} ada.
\eeq
Recalling that the $T$-transformation in field theory corresponds to the shift of Chern-Simions level of background field,
we can see that the factor of 2 in \eqref{eq:U1SU2} corresponds to the exponent 2 in \eqref{eq:Trelation1}.
In the same way, if the gauge group is $SO(3)$ which is broken to $U(1)$, it is natural to embed the $U(1)$ gauge field
as $A=\frac{1}{2}(a, -a)$. Then we get 
\beq
{\rm CS}(A) \to \frac{1}{2 \cdot 4\pi} ada = \frac{1}{2} {\rm CS}(a).
\eeq
This equation corresponds to \eqref{eq:Trelation2}.

In fact, if the group is $SO(3)$ type, then the ${\rm CS}(A)$ is not a properly quantized Chern-Simons invariant.
This is because, in 4d, there can be instantons of instanton number $1/2$,\footnote{
To see this, consider a 4d manifold $ S^2_1 \times S^2_2$.
Then, take the subgroup $U(1) \subset SO(3)$ by $A=\frac{1}{2}(a, -a)$, and include the magnetic fluxes of $f = da$
on each $S^2_1$ and $S^2_2$ as $n_1 = \frac{1}{2\pi}\int_{S^2_1} f$ and $n_2 = \frac{1}{2\pi}\int_{S^2_2} f$. 
One can check that this configuration gives the instanton number $\frac{1}{2}n_1 n_2$.}
and hence 
the integral of ${\rm CS}(A)$ is not well defined in 3d, and only the $2{\rm CS}(A)$ is well defined. This means that when the $A$-cycle is $SO(3)$ type,
only the $(T_2)^2$ is well defined. Therefore, the actual transformation group at the quantum level is a subgroup $\Gamma(2) \subset SL(2,\BZ)_2$
generated by $S_1$ and $(T_2)^2$. These generators $S_1$ and $(T_2)^2$ also preserves the condition that one of the 
cycles $A$ or $B$ is $SU(2)$ type. Under $T_2$, this condition may not be preserved.

From the above discussion of Chern-Simons levels, the field theoretical realization of \eqref{eq:Trelation1} and \eqref{eq:Trelation2} under the explicit breaking 
$su(2)_A \to U(1)_{X_A}  $ by $W =\mu^3$ is clear. Now let us also check the correspondence of the S-transformation \eqref{eq:Srelation} at the field theory level.
Let $\CT$ be a theory with $su(2)$ symmetry. Then $S$-transformed theory is given by
\beq
S_2 \cdot \CT = \CT - su(2) - T[SU(2)],\label{eq:StrasfT}
\eeq
where the center $su(2)$ is a gauge group which is coupled to the $su(2)$ symmetry of $\CT$ and the $su(2)_H$ symmetry of $T[SU(2)]$.
In the language of 3d $\CN=2$ supersymmetry, the $T[SU(2)]$ is given by a $U(1)$ vector multiplet $V$, a neutral chiral field $\phi$,
and two pairs of chiral fields $(E^i,\tilde{E}_i)_{i=1,2}$ with $u(1)_{\rm gauge}$ charge $\pm$ with the superpontial
\beq
W = \phi \tilde{E}_i E^i.
\eeq 
The $su(2)_H$ acts on the index $i$ of $(E^i,\tilde{E}_i)_{i=1,2}$. 
Now, this $T[SU(2)]$ has a $su(2)_C$ symmetry at the quantum level, and this symmetry is the new global $su(2)$ symmetry after the $S$-transformation.
See Appendix~\ref{app:susotypes} for more details.
The Cartan component $\mu^3$ of the moment map operator of this symmetry $su(2)_C$ is given by $\mu^3=\phi$.
Therefore, after the deformation by $\mu^3$, the superpotential becomes
\beq
W = \phi \tilde{E}_i E^i -\phi. 
\eeq
Thus, an F-term condition is $\tilde{E}_i E^i =1$. Then $\tilde{E}_i $ and $ E^i$ get nonzero expectation values as
\beq
\langle E^1 \rangle =\langle \tilde{E}_1 \rangle=1,~~~\langle E^2 \rangle =\langle \tilde{E}_2 \rangle=0.
\eeq
These expectation values break the gauge symmetry $[su(2) \times u(1)]_{\rm gauge}$ down to $u(1)$.
Namely, $(E^i,\tilde{E}_i)$ are charged under $u(2) = [su(2) \times u(1)]_{\rm gauge}$, and only the subgroup $u(1) \subset u(2)$
which acts on $(E^2, \tilde{E}_2)$ is preserved by the expectation values.
Therefore, by the Higgs mechanism, the theory \eqref{eq:StrasfT} becomes
\beq
S_2 \cdot \CT \longrightarrow S_1 \cdot \CT = \CT - u(1),
\eeq
where $u(1)$ is the gauge group which survives the symmetry breaking $[su(2) \times u(1)]_{\rm gauge} \to u(1)$.
The right hand side is just the $S$-transformation of $u(1)$ type. This confirms \eqref{eq:Srelation}.

\section{Symmetry enhancement  }\label{sec:symm}

Using the proposed 6d interpretation of $T^{DGG}[N,X_A]$ in \eqref{Relation-T6d-TDGG}, we will determine the  symmetry enhancement pattern of  the $U(1)_{X_A}$ symmetry associated to the knot based on a topological type of the boundary cycle $A$. 
See the Table \ref{Sym enhancement in DGG} for the summary whose details will be explained in Sec.~\ref{sec:en}.
\begin{table}[h!]
\centering
\begin{tabular}{|c || c | }
\hline
$A \in H_1 \big{(}\partial N,\mathbb{Z}\big{)}$ & Symmetry enhancement of $U(1)_{X_A}$  in $T^{ DGG}[N,X_A]$ 
\\
\hline
\hline
closable &  $U(1)_{X_A} \rightarrow U(1)$  \\\hline
non-closable, $SO(3)$ type   & $U(1)_{X_A} \rightarrow SO(3)$    \\
\hline
non-closable, $SU(2)$ type  & $U(1)_{X_A} \rightarrow SU(2)$   \\
\hline
\end{tabular}
\caption{Symmetry enhancement in $T^{DGG}[N,X_A]$. The definitions of ``$SU(2)/SO(3)$ types'' are explained in Sec.~\ref{sec:DGG}.
The definitions of ``closable/non-closable'' are explained in Sec.~\ref{sec:en}.}
\label{Sym enhancement in DGG}
\end{table}

In Sec.~\ref{sec:SU(3)}, 
we will find infinitely many examples of pair $(N,A)$ and $(N',A')$ whose corresponding DGG theories are identical and both of $A$ and $A'$ are non-closable but 
one of them (say $A$) is $SO(3)$ type while the other ($A'$) is $SU(2)$ type. In that case, combining the table \ref{Sym enhancement in DGG} with a group theoretical argument,  we can argue\footnote{We thank Y. Tachikawa for this argument.} that the DGG theory has enhanced $SU(3)$ symmetry. Using the argument, we prove that $T^{DGG}[M=S^3, K]$ theories for all hyperbolic twist knots $K$ have $SU(3)$-symmetry. We checked the enhancement for several twist knots which gives  non-trivial empirical evidence for the Table \ref{Sym enhancement in DGG}.

\subsection{$SO(3)/SU(2)$ enhancement}\label{sec:en}
From the relation between $T_{\rm irred}^{6d}$ and $T^{DGG}$ in  \eqref{Relation-T6d-TDGG}, we understand the symmetry breaking mechanism of $su(2)_A$  to $ U(1)_{X_A}$. For the symmetry breaking to happen, we need the $su(2)$ moment map operator $\mu$ in $T^{6d}_{\rm irred}$ theory. Otherwise, the $su(2)_A$ is not broken by the mechanism and the resulting $T^{DGG}$ is expected to have an $su(2)$ symmetry. 
Therefore, we need to know when $\mu$ is absent.

This question can be answered by an inspection of the 5d picture discussed in Sec.~\ref{sec:6d}.
In the description there,
the moment map operator comes from the holomorphic $su(2)_C$ moment map of $T[SU(2)]$ theory which is put along a knot $K$. 
After $S^1$ compactification, the $T[SU(2)]$ has 2d $\CN=(4,4)$ supersymmetry. 
In the Language of $\CN=(2,2)$ supersymmetry, there is a twisted chiral operator $\widetilde{\mu}$ and a chiral operator $\mu$,
\footnote{
Let $\phi$ be the neutral scalar chiral field and let $\Sigma$ be twisted chiral field which comes from $u(1)$ vector multiplets in 2d.
Then the Cartan components of $\widetilde{\mu}$ and $\mu$ are given by $\Sigma$ and $\phi$.
However, in the discussion of Sec.~\ref{sec:6d}, we have to exchange the role of chiral and twisted chiral by regarding $\phi$ as twisted chiral and 
$\Sigma$ as chiral, because of the subtle mirror symmetry in the relation between 5d SYM and $T^{6d}$ in \eqref{eq:comm}. 
}
both of which are associated to the Coulomb branch $su(2)_C$ symmetry of $T[SU(2)]$.

Now suppose that the Higgs branch operator $\widetilde{\nu}$ gets a nonzero expectation value.
Then, the nonzero VEV in Higgs branch makes the Coulomb branch fields massive. Therefore, in the low energy limit,
the operators $\mu$ and $\widetilde{\mu}$ become empty;
\begin{align}
\textrm{In $T[SU(2)]$ theory on $\langle \tilde{\nu} \rangle \neq 0$, $\mu$ is absent at low-energy}
\end{align}

In the coupled system (5d SYM+ $T[SU(2)]$), as shown in  \eqref{HolA and nu}, the VEV of the moment operator $\widetilde{\nu}$ is given by 
$\log  \rho_{\rm hol} (A)$  of complexified gauge field $\cA$. 
The $\cA$ is a flat connection in $\chi_0[N]$ defined in \eqref{Dehn-surgery com} and \eqref{Dehn-surgery com-2}. 
Thus the above relation implies  that the $T^{6d}_{\rm irred}$ theory on $\mathbb{R}^2\times S^1$  does not contain  $\mu$ if there is no $PSL(2,\mathbb{C})$ flat-connection in 
$\chi_0[N]$ with trivial $ \rho_{\rm hol}(\CA)$. Let us define 
\begin{align}
\begin{split}
&\textrm{A primitive boundary cycle $A\in H_1 (\partial N, \mathbb{Z})$ is `closable'}
\\
&\textrm{if there is a point in $\chi_0[N]$ with $ \rho_{\rm hol}(A)= 1$\;.} \label{Def : closable cycle}
\end{split}
\end{align}
We remark that in the massless case $m=0$, the eigenvalues of $ \rho_{\rm hol}(A)$ are trivial ($\pm 1$), but $ \rho_{\rm hol}(A)$ may contain off-diagonal components
and hence the above condition is nontrivial.
Then,
\begin{align}
\begin{split}
&\textrm{$A \in H_1 (\partial N, \mathbb{Z})$ is non-closable } 
\\
& \Rightarrow \; \textrm{$\mu$ is absent in $T^{6d}_{\rm irred} [N,A]$ on $\mathbb{R}^2 \times S^1$ for any radius of $S^1$ at low energy} 
\\
& \Rightarrow \; \textrm{$\mu$ is absent in $T^{6d}_{\rm irred} [N,A]$ on $\mathbb{R}^3$ at low energy } 
\\
&\Rightarrow \; T^{DGG}[N,X_A]= T^{6d}_{\rm irred}[N,A] \textrm{ has $su(2)$ symmetry at low energy}\;.
\end{split}
\end{align}
Strictly speaking, the step from $\BR^2 \times S^1$ to $\BR^3$ is nontrivial, but we assume that this step holds.

One  necessary condition for $A$ to be `non-closable' is that the Dehn filled manifold $N_A$ is non-hyperbolic.
\begin{align}
\textrm{If $N_A$ is hyperbolic} \; \Rightarrow \; \textrm{$A$ is closable}\;.
\end{align}
This is because that the flat connection corresponding to the hyperbolic structure on $N_A$ is always contained in $\chi_0[N]$ with trivial $ \rho_{\rm hol}(A)$. 
According to Thurston's hyperbolic Dehn surgery theorem, for given hyperbolic $N$, there are only finite number of primitive boundary cycles $A$ which give non-hyperbolic $N_A$. So, we can conclude that
\begin{align}
|\{\textrm{Set of  primitive `non-closable' boundary cycles $A \in H_1 (\partial N ,\mathbb{Z})$} \}| < \infty\;.
\end{align}
Combining with Table~\ref{Sym enhancement in DGG}, it implies that the $u(1)_{X_A}$ symmetry of $T^{DGG}[N,X_A]$ is not enhanced  to $su(2)$ except for only 
finite  many $A$s. The Thurston's theorem is consistent with our field theoretical consideration in the previous section that the moment map operator $\mu$
generically (although not always) exists; see the discussion in the paragraph containing \eqref{eq:STk}.

One sufficient condition for $A$ to be `non-closable' is that the Dehn filled manifold $N_A$ is Lens-space
\begin{align}
\textrm{If $N_A$ is Lens space} \; \Rightarrow \; \textrm{$A$ is non-closable}\;. \label{non-closable : Lens}
\end{align}
Lens space $L(p,q)$ is defined as
\begin{align}
L(p,q) := \big{(}S^3\backslash (\textrm{unknot}) \big{)}_{p\mu+ q\lambda}
\end{align}
The reason is as follows.
If $A$ is closable, by definition, there should be an irreducible flat connection in $\chi_0[N]$ with trivial $ \rho_{\rm hol}(A)$. Such a  flat connection can be thought as an irreducible flat connection on $N_A$. But if $N_A$ is a Lens space, there can not be any  irreducible flat connection because the fundamental group $\pi_1$ of Lens space is abelian.
Thus the cycle $A$ can not be closable.  When $N_A$ is neither hyperbolic nor a Lens space, no simple criterion to determine the closability has been found. 

An alternative definition of closable/non-closable cycle, which seems to be equivalent to the above definition, is using 3d index which is introduced  in \cite{Dimofte:2011py} as  a topological invariant of  3-manifolds with torus boundaries and is generalized in  Appendix \ref{sec : 3d index} to cover closed 3-manifolds.
\begin{align}
\begin{split}
&\textrm{A primitive boundary cycle $A\in H_1 (\partial N, \mathbb{Z})$ is closable (non-closable)}
\\
&\textrm{if $\mathcal{I}_{N_A}(x) \neq 0$\;($\mathcal{I}_{N_A}(x) = 0$)\;.} 
\end{split}
\end{align}
Here $\mathcal{I}_{N_A}(x)$ is the 3d index on a closed 3-manifold $N_A$. That a primitive boundary cycle $A \in H_1 (\partial N, \mathbb{Z})$ is `non-closable' means that we can not `close' (or eliminate) the co-dimension two defect  along a $K$ on $M$ in a supersymmetric way after sitting on the vacuum $P_{\rm SCFT}$.   As we will study in the next section, there is an operation in SCFT side of 3d/3d correspondence which corresponds to the operation of  `closing  the knot'. If $A$ is non-closable cycle, we expect that the resulting 3d theory $T^{6d}_{\rm irred}[M]$ after taking the closing  knot  operation on $T^{6d}_{\rm irred}[M,K]$ will be a theory with supersymmetry broken.\footnote{Here notice the difference between $T^{6d}_{\rm irred}[M,K]$ and $T^{6d}[M,K]$. 
The  $T^{6d}[M]$ theory after removing knot still have a supersymmetric vacuum because there is  always trivial flat connection on any closed 3-manifold $M$. The trivial flat connection on $M$ disappears  in the moduli space $\CM_{\rm vacua}(T^{6d}_{\rm irred}[M,K] \textrm{ on }\mathbb{R}^2\times S^1)$  after put on the vacua $P_{SCFT}$.  } 
The index $\mathcal{I}_{M=N_A}(x)$  computes the superconformal index of the theory  $T^{6d}_{\rm irred}[M]$ and expected to be zero when  $A$ is non-closable and thus supersymmetry is broken. This is a heuristic argument supporting the equivalence between the two definitions and no rigorous mathematical proof is known. We checked the equivalence for various examples and the equivalence seems to hold possibly except  for exotic cases. As an example, see Table \ref{Boundary cycles of 4-1} for the case when $N=S^3\backslash \mathbf{4_1}=m004$.
\begin{table}[h!]
\centering
\begin{tabular}{|c || c || c |}
\hline
$A \in H_1 (\partial N, \mathbb{Z})$  & \textrm{Closability } & $\cI_{N_{A}}(x)$\\
\hline
\hline
$p\mu + \lambda \; ( |p| \geq 5)$& closable ($\Leftarrow$ $N_A$ is hyperbolic)  & non-trivial power series in $x$ \\\hline
$p\mu + \lambda \; ( |p| = 4)$   & closable  &  divergent  \\
\hline
$p\mu + \lambda \; ( |p|< 4)$ & closable & 1 \\
\hline
$\mu$ & non-closable & 0 \\
\hline
\end{tabular}
\caption{Closability of boundary cycles of $N=S^3\backslash \mathbf{4_1}$. The $\mu$ is meridian and $\lambda$ is longitude. 
We determine the closability using the definition in \eqref{Def : closable cycle} and the $\cD[N,\cT]$ in \eqref{deformation variety-m004} with \eqref{Hol-m004-1} and \eqref{Hol-m004-2}} \label{Boundary cycles of 4-1}
\end{table}

We can further determine the global structure of enhanced symmetry, whether $SO(3)$ or $SU(2)$, from the $SO(3)/SU(2)$ type of $A$. 
When $A$ is non-closable and  of $SU(2)$ type, the compact $U(1)_{X_A}$ symmetry  of $T^{DGG}[N,X_A]$ is embedded into the enhanced $su(2)_A$ symmetry via $\mathbf{2}_{su(2)_A}\rightarrow (\pm 1)_{U(1)_{X_A}}$. This is manifest from  the relations given in eq.~\eqref{Hol(A) in terms of edges} and \eqref{(XA,PB)}  between the variable $X_A$, associated to the $U(1)_{X_A}$, and  the $PSL(2,\mathbb{C})$ holonomy variables $a$, associated to the $su(2)_A$. 
Namely, $\mathbf{2}_{su(2)}$ has properly quantized $U(1)_{X_A}$ charges and the theory can have operators charged under half integer spin representations of $su(2)_A$ which means that the symmetry is $SU(2)$. Similarly we can see that only operators in integer spin representation are allowed when $A$ is of $SO(3)$ type.
See also Appendix~\ref{app:susotypes} for more justifications from different arguments. 

In general, it is not easy to determine the $SO(3)/SU(2)$ type of a given primitive boundary cycle $A$ in $H_{1}(\partial N, \mathbb{Z})$. When $N$ is a knot complement in a homological sphere, there is a canonical choice of the basis of $H_{1}(\partial N, \mathbb{Z})$, meridian $(\mu)$ and longitude ($\lambda$). Meridian cycle is defined to be the circle around the knot and longitude cycle is determined by the condition that $\lambda \in \textrm{Ker}\big{(}i_* : H_1 (\partial N, \mathbb{Z}) \rightarrow H_1 (N, \mathbb{Z})\big{)}$.  Then, $\lambda$ is of $SU(2)$-type by definition while $\mu$ is  always of $SO(3)$-type.  More generally, when $N$ is a knot complement in a $\mathbb{Z}_2$-homological sphere ($p$ and $q$ are coprime)
\begin{align}
p\mu+q \lambda \textrm{ is of }
 \begin{cases}
 \textrm{$SU(2)$-type}\;,\; \textrm{for even $p$ }\;,
 \\
\textrm{$SO(3)$-type}\;,\;\textrm{for odd $p$}\;. \label{SO(3)/SU(3) type in Z2-homology sphere}
\end{cases} 
\end{align}
Here $\mu\in H_1 (\partial N, \mathbb{Z})$ is the meridian cycle and  $\lambda$ is a boundary cycle  in $ \textrm{Ker}\big{(}i_* : H_1 (\partial N, \mathbb{Z}) \rightarrow H_1 (N, \mathbb{Z}_2)\big{)}$.  The choice of $\lambda$ is not unique but can be shifted by $2\mu$.

\paragraph{Example : $N=S^3\backslash \mathbf{4_1}= m004$ and $A =\mu$} In the case, the merdian cycle $\mu$ is non-closable and of $SO(3)$-type and we expect  $SO(3)$ symmetry enhancement of $u(1)_{X_{\mu}}$ in $T^{DGG}[m004, X_{\mu}]$ whose Lagrangian is give in eq.~\eqref{TDGG[m004]}. In the next section, we  argue that $u(1)_{X_\mu}$ is actually enhanced to $SU(3)$ which contain the $SO(3)$ as a subgroup.

\paragraph{Example : $N= \overline{(S^3\backslash \mathbf{5^2_1})_{3\mu_1 - 2 \lambda_1}}= m007$ and $A =\mu_2$} As another example, we consider a knot complement called $m007$ in SnapPy's census. The knot complement can be obtained by performing Dehn filling on one component of Whitehead link complement.  
 \begin{figure}[h]
\begin{center}   \includegraphics[width=.35\textwidth]{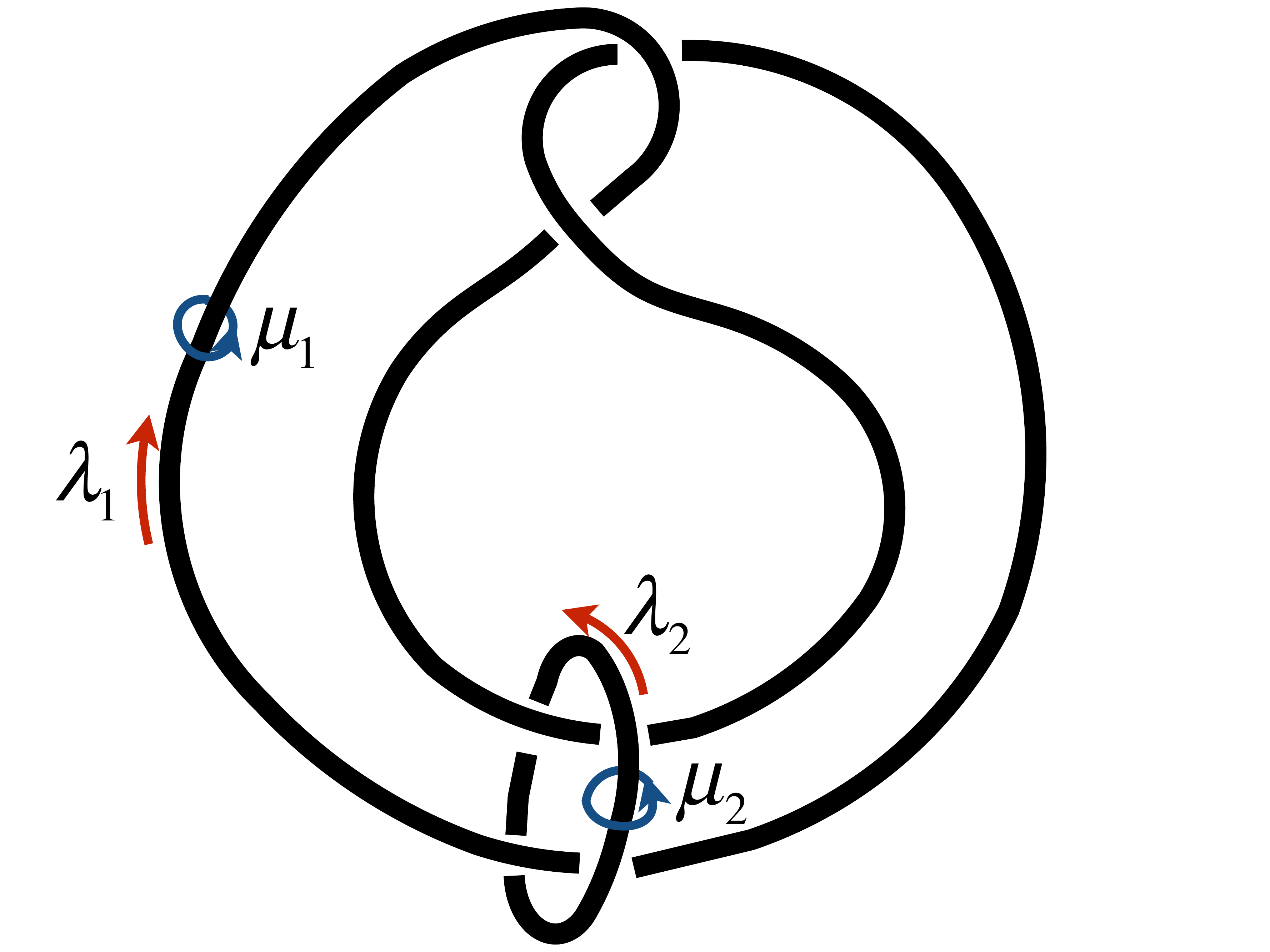}
   \end{center}
   \caption{ Whitehead link ($\mathbf{5}^2_1$). It is one of the simplest (having smallest volume $\simeq 3.664$) two-component hyperbolic link.}   
   \label{fig:Whitehead}
\end{figure}
Whitehead link is denoted by $\mathbf{5}^2_1$, the 1st link with 2 components and 5 crossings, as shown in Fig.~\ref{fig:Whitehead}. 
The orientation of the link complement ($S^3\backslash \mathbf{5}^2_1$) is chosen as the one induced from  an ideal triangulation in \eqref{Gluing eqns for Whitehead}.     We always choose a particular orientation of each ideal tetrahedron in an ideal triangulation which is reflected in the choice of  CS level sign of $T_{\Delta}$ in \eqref{DGG : step 1}. Then, the Dehn filled manifolds  have natural orientation induced from the link complement. The overline in the equation $N= \overline{(S^3\backslash \mathbf{5^2_1})_{3\mu_1 - 2 \lambda_1}}$ means that
$N=m007$ has opposite orientation to the one induced from $S^3\backslash \mathbf{5}^2_1$ when the orientation of $N$ is chosen to be the one induced from an ideal triangulation  in \eqref{triangulation of m007}.
The Dehn filling also gives an  induced basis of $H_1 (\partial N, \mathbb{Z})= \langle \mu_2, \lambda_2 \rangle$ on $N$ from the basis choice of $H_1 (\partial (S^3\backslash \mathbf{5^2_1}),\mathbb{Z}) = \langle \mu_1, \mu_2 , \lambda_1, \lambda_2 \rangle$. From the topological fact that
\begin{align}
N_A = (S^3 \backslash \mathbf{5^2_1})_{3\mu_1- 2 \lambda_1, \mu_2} =  (S^3\backslash ({\rm unknot}))_{3\mu- 2 \lambda}= L(3,-2)\;,
\end{align}
we see that $A$ is non-closable according to \eqref{non-closable : Lens}.  The $N$ is a knot complement in a $\mathbb{Z}_2$-homological sphere, $N_A = L(3,-2)$, and according to \eqref{SO(3)/SU(3) type in Z2-homology sphere} $A$ is of $SO(3)$ type.  
So from Table \ref{Sym enhancement in DGG}, we expect the $u(1)_{X_{\mu_2}}$ in $T^{DGG}[m007, X_{\mu_2}]$ is  enhanced to $SO(3)$. Now, let us check the enhancement from explicit construction of the DGG theory. 
According to SnapPy, the knot complement can be triangulated by 3 ideal tetrahedra and the corresponding gluing data are (we choose $B=4\mu_2-\lambda_2$)
\begin{align}
\begin{split}
&C_1 = Z_1 +2Z_2+Z_3\;, \quad C_2 = Z_1+Z_1''+Z_2'+Z_3+Z_3''\;,
\\
&C_3 = 2Z_1'+Z_1''+Z_2'+2Z_2''+2Z_3'+Z_3''\;,
\\
& a_{\mu_2 } = -Z_1 -Z_1''-Z_2+Z_3''+i\pi\;,
\\
&b_{4\mu_2 -\lambda_2} = 2 (- i \pi +Z_1+Z_2) \label{triangulation of m007}
\end{split}
\end{align}
Since $(A,B)$ are of $(SO(3),SU(2))$-type, we choose
\begin{align}
X_{\mu_2} =  a_{\mu_2 }\;, \quad P_{4\mu_2  - \lambda_2} = \frac{1}2 b_{4\mu_2 - \lambda_2}\;.
\end{align}
Then, the symplectic matrix in \eqref{g_N} for this example is 
\begin{align}
\begin{split}
& g_{m007} = \left(
\begin{array}{cccccc}
 -1 & -1 & 0 & -1 & 0 & 1 \\
 1 & 2 & 1 & 0 & 0 & 0 \\
 1 & -1 & 1 & 1 & -1 & 1 \\
 1 & 1 & 0 & 0 & 0 & 0 \\
 0 & 0 & 0 & 0 & 0 & 1 \\
 0 & 1 & 0 & 0 & 0 & 0 \\
\end{array}
\right)\;.
\end{split} 
\end{align}
The matrix can be decomposed into $g_{m007} = g^s_{J_{m007}} g^t_{K_{m007}} g^{gl}_{U_{m007}}$ with \eqref{T,S and GL type}
\begin{align}
U_{m007} = \left(
\begin{array}{ccc}
 1 & 1 & 0 \\
 0 & 2 & 1 \\
 0 & 1 & 0 
\end{array}
\right)\;, \quad K_{m007} = \left(
\begin{array}{ccc}
 1 & 0 & 0 \\
 0 & 0 & 0 \\
 0 & -1 & 3
\end{array}
\right)\;, \quad J_{m007} = \left(
\begin{array}{ccc}
 1 & 0 & 0 \\
 0 & 0 & 0 \\
 0 & 0 & 1
\end{array}
\right)\;.
\end{align}
Following each steps in eq.~\eqref{DGG : step 1},\eqref{DGG : step 2} and \eqref{DGG : final}, the Lagrangian for $T^{DGG}[m007,X_{\mu_2};P_{4\mu_2 -\lambda_2}]$ is given by 
\begin{align}
\begin{split}
&\mathcal{L}_{T^{DGG}[m007,X_{\mu_2};P_{4\mu_2 -\lambda_2}]}
\\
&=  \int d^4\theta \left( \frac{1}{4\pi } \big{(}\frac{3}2 \Sigma_2  V_2+  2V_{X}\Sigma_1  +2V_{C} \Sigma_2\big{)} + \big{(} \Phi_{1}^\dagger e^{V_1-V_2} \Phi_1+\Phi_{2}^\dagger e^{V_2} \Phi_2+\Phi_{3}^\dagger e^{-V_1-V_{2}} \Phi_3 \big{)} \right)
\\
& \quad + \frac{1}2 \int d^2 \theta  \Phi_1 \Phi_2^2 \Phi_3 +(c.c) \;.
\end{split}
\end{align}
In the Lagrangian, $V_1$ and  $V_2$ are dynamical $u(1)$ vector multiplets. The superpotential term comes from an easy internal edge $C_1$, $O_{C_1} = \Phi_1 \Phi_2^2 \Phi_3$. The theory has $u(1)_{X_\mu}$ and $u(1)_C$ whose background vector multiplets are $V_X$ and $V_C$ respectively. Applying the mirror symmetry in eq.~\eqref{3d mirror-1} and \eqref{3d mirror-2} with the following replacement
\begin{align}
\Phi_A \rightarrow \Phi_1, \Phi_B \rightarrow \Phi_3, W \rightarrow V_X+ V_2, U \rightarrow -2 V_2\;.
\end{align}
we have
\begin{align}
\begin{split}
&\mathcal{L}_{T^{DGG}[m007,X_{\mu_2};P_{4\mu_2 -\lambda_2}]}
\\
&=  \frac{1}{4\pi }\int d^4\theta \big{(}\frac{3}2 \Sigma_2  V_2 +2V_{C} \Sigma_2 +\Phi_{2}^\dagger e^{V_2} \Phi_2+ (T_p)^\dagger e^{V_X+ V_2} T_p +(T_m)^\dagger e^{-V_X+V_2} T_p +M^\dagger e^{-2V_2} M \big{)}
\\
& \quad + \bigg{(} \int d^2 \theta M ( \frac{1}2\Phi_2^2 + T_m T_p)  +(c.c) \bigg{)}\;.
\end{split}
\end{align}
In the dual picture, the $SO(3)$ symmetry is manifest after the redefinition of chiral fields as
\begin{align}
\begin{split}
&\phi_1 := \Phi_2, \phi_2 := \frac{1}{\sqrt{2}} (T_m +T_p),\phi_3 := \frac{i}{\sqrt{2}} (T_m -T_p)\;,
\\
&\Rightarrow M(\frac{1}2\Phi_2^2+ T_m T_p )=\frac{1}2 M( \phi_1^2 +\phi_2^2 +\phi_3^2 )\;.
\end{split}
\end{align}
The $u(1)_X$ is in the Cartan of this $SO(3)$.

\subsection{$SU(3)$ enhancement}\label{sec:SU(3)}
From the point of view of 6d $\CN=(2,0)$ theories, we only expect that the symmetry associated to codimension-2 defects (which are knots in 3-manifolds)
are $su(2)$. However, we will see that there are many theories which have larger symmetry enhancement.

We consider a pair of $(N,A;B)$ and $(N',A';B')$ such that 
\begin{align}
\begin{split}
&1) \;\textrm{$A$ is of $SO(3)$ type  while $A'$ is of $SU(2)$ type}\;,
\\
&2) \;\cI_{N}^{(A,B)} (m,e;x)= \cI_{N'}^{(A',B')} (m,e;x)\;,
\\
&3)\; \textrm{Both of $A$ and $A'$ are  non-closable cycles\;.} \label{Def: SU(3) pair}
\end{split}
\end{align}
which we call $SU(3)$-enhancement pair. The $\cI_{N}^{(A,B)}$ in 2) denotes a topological invariant called 3d index, see Appendix \ref{sec : 3d index}. For such a  pair,  we claim  that 
\begin{align}
\begin{split}
&\textrm{I. Two theories $T^{DGG}[N,X_A]$ and $T^{DGG}[N',X_{A'}]$ are identical}
\\
&\quad \;\textrm{possibly modulo a topological sector}
\\
&\textrm{II. The theory has enhanced $SU(3)$ flavor symmetry where $SO(3)_{A}$  and $SU(2)_{A'}$ are}
\\
&\quad \;\textrm{embedded into the $SU(3)$ in a way that $\mathbf{3}_{SU(3)}\rightarrow \mathbf{3}_{SO(3)_{A}}$ and $\mathbf{3}_{SU(3)}\rightarrow (\mathbf{2}\oplus \mathbf{1})_{SU(2)_{A'}}$}\;. \label{claim : SU(3) enhancement}
\end{split}
\end{align}
The 3d index $\CI_{N}^{(A,B)}(x)$ is equivalent to the superconformal index of $T^{DGG}[N,X_A]$ in charge basis. For hyperbolic  complement $N$, the index is  a non-trivial power series in $x$. The non-trivial match of the superconformal indices in the 2nd condition strongly suggests that two theories are actually equivalent  possibly up to a topological sector.  

From 1st and 3rd conditions in \eqref{Def: SU(3) pair}, the theory $T^{DGG}[N,X_A]=T^{DGG}[N',X_{A'}]$ has both of $SU(2)_A$ and $SO(3)_{A'}$ symmetry where the $U(1)_{X_A}=U(1)_{X_A'}$ is embedded into them as $\mathbf{2}_{SU(2)} = (\pm 1)_{U(1)}$ and $\mathbf{3}_{SO(3)} =(\pm 1, 0)_{U(1)}$ respectively. The  only  consistency way of this happening is that the theory has a $SU(3)$ symmetry into which the $SU(2)_A$ and $SO(3)_{A'}$ are embedded as  II in \eqref{claim : SU(3) enhancement}.
The reason is that the $SU(2)_A$ enhancement requires that there are conserved currents with charge $\pm 2$ under $U(1)_{X_A}$ from off-diagonal components of $SU(2)_A$,
while the $SO(3)_{A'}$ enhancement requires that there are conserved currents with charge $\pm 1$ under $U(1)_{X_A'}$.
Then the conserved currents with charge $\pm 1$ from $SO(3)_{A'}$ is a doublet of $SU(2)_A$. 
A minimal completion of such a situation to a Lie algebra is to embed the symmetries to the $SU(3)$ algebra.

One may wonder if there exits such a pair. Surprisingly, we can find infinitely many examples of these pairs. A class of  examples is 
\begin{align}
\begin{split}
&N= (S^3 \backslash \mathbf{5}^2_1)_{\mu_1+ k \lambda_1}\;, \quad  A = \mu_2\;, \quad B= 2\mu_2 +\lambda_2
\\
&N'= \overline{(S^3\backslash \mathbf{5}^2_1)_{(4k+1)\mu_1 - k \lambda_1}}\;, \quad  A' = 2\mu_2 - \lambda_2\;, \quad B'= \mu_2 - \lambda_2 \;. \label{SU(3) pair : twist knot}
\end{split}
\end{align}
\begin{figure}[h]
\begin{center}   \includegraphics[width=.50\textwidth]{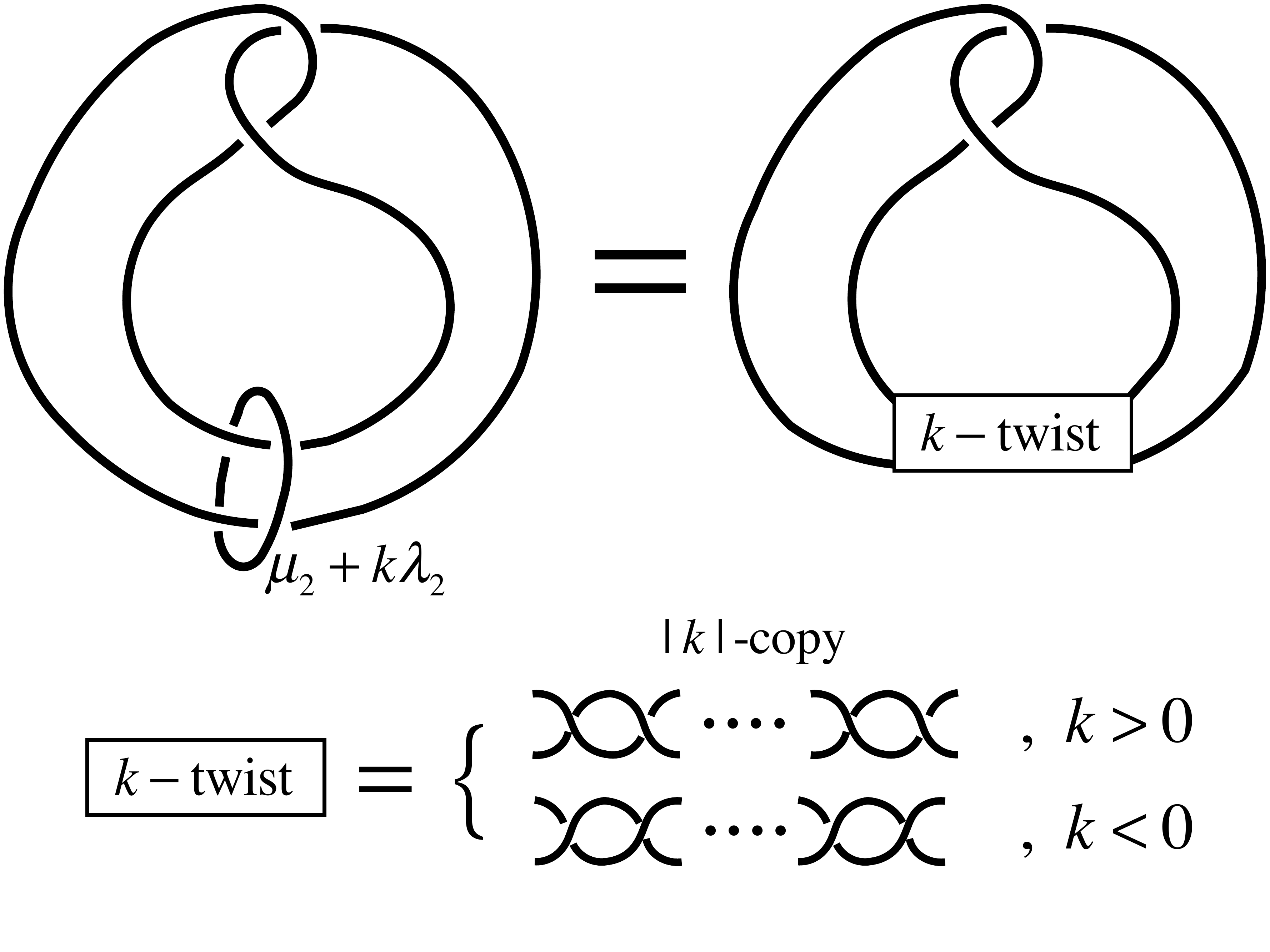}
   \end{center}
   \caption{A rational surgery calculus \cite{rolfsen1984rational} shows that $(S^3\backslash \mathbf{5^2_1})_{\mu_2 +k \lambda_2}$ is a twist knot $K_k$. For example, $K_{k=1}= \mathbf{4_1}, K_{k=-2}=\mathbf{5_2}$ and $K_{k=2}= \mathbf{6}_1$.  } \label{fig:1/k-surgery-Whitehead}
\end{figure}
As shown in fig.~\ref{fig:1/k-surgery-Whitehead}, the $N$ above are nothing but twist knots which will be denoted as $K_k$.  Let us check that the pair satisfy the 3 conditions in \eqref{Def: SU(3) pair}. First, note that both of $N$ and $N'$ can be considered as a knot complement in $\mathbb{Z}_2$-homological spheres, $L(1,k)=(N)_{\mu_2}$ and $L(4k-1,-k)=(N')_{\mu_2}$ respectively.  Applying  \eqref{SO(3)/SU(3) type in Z2-homology sphere} with $\mu=\mu_2$ and $\lambda= \lambda_2$, we can conclude that $A/A'$ is of  $SO(3)/SU(2)$ type.  Now let us check the 2nd condition in  \eqref{Def: SU(3) pair}. 
Combining  the $\mathbb{D}_8$-symmetry  \eqref{D_8 of Whitehead} of the Whitehead link index $\cI_{\mathbf{5}^2_1}$  \eqref{Whitehead index} and the following polarization  transformation rules of 3d index 
\begin{align}
\begin{split}
&\cI_{S^3\backslash \mathbf{5}^2_1}^{(- \lambda_1, \mu_2;\mu_1,2\mu_2+ \lambda_2)} (m_1 , m_2, e_1, e_2;x) =\cI_{\mathbf{5^2_1}} (e_1, m_2, -m_1, e_2-m_2;x)
\\
&\cI_{S^3\backslash \mathbf{5}^2_1}^{(4\mu_1 - \lambda_1, 2 \mu_2 - \lambda_2 ;\mu_1, \lambda_2-\mu_2)}  (m_1, m_2, e_1, e_2)=\cI_{\mathbf{5^2_1}} (e_1,2 m_2+e_2, 2e_1-m_1, m_2+e_2;x)
\\
\end{split}
\end{align}
and the following matrix multiplication ($S_2$ is a generator of the $\mathbb{D}_8$ in \eqref{D_8 of Whitehead}) 
\begin{align}
\begin{split}
&\left(\begin{array}{c}e_1 \\ 2m_2 +e_2\\ 2e_1-m_1 \\ m_2+e_2\end{array}\right)_{(e_1,e_2)\rightarrow (-e_1,-e_2)}= S_2  \cdot \left(\begin{array}{c}e_1 \\ m_2 \\ -m_1 \\ e_2-m_2 \end{array}\right)\;
\end{split}
\end{align}
we  have following identity 
\begin{align}
\cI_{S^3\backslash \mathbf{5}^2_1}^{(- \lambda_1, \mu_2;\mu_1,2\mu_2+ \lambda_2)} (m_1 , m_2, e_1, e_2;x) =  \cI_{S^3\backslash \mathbf{5}^2_1}^{(4\mu_1 - \lambda_1, 2 \mu_2 - \lambda_2 ;\mu_1, \lambda_2-\mu_2)} (m_1 , m_2, -e_1, -e_2;x)\;.
\end{align}
Applying the Dehn filling  formula in eq.~\eqref{3d index-2} to the above equality,
\begin{align}
\begin{split}
&\cI_{(S^3\backslash \mathbf{5}^2_1)_{\mu_1 + k \lambda_1}}^{( \mu_2;2\mu_2+ \lambda_2)} ( m_2,  e_2;x) =  \cI_{(S^3\backslash \mathbf{5}^2_1)_{(4k+1)\mu_1 - k \lambda_1}}^{( 2 \mu_2 - \lambda_2 ;\lambda_2-\mu_2)} ( m_2, -e_2;x)\;, 
\\
&\Rightarrow  \cI_{(S^3\backslash \mathbf{5}^2_1)_{\mu_1 + k \lambda_1}}^{( \mu_2;2\mu_2+ \lambda_2)} ( m_2,  e_2;x) =  \cI_{\overline{(S^3\backslash \mathbf{5}^2_1)_{(4k+1)\mu_1 - k \lambda_1}}}^{( 2 \mu_2 - \lambda_2 ;\mu_2-\lambda_2)} ( m_2, e_2;x)\;, \quad \textrm{for all $k\in \mathbb{Z}$}\;.
\end{split}
\end{align}
we confirm 2) in  \eqref{Def: SU(3) pair}. 
In the above, we use the transformation rule of 3d index under the orientation reversal in \eqref{Index under orientation reversal}.
Finally, from the following topological facts \cite{martelli2002dehn}
\begin{align}
\begin{split}
&N_A = (S^3\backslash \mathbf{5}^2_1)_{\mu_1+k \lambda_1, \mu_2}= L(1,k)\;, 
\\
&  (N')_{A'}=(S^3\backslash \mathbf{5}^2_1)_{(4k+1)\mu_1- k \lambda_1,2\mu_2- \lambda_2} = L(-8k-2,-2k-1)\;,
\end{split}
\end{align}
we see that both of $A$ and $A'$ are non-closable cycles according to \eqref{non-closable : Lens}. So we confirm that the pair, $(N,A)$ and $(N',A')$, in \eqref{SU(3) pair : twist knot} satisfy all the conditions in \eqref{Def: SU(3) pair} and the corresponding DGG theory is expected to have $SU(3)$ flavor symmetry.  We will  check the ehancement explicitly for  $k=1,-2$ by explicitly constructing $T^{DGG}[N,A]$ and $T^{DGG}[N',A']$.

\paragraph{For $k=1$} In the case,
\begin{align}
\begin{split}
&(S^3\backslash \mathbf{5}^2_1)_{\mu_1 + \lambda_1} = (S^3\backslash \mathbf{4}_1) = m004\;,
\\
&\overline{(S^3\backslash \mathbf{5}^2_1)_{5\mu_1 -  \lambda_1}} = (\textrm{Sister of }S^3\backslash \mathbf{4}_1) = m003\;. \label{m003/m004 from whitehead}
\end{split}
\end{align}
$m003$ is a knot complement called  sister of figure-eight knot complement. Both 3-manifolds have the same hyperbolic volume and  are the smallest hyperbolic 3-manifolds with one cusp torus boundary. From  ideal triangulations of $m003$ and $m004$  given below, their orientation are fixed. The equality in the above  means not only that the two manifolds are homeomorphism but also that they have the same orientation, i.e. the orientation of $m004$ is same as the orientation induced from a Dehn filling on $S^3\backslash \mathbf{5^2_1}$, whose orientation is induced from an ideal triangulation in \eqref{Gluing eqns for Whitehead}.

According to SnapPy, both can be ideally triangulated by two tetrahedra and  have common internal edge variables given in \eqref{internal edges of 4_1} while boundary variables are different by a factor $2$ or $1/2$
\begin{align}
\begin{split}
& a_{\mu_2} = Z_1-Z_2\;, \quad  b_{\lambda_2+2 \mu_2} = 4Z_1-2Z_1'-2Z_2\;,  \quad \textrm{ for $m004$}\;. 
\\
& a_{2\mu_2-\lambda_2}= 2 Z_1-2 Z_2\;, \quad b_{\mu_2 -\lambda_2} = 2Z_1-Z_1'-Z_2\;,   \quad \textrm{ for $m003$}\;. 
\end{split}
\end{align}
For DGG's construction, we choose
\begin{align}
\begin{split}
& X_{\mu_2}= a_{\mu_2}\;, \quad P_{2\mu_2+\lambda_2}=\frac{1}2 b_{2\mu_2+\lambda_2} \;,   \quad \textrm{ for $m004$}\;. 
\\
& X_{2\mu_2-\lambda_2}= \frac{1}2 a_{2\mu_2-\lambda_2}\;, \quad P_{\mu_2}= b_{\mu_2-\lambda_2} \;,   \quad \textrm{ for $m003$}\;. 
\end{split}
\end{align}
Thus, both DGG theories are identical and described by the Lagrangian in \eqref{TDGG[m004]} up to background CS level for $U(1)_{X}$ which is irrelevant in symmetry enhancement.
We reproduce the Lagrangian here with the modified background CS level;
\begin{align}
\begin{split}
&\mathcal{L}_{T^{DGG}[m004, X_{\mu_2} ;P_{2 \mu_2 +\lambda_2}]} (V_{X}, V_C)  = \mathcal{L}_{T^{DGG}[m003, X_{2\mu_2-\lambda_2} ;P_{\mu_2-\lambda_2}]} (V_{X}, V_C) 
\\
&=\frac{1}{4\pi} \int d^4 \theta \left(- \frac{1}2 \Sigma_C V_X + \Sigma (2V_C + 3 V_X) + \Sigma_X V_X \right) + \int d^4 \theta \left( \Phi_1^\dagger e^{V+\frac{V_X}2}\Phi_1 +\Phi_2^\dagger e^{V- \frac{V_X}2}\Phi_2 \right) \;.\label{TDGG[m004]reproduced}
\end{split}
\end{align}
So the theory is 
\begin{align}
\begin{split}
&T^{DGG}[m004,\mu_2] = T^{DGG}[m003,2\mu_2- \lambda_2] 
\\
&=\textrm{A $U(1)$  vector multiplet coupled to 2 chirals of charge $+1$} \;.
\end{split}
\end{align}

The theory has manifest $u(1)_{\rm top} \times su(2)_{\rm manifest}$ where $u(1)_{\rm top} $ is the topological monopole charge of the $u(1)_{\rm gauge}$ gauge symmetry,
and $su(2)_{\rm manifest}$ acts on the two chiral fields.
This $u(1)_{\rm top} \times su(2)_{\rm manifest}$ will be enhanced to $SU(3)$.   

The $u(1)_{C}$ flavor symmetry associated to the background field $V_C$
corresponds to the topological symmetry $u(1)_{\rm top}$ and will be embedded to $SU(3)$ as 
\begin{align}
u(1)_C = u(1)_{\rm top}= T_8 := \textrm{diag}(-1/3, -1/3, 2/3) \in SU(3)\;.
\end{align}
This is because the ``off-diagonal components'' of $SU(3)$ (which are not in $u(1)_{C}\times su(2)_{\rm manifest}$) are 
provided by monopole operators with monopole charge $\pm 1 = \pm (1/3 - (-2/3))$. 

On the other hand, the $V_X$ is coupled to the system as follows. Let
\begin{align}
T_3:=\textrm{diag}(1/2,-1/2,0)\;.
\end{align}
be the Cartan generator of the manifest $su(2)_{\rm manifest}$. The $V_X$ is coupled to the chiral fields via this generator $T_3$.
Also, notice that $V_X$ is coupled to the monopole current $\Sigma$ with coefficients $3/2$. 
Therefore the $u(1)_{X} $ is embedded in $SU(3)$ as 
\begin{align}
u(1)_X =  \frac{3}2 T_8+T_3 = \textrm{diag} (0,-1,1) \in SU(3)\;.
\end{align}
This $u(1)_X$ must be enhanced to $su(2)_X$ because the $A$-cycle is non-closable. 

Notice that this $su(2)_X$ is different from the manifest $su(2)_{\rm manifest}$ symmetry.
Therefore, if $u(1)_X$ is enhanced to $su(2)_X$, then the $u(1)_{\rm top}\times su(2)_{\rm manifest}$ must be enhanced to $SU(3)$.
This agrees with our general discussion that this theory has enhanced $SU(3)$ symmetry.

The superconformal index of theory is
\begin{align}
\begin{split}
&\cI_{m003/m004}(u_1,u_2;x) 
\\
&=\sum_{(e_1,e_2)\in \mathbb{Z}^2} (-x^{\frac{1}2})^{e_2}  \cI_{\Delta }(-e_2,-e_1+2e_2;x)\cI_{\Delta}(-e_2,e_1-e_2;x)  u_1^{e_1}u_2^{e_2}\;.
\end{split}
\end{align}
Here $u_1$ and $u_2$ fugacity variable for $u(1)_{X}$ and $u(1)_{C}$ symmetry respectively. The index  depends on  the  choice of $R$-charge mixing between $u(1)_R$ and $u(1)_C$. In the above expression, we in particularly choose\footnote{In general, R-charge of BPS monopole operators $V_+, V_-$ of charge $\pm 1$ are related to  the R-charges of chiral multiplets $\Phi_a$ as  $R(V_\pm ) = \frac{1}2\sum q_a  (1-R(\Phi_a))$ where $q_a$ is  the  $u(1)_{\rm gauge}$  charge of  $\Phi_a$.}, 
\begin{align}
R(\Phi_a) =\frac{1}3,\quad   R(V_\pm) = \frac{2}3\;.
\end{align}
Here  $V_\pm$ denote a BPS monopole operator of charge $\pm 1$.
Then the index show the  $SU(3)$ structure :
\begin{align}
\begin{split}
&\cI_{m003/m004}(u_1, u_2;x)  =1- (\chi_{1,1}) x-(\chi_{0,0}+\chi_{1,1}) x^2+(\chi_{2,2}-\chi_{0,0}-\chi_{1,1})x^3
\\
&\qquad \quad +(\chi_{3,0}+\chi_{2,2}+\chi_{0,3}-\chi_{0,0} )x^4+(\chi_{0,3}+2\chi_{1,1}+2 \chi_{2,2}+\chi_{3,0}) x^5+\ldots
\end{split}
\end{align}
Here $\chi_{m,n}(u_1, u_2)$ is the character of $SU(3)$-representation with Dynkin labels $(m,n)$. 
The correct IR R-charge mixing should be determined by F-maximization, but the non-trivial appearance of the $SU(3)$ in a particular choice strongly suggests that the choice gives the correct R-charge assignment. 
The first non-trivial terms comes form  operators listed in the Table below. 
\begin{table}[h!]
\centering
\label{leading operators in T[m003/m004]}
\begin{tabular}{|c | c | c | c | }
\hline
 & $u(1)_C$ & $u(1)_X$  & SCI contribution \\\hline
$\phi_a (\psi^*_+)_b$ & 0  & $\pm \frac{1}2 \pm \frac{1}2$   & $ -(2+u_1 + \frac{1}{u_1}) x$ \\\hline
$V_-  (\phi_a)$ & $1$ & $(1, 2)$ &    $- (u_1 u_2  + u_1^2 u_2) x $\\\hline
$V_+ (\phi_a )$ &$-1$& $(-1,-2)$ &  $- (\frac{1}{u_1^2 u_2} + \frac{1}{u_1 u_2})x$ \\
\hline 
\end{tabular}
\label{leading operators in T[m003/m004]}
\end{table}
In the table,  $\phi_a$ and $(\psi_{\pm})_a$ denote the scalar and fermionic fields in chiral field $\Phi_{a}$ respectively with $a=1,2$.  $V_\pm (\ldots)$ denote a gauge invariant BPS monopole operator of charge $
\pm 1$ dressed by matter fields $(\ldots)$.  %
 All these operators have quantum numbers $(R,j_3,\Delta)= (1, \frac{1}2, \frac{3}2)$ and form descents of conserved current multiplet for $SU(3)$ flavor symmetry.  In 3d $\cN=2$ SCFT, a conserved current multiplet of flavor group $F$ consists of following operators in the adjoint representation of $F$:
 \begin{align}
 [0]^{(0)}_{\Delta=1} \xrightarrow[\text{}]{\; Q,\tilde{Q}\; } [\frac{1}2]^{(1)}_{\Delta = \frac{3}2} \oplus [\frac{1}2]^{(-1)}_{\Delta = \frac{3}2} \xrightarrow[\text{}]{\; Q,\tilde{Q}\; }  [1]^{(0)}_{\Delta=2} \oplus [0]^{(0)}_{\Delta=2}
 \end{align}
 Here operators are denoted by its quantum number $[j]^{(r)}_{\Delta}$ where $j$ denote a spin of space-time rotational symmetry $su(2)$ in a normalization such that $[1/2]$ corresponds to the fundamental representation. 

\paragraph{For $k=-2$} In the case, 
\begin{align}
\begin{split}
&(S^3\backslash \mathbf{5}^2_1)_{\mu_1 - 2 \lambda_1}  = \overline{S^3 \backslash \mathbf{5}_2}= \overline{m015};,
\\
&\overline{(S^3\backslash \mathbf{5}^2_1)_{-7\mu_1 +2 \lambda_1}} = \overline{m017}\;. \label{m015/m017 from whitehead}
\end{split}
\end{align}
Both manifolds have the same hyperbolic volume $\textrm{vol}(m015)=\textrm{vol}(m017) = 2.82812\ldots$. 
According to SnapPy, both can be ideally triangulated by three tetrahedra and  have common internal edge variables 
\begin{align}
\begin{split}
&C_1 = Z_1'+Z_1''+2Z_2+Z_3'+Z_3''\;, \quad C_2 =Z_1+Z_1'+Z_2'+Z_3+Z_3'\;, 
\\
&C_3 = Z_1+Z_1''+Z_2'+2 Z''_2+Z_3+Z_3''\;.
\end{split}
\end{align}
while boundary variables are different by a factor $2$ or $1/2$
\begin{align}
\begin{split}
& a_{\mu_2}= -Z_1 +Z_2\;, \quad b_{-2 \mu_2-\lambda_2} = 2(2Z_1-Z_1''-2Z_2+Z_3'')\;,   \quad \textrm{ for $m015$}\;. 
\\
& a_{2\mu_2-\lambda_2}= -2Z_1 +2Z_2\;, \quad b_{ \lambda_2 - \mu_2} =2Z_1-Z_1''-2Z_2+Z_3''\;,   \quad \textrm{ for $m017$}\;. 
\end{split}
\end{align}
We choose
\begin{align}
\begin{split}
& X_{\mu_2}= a_{\mu_2}\;, \quad P_{-2\mu_2-\lambda_2}=\frac{1}2 b_{-2\mu_2-\lambda_2} \;,   \quad \textrm{ for $m015$}\;. 
\\
& X_{2\mu_2-\lambda_2}= \frac{1}2 a_{2\mu_2-\lambda_2}\;, \quad P_{\mu_2}= b_{\lambda_2-\mu_2} \;,   \quad \textrm{ for $m017$}\;. 
\end{split}
\end{align}
Then, the $Sp(6,\mathbb{Z})$+(affine-shifts) are
\begin{align}
\begin{split}
&\left(\begin{array}{c} X \\ C_1 \\ C_2 \\ P \\ \Gamma_1 \\ \Gamma_2\end{array}\right) = g_{m0015/m017} \cdot  \left(\begin{array}{c} Z_1 \\ Z_2 \\ Z_3 \\  Z_1'' \\ Z_2'' \\ Z_3''\end{array}\right) +  i \pi \nu_{m015/m017}
\\
&g_{m015/m017}= \left(
\begin{array}{cccccc}
 -1 & 1 & 0 & 0 & 0 & 0 \\
 -1 & 2 & -1 & 0 & 0 & 0 \\
 0 & -1 & 0 & -1 & -1 & -1 \\
 2 & -2 & 0 & -1 & 0 & 1 \\
 0 & 0 & 0 & 0 & 0 & -1 \\
 0 & 1 & 0 & 0 & 0 & 0 \\
\end{array}
\right)\;, \quad \nu_{m015/m017}= \left(\begin{array}{c} 0 \\ -2 \\ -3 \\  0 \\ 0 \\ 0\end{array}\right) 
\end{split}
\end{align}
The matrix $g_{m015/m017}$ can be decomposed into $g_{m015/m017} = g^s_{J_{m015}} g^t_{K_{m015}} g^{gl}_{U_{m015}}$ \eqref{T,S and GL type}
\begin{align}
&U_{m015}  = \left(
\begin{array}{ccc}
 -1 & 1 & 0   \\
 -1 & 2 & -1   \\
 0 & 1 & 0   \\
\end{array}
\right)\;, \quad 
K_{m015}  = \left(
\begin{array}{ccc}
 -2 & 0 & 0   \\
 0 & 0 & 0   \\
 0 & 0 & 1   \\
\end{array}
\right)\;, \quad 
J_{m015} = \left(
\begin{array}{ccc}
 0 & 0 & 0   \\
 0 & 0 & 0   \\
 0 & 0 & 1   \\
\end{array}
\right)\;.
\end{align}
Using the decomposition, we have 
\begin{align}
\begin{split}
&\mathcal{L}_{T^{DGG}[m015, X_{\mu_2};P_{-2\mu_2- \lambda_2}]} (V_X, V_{C_1}, V_{C_2}) = \mathcal{L}_{T^{DGG}[m017, X_{2\mu_2- \lambda_2};P_{\lambda_2 - \mu_2}]} (V_X, V_{C_1}, V_{C_2})
\\
&=  \frac{1}{4\pi }\int d^4\theta \big{(}- 3\Sigma_{X} V_{X} + \Sigma_{X} V_{C_1} -\frac{1}2 (\Sigma_{C_1}- \Sigma)(V_{C_1}-V) +2  \Sigma V_{C_2} \big{)}
\\
& \quad +\int d^4\theta \big{(}\Phi_{1}^\dagger e^{V-V_{X}} \Phi_1+\Phi_{2}^\dagger e^{V} \Phi_2+\Phi_{3}^\dagger e^{V+V_{X}-V_{C_1}} \Phi_3 \big{)}\;. \label{TDGG[m015]}
\end{split}
\end{align}
Here $V$ is a dynamical $u(1)$ vector multiplet. Since $C_1$ and $C_2$ are hard internal edges and we can not add $O_{C_1}$ and $O_{C_2}$ to superpotential. 
So, the DGG theory is
\begin{align}
\begin{split}
&T^{DGG}[m015,\mu_2] = T^{DGG}[m017,2\mu_2- \lambda_2] 
\\
&=\textrm{A $u(1)_{-1/2}$  vector multiplet coupled to 3 chirals of charge $+1$} \;.
\end{split}
\end{align}
The theory has manifest $SU(3)$ flavor symmetry rotating 3 chrials as expected.

\section{Dehn filling in 3d/3d correspondence}\label{sec : Dehn-filling}

In this section, we generalize the DGG's construction to obtain $T^{6d}_{\rm irred}[M]$ for closed 3-manifolds $M$ by incorporating Dehn filling operation. Refer to \cite{Pei:2015jsa,Gadde:2013sca,Gukov:2017kmk, Alday:2017yxk} for previous discussions on Dehn filling operation in the context of 3d/3d correspondence and the construction of 3d theory, which we will denote $T^{6d}_{(\rm  irred)^c}[M]$, labelled by Seifert manifolds $M$.\footnote{Their theory $T^{6d}_{(\rm  irred)^c}[M]$ is different from  our $T^{6d}_{\rm irred}[M]$ theory. For example, when $M$ is a Lens space, their $T^{6d}_{(\rm  irred)^c}[M]$  is a non-trivial SCFT while  supersymmetry is broken in our $T^{6d}_{\rm  irred}[M]$ theory. As discussed in section~ \ref{sec:6d}, we need to specify a point $P \in \CM_{\rm vacua}(T^{6d}[M] \textrm{ on } \mathbb{R}^3)$ to obtain a 3d effective theory. Their theory may correspond to different choice of $P$ other than $P_{\rm SCFT}$ in \eqref{Relation-T6d-TDGG:intro}.   }  


\subsection{Dehn filling on $T^{6d}_{\rm irred}[N]$}
%

For a hyperbolic knot complement $N$ and  a primitive boundary  cycle $(pA+qB) \in H_1 (\partial N, \mathbb{Z})$,
\begin{align}
T^{6d}_{\rm irred}[N_{pA+qB}] = (\textrm{Giving a nilpotent vev to  $\mu$ of  $T^{6d}_{\rm irred}[N,pA+qB]$})\;.
\end{align}
The fact that the closing of codimension-2 defects (i.e., knots) corresponds to giving the nilpotent vev to $\mu$
is standard in 4d class S theories. See e.g., \cite{Tachikawa:2013kta} and references therein.
This is also in accord with our terminology `closable/non-closable', because non-closable cycles have empty $\mu$ and hence
it is not possible to do the above operation while preserving supersymmetry. See Table~\ref{table:closing} below.
\begin{table}[h!]
\centering
\begin{tabular}{|c || c | }
\hline
$pA+qB \in H_1 \big{(}\partial N,\mathbb{Z}\big{)}$ & $T^{6d}_{\rm irred}[N_{pA+qB}]$
\\
\hline
\hline
non-exceptional ($\Rightarrow$ closable) &  non-trivial SCFT  \\\hline
exceptional and closable & Gapped theory (possibly with decoupled free chirals)  \\
\hline
non-closable & SUSY broken   \\
\hline
\end{tabular}
\caption{Basic property of $T^{6d}_{\rm irred}[N_{pA+qB}]$. A primitive boundary cycle $pA+qB$ is called exceptional if $N_{pA+qB}$ is non-hyperbolic.}\label{table:closing}
\end{table}

If the $A$  is non-closable and $q=1$, the above relation can be simplified as follows.
The theory $T^{6d}_{\rm irred}[N, pA+B] $ is related to the theory $T^{6d}_{\rm irred}[N,A;B] $ by the transformation $ST^p \in SL(2,\BZ)$ as
\beq
T^{6d}_{\rm irred}[N, pA+B] = T^{6d}_{\rm irred}[N,A;B] - su(2)_p - T[SU(2)],
\eeq
where $su(2)_p$ is gauging the $su(2)$ symmetry of $T^{6d}_{\rm irred}[N,A;B] $ and the $su(2)_H$ symmetry of $T[SU(2)]$.
The Chern-Simons level of this group is $p+[\text{original value}]$, where $[\text{original value}]$ means the contribution 
of $T^{6d}_{\rm irred}[N,A;B] $ to the Chern-Simons level. 
To specify this contribution, we have to specify not only the $A$-cycle, but also the $B$-cycle. This is the reason why we are writing $B$ explicitly in
the notation $T^{6d}_{\rm irred}[N,A;B] $.

Now, the operator $\mu$ comes from the moment map operator of $T[SU(2)]$ associated to the $su(2)_C$ symmetry which is not gauged.
If we give a nilpotent vev to this operator $\mu$, the $T[SU(2)]$ becomes massive and flows to an empty theory in the low energy limit up to the 
Goldstone multiplets associated to the symmetry breaking of $su(2)_C$ by the vev~\cite{Gaiotto:2008ak}.
Neglecting those Goldstone multiplets, the $T[SU(2)]$ disappears and hence we get
\begin{align}
\begin{split}
T^{6d}_{\rm irred}[N_{pA+B}] &= (\textrm{Gauging $su(2)$ of $T^{6d}_{\rm irred}[N,A;B]$ with additional CS level $p$})\;, 
\\
&= (\textrm{Gauging $su(2)$ of $T^{DGG}[N,A;B]$ with additional CS level $p$})\;.
\label{integral Dehn filling = gauging}
\end{split}
\end{align}
where we have assumed that $A$ is non-closable and hence $T^{6d}_{\rm irred}[N,A;B] = T^{DGG}[N,A;B]$.

As  examples, we  consider closed 3-manifolds obtained from $m003/m004/m015$ by performing a Dehn filling.  Combing \eqref{TDGG[m004]reproduced},\eqref{TDGG[m015]} and \eqref{integral Dehn filling = gauging}, we have\footnote{Taking account of orientation reversal in \eqref{m015/m017 from whitehead},  the boundary 1-cycle basis $(\mu_2, \lambda_2)$ of the $S^3\backslash \mathbf{5}_2$ induced from  the basis of  $S^3\backslash \mathbf{5}^2_1$  can be identified with $(\mu, -\lambda)$=(meridian, -(longitude)) of the knot complement.  For $S^3 \backslash \mathbf{4}_1$ case, on the other hand, the $(\mu_2, \lambda_2)$ can be identified with $(\mu, \lambda)$ without sign change. Note that there is no orientation reversal in \eqref{m003/m004 from whitehead}. }
\begin{align}
\begin{split}
&T^{6d}_{\rm irred}[(m003)_{ p(2\mu_2-\lambda_2)+(\mu_2-\lambda_2)} = \overline{(S^3\backslash \mathbf{5}^2_1)_{5\mu_1 -  \lambda_1,~ p(2\mu_2-\lambda)+(\mu_2-\lambda_2)}} ]
\\
& = \frac{\textrm{A $u(1)_0$ vector coupled to 2 chrials of charge +1}  }{SU(2)_{p+1/2}}\;,
\\
&T^{6d}_{\rm irred}[(m004)_{ p \mu_2 + (\lambda_2+2 \mu_2)} =(S^3\backslash \mathbf{4}_1)_{(p+2) \mu + \lambda} ]
\\
& = \frac{\textrm{A $u(1)_0$ vector coupled to 2 chrials of charge +1}  }{SO(3)_{p+2}}\;,
\\
&T^{6d}_{\rm irred}[(m015)_{ p \mu_2 + (-2\mu_2- \lambda_2)} =(S^3\backslash \mathbf{5}_2)_{(p-2)\mu + \lambda}]
\\
& = \frac{\textrm{A $u(1)_{-1/2}$ vector coupled to 3 chrials of charge +1}  }{SO(3)_{p-6}}\;.
 \label{Theory for m003/m004 Dehn-filled}
\end{split}
\end{align}
Here the notation $/G_k$ means that we couple a vector multiplet of group $G$ with the Chern-Simons level $k$. The theories in the numerator has $SU(3)$ symmetry at IR as argued in sec.~\ref{sec:SU(3)} and we are gauging its $SO(3)/SU(2)$ subgroup. Since the $u(1)_X$ is embedded to the $SU(2)$ (resp. $SO(3)$) in a way that $\mathbf{2}_{su(2)} = (\pm 1)_{u(1)_X}$ (resp. $\mathbf{2}_{su(2)} = (\pm \frac{1}2)_{u(1)_X}$), the CS level $+1$  for $u(1)_X$ in \eqref{TDGG[m004]reproduced} corresponds to CS level $1/2$ (resp. 2) for the $su(2)$. Similarly the CS level $-3$ for $u(1)_X$ in \eqref{TDGG[m015]} corresponds to CS level $-6$ for the $su(2)$. A parity operation filps the  signs of  CS levels of the $T^{6d}_{\rm irred}$ theories. The parity operation corresponds to orientation reversal on the internal 3-manifold. It is compatible with following topological facts
\begin{align}
\begin{split}
 &(m003)_{p(2\mu_2-\lambda_2)+(\mu_2-\lambda_2)}  =\overline{(m003)_{(-p-1)(2\mu_2-\lambda_2)+(\mu_2-\lambda_2)}}\;,
 \\
 &
 (m004)_{ p \mu_2 + (\lambda_2+2 \mu_2)}  =\overline{ (m004)_{ (-p-4) \mu_2 + (\lambda_2+2 \mu_2)} }\;.
 \end{split}
\end{align}
After gauging $SU(2)/SO(3)$ subgroup of $SU(3)$, the resulting theory generically has following flavor symmetry 
\begin{align}
\begin{split}
&T^{6d}_{\rm irred}[(S^3\backslash \mathbf{5}^2_1)_{5\mu_1 -  \lambda_1,~ p(2\mu_2-\lambda)+(\mu_2-\lambda_2)}] \; \textrm{has $u(1)$ flavor symmetry }
\\
&T^{6d}_{\rm irred}[(S^3\backslash \mathbf{4}_1)_{(p+2)\mu_2 +\lambda_2}] \; \textrm{has no flavor symmetry }
\\
&T^{6d}_{\rm irred}[(S^3\backslash \mathbf{5}_2)_{(p-2)\mu + \lambda}]  \; \textrm{has $u(1)$ flavor symmetry }
\end{split}
\end{align}
The $u(1)$ for the 3rd case comes from the topological symmetry of $u(1)_{-1/2}$ gauge symmetry of the theory in the numerator. 
The above is correct when $|p|$ is large enough where the semiclassical analysis is reliable. When $|p|$ is small,  the theories could have accidental symmetries. Actually from following topological fact (see Figure.~\ref{fig:rational-surgery-moves}),
\begin{align}
(S^3\backslash \mathbf{4}_1)_{ -5 \mu +\lambda } = (S^3\backslash \mathbf{5}_2)_{5\mu +\lambda}
\end{align}
we can conclude that $T^{6d}_{\rm irred}[(S^3\backslash \mathbf{4}_1)_{(p+2)\mu_2 +\lambda_2}]$ has accidental $u(1)$ symmetry for $p=3$ and $p=-7$. We will come back to this point in sec~\ref{sec: Duality/Surgery}.

\subsection{Small  hyperbolic manifolds} \label{sec : small-manifolds}
Let us discuss the case of closed 3-manifolds $M=$Weeks, a oriented hyperbolic closed 3-manifold with smallest hyperbolic volume. This was already discussed in \cite{Gang:2017lsr} and here we supply a little bit more details.
The Weeks manifold is obtained by performing a Dehn filling operation on   $m003$,
\beq
\overline{(S^3\backslash \mathbf{5}^2_1)_{5\mu_1 -  \lambda_1, -5 \mu_2 +2 \lambda_2}} = (m003)_{-5\mu_2 +2\lambda_2} = \textrm{Weeks}\;.
\eeq
Corresponding 3d gauge theory is the theory in the second line of eq.~\eqref{Theory for m003/m004 Dehn-filled} with $p=-3$. 
The theory in the numerator has $SU(2)_X$ symmetry which is a subgroup of the $SU(3)$.  The $SU(2)_X$ symmetry is different from the manifest $SU(2)_{\rm manifest}$ rotating two chirals. But using the Weyl symmetry of the $SU(3)$, the symmetry $SU(2)_X$ and $SU(2)_{\rm manifest}$ can be exchanged with each other. Therefore, we can take $SU(2)_X$ to be the manifest $SU(2)_{\rm manifest}$. We will just denote it as $SU(2)$ in the following.
Then by \eqref{Theory for m003/m004 Dehn-filled},
\beq
T^{6d}_{\rm irred}[\text{Weeks}] =\text{(Two chiral fields coupled to $U(1)_0 \times SU(2)_{-5/2}$)}.
\eeq

\paragraph{AF duality}
Now, we can further simplify this theory to a much simpler theory~\cite{Gang:2017lsr}.
There is a duality found by Aharony and Fleischer (AF)~\cite{Aharony:2014uya}
\beq
\text{(Two chiral fields coupled to $ SU(2)_{-5/2}$)}=\text{(One chiral field gauged by $ U(1)_{+3/2}$)}.
\eeq
To apply this duality, we need to know the relation between the flavor $U(1)$ symmetries of both sides of this equation and their background Chern-Simons levels.

Here we supply the details promised in \cite{Gang:2017lsr}.
In the AF duality, the $U(1)$ charge acting on two chiral fields with charge 1 on the left hand side corresponds to the 
topological charge of the $U(1)_{+3/2}$ gauge field multiplied by 2. The reason is that the $-1 \in U(1)$ acting on the two chiral fields
can be compensated by the $-1 \in SU(2)_{-5/2}$ gauge transformation, and hence all gauge invariant operators have even charge on the left hand side.
So the relation is
\beq
&SU(2)_{-5/2}-\text{two chirals }-U(1)^{\rm bkg}_0  \nonumber \\
\Longleftrightarrow~&\text{single chiral}  -  U(1)_{3/2}  \overset{ \times 2}{-}   U(1)^{\rm bkg}_n . 
\label{eq:AF}
\eeq
where $U(1)^{\rm bkg}$ is the global symmetry with background field, $n$ is the background Chern-Simons level,
and $ \times 2 $ means that the $U(1)^{\rm bkg}_n$ is coupled to the topological current of $U(1)_{3/2}$ multiplied by two.

We want to determine the value of $n$.
This can be done as follows. Let $\sigma^{\rm bkg}$ be the real scalar for the background $U(1)^{\rm bkg}$ vector multiplet, or 
in other words, the real mass associated to this symmetry. 
We choose the sign of it such that after integrating out the two chiral fields on the left hand side, the left hand side flows to
\beq
SU(2)_{-3} \oplus U(1)^{\rm bkg}_{-1}.
\eeq
where $\oplus$ means that the two factors $SU(2)_3$ and $U(1)^{\rm bkg}_1$ are completely decoupled.
Here we need to remark the important point. The $SU(2)_{-3}$ is the 3d $\CN=2$ gauge theory at the level $-3$. This theory contains the gaugino,
and by integrating out the gaugino, we get a pure topological Chern-Simons theory as
\beq
SU(2)_{-3} = SU(2)^\text{topo CS}_{-1}
\eeq
Namely, the gaugino reduces the level by $2 = h^\vee _{su(2)}$.
Therefore, the low energy limit is
\beq
SU(2)^\text{topo CS}_{-1} \oplus U(1)^{\rm bkg}_{-1}
\eeq

The effect of $\sigma^{\rm bkg}$ on the right-hand-side of \eqref{eq:AF} is to give the dynamical $U(1)$ an FI parameter. 
We want the dynamical gauge group $U(1)$ to be not Higgsed so that we can match it with the $SU(2)^\text{topo CS}_{-1}$ later. 
Then, the D-term condition implies that the dynamical real scalar $\sigma$ gets a vev proportional to $\sigma^{\rm bkg}$
because the Lagrangian contains $3/2D \sigma + 2 \cdot 2D  \sigma^{\rm bkg}$ and we need to impose stationary condition for $D$.
The vev of $\sigma$ gives the chiral field a mass term. The sign of the mass is anticipated by the fact that it must make the low energy CS level of the dynamical field
as $U(1)_{-2}$. This is because the only consistent way for the duality to work in low energy is to use the duality of topological CS theory given by 
\beq
SU(2)^\text{topo CS}_{-1} = U(1)^\text{topo CS}_{-2} \sim  U(1)^\text{topo CS}_{2}
\eeq
where we have used the fact that the gaugino plays no role in $U(1)$ and hence $U(1)_{-2}=U(1)^\text{topo CS}_{-2}$.
First equality is well-known (see, e.g., \cite{DiFrancesco:1997nk} for the corresponding statement in Wess-Zumino-Witten models which are related to
topological Chern-Simons theories~\cite{Witten:1988hf}.).
The second equality $ U(1)^\text{topo CS}_{2} \sim  U(1)^\text{topo CS}_{-2}$ is more precisely given by 
$U(1)_2 \times U(1)_{-1} =U(1)_{-2} \times U(1)_{1}$~\cite{Seiberg:2016rsg, Tachikawa:2016cha, Tachikawa:2016nmo}
and we have neglected $U(1)_{\pm 1}$ because these theories have only one state in the Hilbert space on any space (and they are called invertible field theory),
and our argument is not careful enough to detect those invertible field theories.

After integrating out the chiral field, 
the right-hand-side of \eqref{eq:AF} is given by the Lagrangian
\beq
\frac{1}{4\pi } \left(  2 V\Sigma + 2 \cdot 2   V\Sigma^{\rm bkg} +n V^{\rm bkg} \Sigma^{\rm bkg}   \right).
\eeq
where in the second term, the factor of $2$ have taken into account the fact that $\U(1)^{\rm bkg}$ is coupled to the topological current of $\U(1)$ by charge $2$.
We shift the dynamical gauge field as $V \to V+V^{\rm bkg}$ to get
\beq
\frac{1}{4\pi } \left(  2 V\Sigma + (n-2) V^{\rm bkg} \Sigma^{\rm bkg}   \right).
\eeq
This means that the low energy theory is given by
\beq
U(1)^\text{topo CS}_{2} \oplus U(1)^{\rm bkg}_{n-2}.
\eeq
Therefore by comparing the low energy limit of the left and right hand side of \eqref{eq:AF}, we get
\beq
-1=n-2  \Longrightarrow n=1.
\eeq

\paragraph{Weeks theory}
Now let us gauge $U(1)^{\rm bkg}$ (but we use the same name for simplicity). 
The left hand side of \eqref{eq:AF} after gauging  $U(1)^{\rm bkg}$ is precisely the theory $T^{6d}_{\rm irred}[\text{Weeks}]$.

Let us see the right hand side.
The Chern-Simons action of the right-hand-side of \eqref{eq:AF} after putting $n=1$ is given by
\beq
\frac{1}{4\pi } \left(  \frac{3}{2}V\Sigma + 2 \cdot 2   V\Sigma^{\rm bkg} + V^{\rm bkg} \Sigma^{\rm bkg}   \right).
\eeq
where in the second term, the factor of $2$ have taken into account the fact that $U(1)^{\rm bkg}$ is coupled to the topological current of $U(1)_{3/2}$ by charge $2$.
If we integrate out $V^{\rm bkg}$, or in other words, by making the shift $V^{\rm bkg}  \to V^{\rm bkg}  + 2V$ and neglecting the decoupled $U(1)_{1}$ theory,
we get
\beq
\frac{1}{4\pi } \left( ( \frac{3}{2} -4) V\Sigma   \right) =\frac{1}{4\pi } \left( - \frac{5}{2}  V\Sigma   \right).
\eeq

We conclude that the theory $T^{6d}_{\rm irred}[\text{Weeks}]$ is given by
\beq
T^{6d}_{\rm irred}[\text{Weeks}] = (\text{One chiral field coupled to $U(1)_{-5/2}$}).
\eeq
This is one of the small theories discussed in \cite{Gang:2017lsr}.

\section{3d $\cN=2$ Dualities from Surgery calculus} \label{sec: Duality/Surgery}

\begin{figure}[h]
\begin{center}   
\includegraphics[width=.45\textwidth]{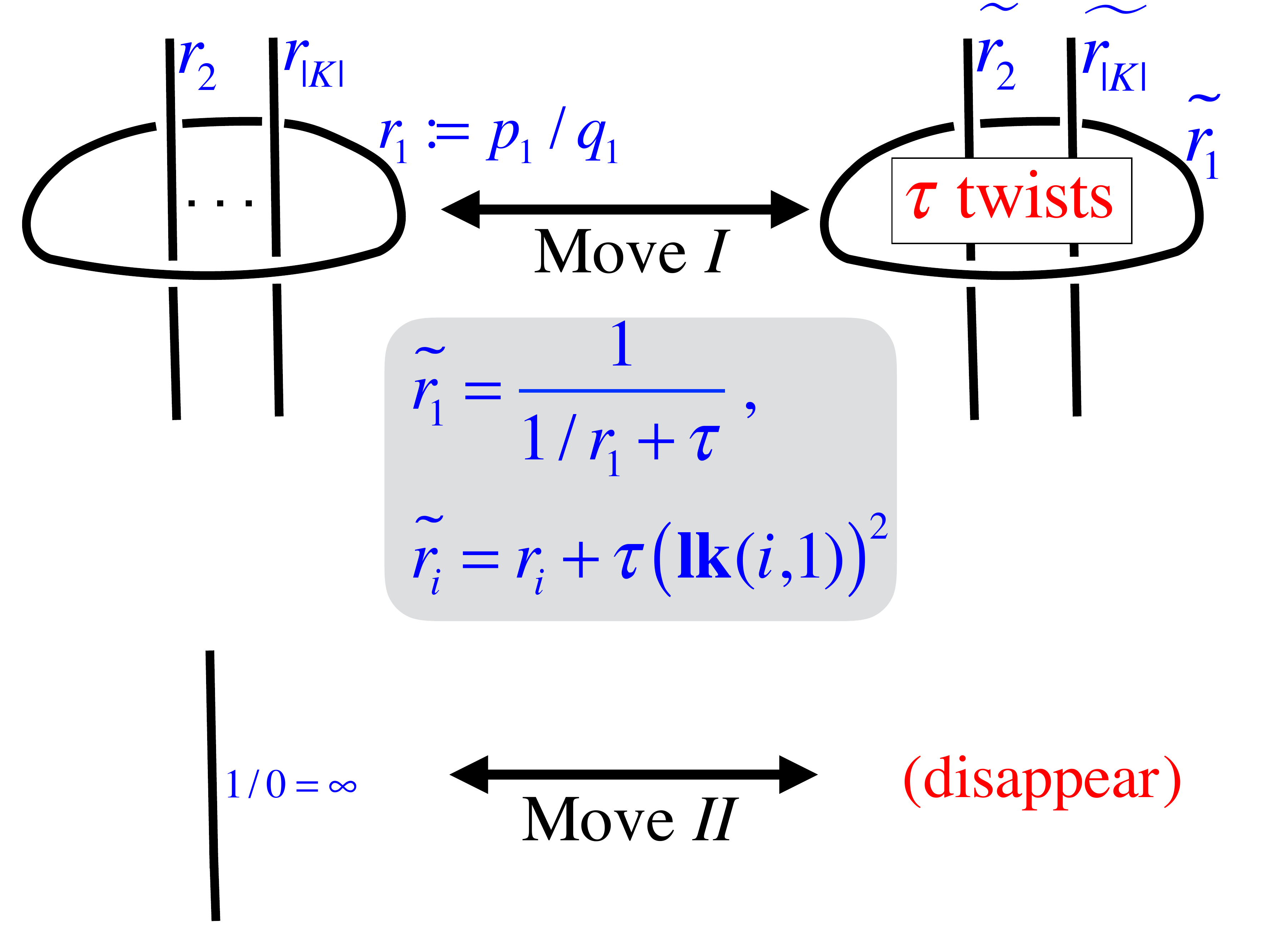}\quad
\includegraphics[width=.45\textwidth]{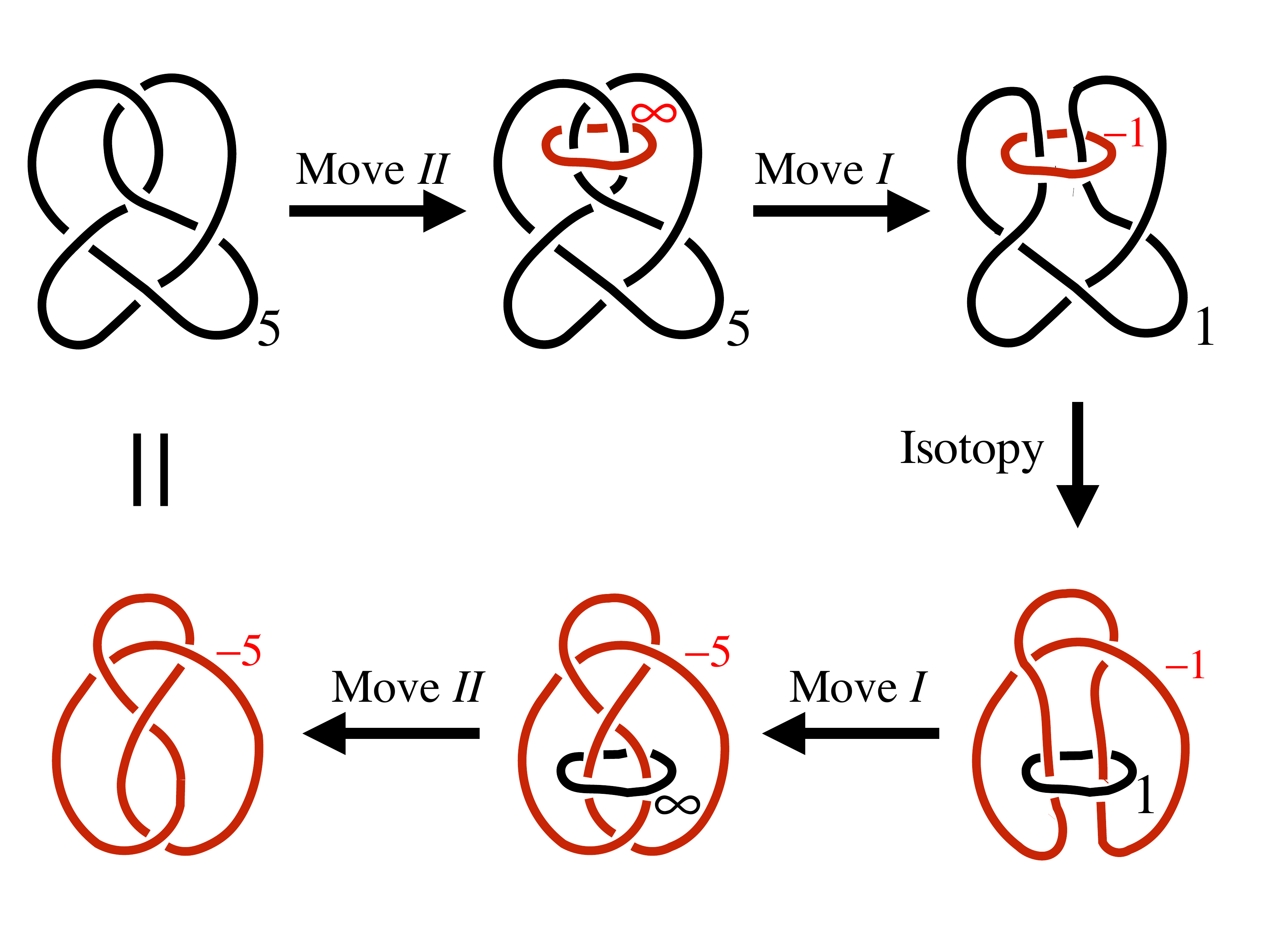}
   \end{center}
   \caption{Left : Basic moves in rational surgery calculus \cite{rolfsen1984rational}. The rational number $r_i$ next to $i$-th component of link $K$ represent Dehn filling slope and $\mathbf{lk}(i,1)$ denotes the linking number between $i$-th component and 1st component. Right: A sequence of basic moves showing $(S^3\backslash \mathbf{4_1})_{-5\mu+\lambda} = (S^3 \backslash \mathbf{5_2})_{5\mu+\lambda}$  } \label{fig:rational-surgery-moves}
\end{figure}

The DGG's construction is based on an ideal triangulation of a knot complement $N$. Different ideal triangulations give different field theory descriptions of $T^{DGG}[N]$ theory related by a duality. 
In our  construction $T^{6d}_{\rm irred}[M]$ for a closed 3-manifold $M$, we use a Dehn filling description of the closed 3-manifold, $M=N_A$, with a hyperbolic knot complement $N$ and a primitive boundary cycle $A$.   The construction can be straightforwardly generalized to the case when $M$ can be given by Dehn fillings on a link complement.  
  According to the Lickorish-Wallace theorem \cite{Lickorish}\cite{Wallace}, every closed orientable 3-manifold $M$ can be obtained by performing Dehn surgery along a link $L$ in $S^3$. 
\begin{align}
\begin{split}
M &= (S^3\backslash L)_{p_1 \mu_1 +q_1\lambda_1, p_2 \mu_2 +q_2\lambda_2,\ldots, p_l \mu_l +q_l\lambda_l}\;, \quad \textrm{$l=|L|$ : $\sharp$ of components of a  link $L$}\;.
\end{split}
\end{align}
The Dehn surgery representation is not unique and there are different choices of $(L,\{p_i,q_i\})$ and $(L',\{p_i',q_i'\})$ which give a same closed 3-manifold. Different Dehn surgery presentations of $M$ give different gauge theory descriptions of $T^{6d}_{\rm irred}[M]$ related by a 3d duality.   A rational surgery calculus \cite{rolfsen1984rational} studies the equivalence relation among the choices. Every pair of equivalent Dehn surgery representations are known to be related by a sequence of basic local moves depicted in figure ~\ref{fig:rational-surgery-moves}.  The basic moves may corresponds to basic 3d $\mathcal{N}=2$ dualities among $T^{6d}_{\rm irred}[M]$. Identifying the basic dualities would be interesting and we leave it as future work. 
\paragraph{Example :} As depicted in figure \ref{fig:rational-surgery-moves}, topologically $(S^3\backslash \mathbf{4_1})_{-5\mu+\lambda} = (S^3 \backslash \mathbf{5_2})_{5\mu+\lambda}$.  Combining the topological fact with  eq.~\eqref{Theory for m003/m004 Dehn-filled}, we have a  3d duality between following twos
\begin{align}
\begin{split}
&T^{6d}_{\rm irred}[(S^3 \backslash \mathbf{4_1})_{-5\mu+\lambda}] = \frac{\textrm{A $u(1)_0$ vector coupled to 2 chrials of charge +1}  }{SO(3)_{-5}} \quad \textrm{and}
\\
&T^{6d}_{\rm irred}[(S^3 \backslash \mathbf{5_2})_{5\mu+\lambda}] = \frac{\textrm{A $u(1)_{-1/2}$ vector coupled to 3 chrials of charge +1}  }{SO(3)_{1}}\;. \label{duality-surgery-example}
\end{split}
\end{align}
The theory in the numerator of the first line  is the theory $T^{6d}_{\rm irred} [S^3 \backslash \mathbf{4_1}, \mu]= T^{DGG}[ S^3\backslash \mathbf{4_1}, \mu]$ which is claimed to have $SU(3)$ in section \eqref{sec:SU(3)}.  The theory in in the first line is obtained by gauging  $SO(3)$ subgroup of the $SU(3)$ flavor symmetry with CS level $-5$. 

\acknowledgments
We are grateful to Yuji  Tachikawa for very helpful discussions and for collaboration during part of this project.
The work of DG was supported by Samsung
Science and Technology Foundation under Project Number
SSTBA1402-08. The work of KY is supported in part by
the WPI Research Center Initiative (MEXT, Japan), and also
supported by JSPS KAKENHI Grant-in-Aid (17K14265).


\appendix
\section{3d index $\cI_{N}^{(A,B)}(m,e;x)$ and $\cI_{N_{pA+q B}} (x)$ }\label{sec : 3d index}

3d index \cite{Dimofte:2011py} is an invariant associated to an knot complement $N$ and a choice of  basis $(A,B)$ of $H_1(\partial N, \mathbb{Z})$. It is defined with respect to a choice of an ideal triangulation $\cT$ of $N$ with positive angle structure. But it is invariant under the local 2-3 move of triangulation and believed to be independent on the choice of $\cT$. After reviewing the definition based on an ideal triangulation, we generalized 3d index to be applicable to closed 3-manifolds by incorporating Dehn filling. The 3d index for a 3-manifold $M$ computes the superconformal index of $T^{6d}_{\rm irred}[M]$ theory, which is defined as follows
\begin{align}
\textrm{Tr}(-1)^{R} x^{\frac{R}2 + j_3}\;.
\end{align}
Here the trace is taken over all local operators in the 3d SCFT and $R$ and $j_3$ are the Cartans of $u(1)$ R-symmetry and  $SO(3)$ Lorentz spin respectively. 

\paragraph{3d index on knot complements} For given choice of an ideal triangulation, with $k$-tetrahedra, of a knot complement $N$ and the basis boundary cycle $(A,B)$, we can associate $Sp(2k,\mathbb{Z})$ matrix $g_N$ and an integer-valued vector $\nu_N$ of size $2k$ as in eq.~\eqref{g_N}. Then, the 3d index is defined by  \cite{Dimofte:2011py}
\begin{align}
\begin{split}
&\cI_{N}^{(A,B)} (m,e;x) 
\\
&=\sum_{(e_2,\ldots,e_{k}) \in \mathbb{Z}^{k-1}} \left( (-x^{\frac{1}2})^{\langle \nu_N, \gamma \rangle } \prod_{i=1}^k  \cI_{\Delta} \big{(}(g_N^{-1}  \gamma)_i , (g_N^{-1}  \gamma)_{k+i} ;x \big{)} \right)\bigg{|}_{m_1 \rightarrow m, \;e_1\rightarrow e, \;m_{I>1} \rightarrow 0} \;, 
\\
&\textrm{where } \gamma := (m_1,\ldots, m_k, e_1, \ldots, e_k)^T\; \textrm{and}\; \langle \nu_N, \gamma \rangle := \sum_{i=1}^k (\nu_N)_{k+i}m_i-(\nu_N)_i e_i\;.
\label{3d index-1}
\end{split}
\end{align}
The tetrahedron index  $\cI_{\Delta}(m,e;x)$ in charge basis is given by \cite{Dimofte:2011py}%
\begin{align}
\begin{split}
&\sum_{e\in \mathbb{Z}}\cI_{\Delta}(m,e;x)u^e = \prod_{r=0}^{\infty} \frac{1-x^{r-\frac{m}{2}+1}u^{-1}}{1- x^{r -\frac{m}{2}} u}\;, 
\\
&\textrm{or more explicitly} 
\\
&\mathcal{I}_{\Delta} (m,e;x) = \sum_{n=[e]}^\infty \frac{(-1)^n x^{\frac{1}2 n (n+1)-(n+\frac{1}2 e)m}}{(x)_n(x)_{n+e}} \;.
\label{tetrahdron index}
\end{split}
\end{align}
where $[e]:=\frac{1}2 (|e|-e)$ and $(x)_n:= (1-x)(1-x^2)\ldots (1-x^n)$.
For example,
\begin{align}
\cI_{\Delta}(0,0;x) = 1-x-2x^2-2x^3-2x^4+x^6+\ldots
\end{align}
The index satisfies following identities
\begin{align}
\begin{split}
&\cI_{\Delta}(m,e;x)=\cI_\Delta (-e,-m;x)\;, 
\\
\textrm{Triality : }& \cI_{\Delta}(m,e;x)=  (-x^{\frac{1}2})^{-e}\cI_{\Delta}(e,-e-m;x)= (-x^{\frac{1}2})^m \cI_{\Delta}(-e-m,m;x)\;.
\end{split}
\end{align}
Under the orientation change, the index transforms as follows
\begin{align}
\cI_{\overline{N}}^{A,-B} (m,e;x) = \cI_{N}^{A,B} (m,-e;x)\;. \label{Index under orientation reversal}
\end{align}
\paragraph{Index as Chern-Simons ptn} The index can be thought as a $SL(2,\mathbb{C})$ CS ptn on $N$ with quantized level $k=0$. The complex CS theory has two levels, $k$ and $\sigma$
\begin{align}
\begin{split}
&S_{CS}(\cA, \bar{\cA}; \hbar , \tilde{\hbar}) = \frac{i}{2\hbar} \int \textrm{Tr}\bigg{(} \cA \wedge d \cA  + \frac{2}3 \cA^3 \bigg{)} +  \frac{i}{2\tilde{\hbar}} \int \textrm{Tr} \bigg{(} \bar{\cA} \wedge d \bar{\cA}  + \frac{2}3 \bar{\cA}^3 \bigg{)}
\\
& \hbar = \frac{4\pi i }{k +i \sigma}\;, \quad  \tilde{\hbar} = \frac{4\pi i }{k -i  \sigma} \;,\;\quad  \textrm{where }k \in \mathbb{Z} \textrm{ and } \sigma \in \mathbb{R}\;.
\end{split}
\end{align}
For $k=0$, the $\hbar = - \tilde{\hbar}$ is real and is related to the variable $x$ in the index as follows
\begin{align}
e^{\hbar} = x\;.
\end{align}
The index is a wave function in a Hilbert $\cH_{k=0}(\partial N) = \cH_{k=0}(\mathbb{T}^2)$, Hilbert space of the complex CS theory on a torus. Classically, the phase space on $\mathbb{T}^2$ is parameterized by exponentiated holonomy variables $( e^{a/2} , e^{b/2})$ along boundary $(A,B)$ cycle respectively and their complex conjugates. 
Quantum mechanically these variables are promoted to operators acting on the Hilbert-space $\cH_{k=0}(\mathbb{T}^2)$, $(e^{a/2},e^{b/2},e^{\bar{a}/2},e^{\bar{b}/2}) \rightarrow (e^{\hat{a}/2},e^{\hat{b}/2},e^{\hat{\bar{a}}/2},e^{\hat{\bar{b}}/2}) $
\begin{align}
\begin{split}
&\mathcal{O} [\cH_{k=0} (\mathbb{T}^2)] 
\\
&= \langle e^{\hat{a}/2}, e^{\hat{b}/2}, e^{\hat{\bar{a}}/2}, e^{\hat{\bar{b}}/2} \; :\; e^{\alpha/2} e^{\beta/2} = x^{-\frac{1}2} e^{\beta/2} e^{\alpha/2},\; e^{\bar{\alpha}/2} e^{\bar{\beta}/2} = x^{\frac{1}2} e^{\bar{\beta}/2} e^{\bar{\alpha}/2},
\\
&\qquad \qquad \qquad \qquad \qquad \quad  \;[e^{\hat{a}/2}, e^{\hat{\bar{a}}/2}] = [e^{\hat{a}/2}, e^{\hat{\bar{b}}/2}] =[e^{\hat{b}/2}, e^{\hat{\bar{a}}/2}] = [e^{\hat{b}/2}, e^{\hat{\bar{b}}/2}]= 0 \rangle\;.
\end{split}
\end{align}
The algebra acts on $\cH_{k=0}(\mathbb{T}^2)$ as follows \cite{Dimofte:2011py}
\begin{align}
\begin{split}
&\textrm{Basis of }\cH_{k=0} (\mathbb{T}^2) = \{ |m,e\rangle \}_{m,e\in \mathbb{Z}}\;,
\\
&\sum_{(m,e) \in \mathbb{Z}^2} |m,e\rangle \langle m,e| = 1 \quad (\textrm{completeness relation})\;.
\\
&\langle m,e | e^{\hat{a}/2} = \langle m, e-1 | x^{\frac{m}4}\;, \quad \langle m,e | e^{\hat{\bar{a}}/2} = \langle m, e+1 | x^{\frac{m}4}\;,
\\
&\langle m, e| e^{\hat{b}/2} = \langle m+1, e| x^{\frac{e}4}\;, \quad \langle m, e| e^{\hat{\bar{b}}/2} = \langle m-1, e| x^{\frac{e}4}\;.  \label{Basis(m,e)}
\end{split}
\end{align}
Quantum mechanically, we  associate a vector $|N \rangle \in \cH_{k=0}(\partial N)$ to $N$
\begin{align}
\begin{split}
\partial N \; &\rightsquigarrow  \cH_{k=0}(\partial N) 
\\
 \cap \;&  \qquad \quad \text{\begin{rotate}{90}$\in$\end{rotate}}
\\
N  \; &\rightsquigarrow \;\; |N\rangle 
\end{split}
\end{align}
Then, the 3d index can be interpreted as
\begin{align}
\cI_{N}^{(A,B)}(m,e) =
\begin{cases} \langle 2m,e|N\rangle
\;, \quad \textrm{when $A$ is of $SU(2)$ type}
\\
\langle m, 2e |N \rangle \;, \quad \textrm{when $A$  is of $SO(3)$ type}  \label{3d index as overlap}
\end{cases} 
\end{align}
\paragraph{Quantum Dehn filling on index, $\cI_{pA+qB}(x)$} Mimicking  the $k=1$ case in \cite{Bae:2016jpi,Gang:2017cwq} , we  give a Dehn filling operation on the index $(k=0)$. The $SL(2,\mathbb{C})$ CS wave function on a solid torus $D_2 \times S^1 = S^3\backslash ({\rm unknot})$ is annihilated by following operators (a pair of quantum A-polynomial for unknot)
\begin{align} 
(e^{\hat{b}}+1 + x^{\frac{1}2} e^{\hat{b}/2} +x^{-\frac{1}2} e^{\hat{b}/2})|D_2 \times S^1\rangle = (e^{\hat{\bar{b}}/2}+1 + x^{-\frac{1}2} e^{\hat{\bar{b}}/2} +x^{\frac{1}2}e^{\hat{\bar{b}}/2})|D_2 \times S^1\rangle =0\;.
\end{align}
Here the boundary cycle $B$ corresponds to the shrinkable cycle in $D_2 \times S^1$,
\begin{align}
B \subset H_1 \big{(}\partial (D_2),\mathbb{Z} \big{)} \subset H_1\big{(}\partial (D_2 \times S^1),\mathbb{Z}\big{)}\;.
\end{align}
In the classical limit $x = e^{\hbar}\rightarrow 1$, the operator equation become $(e^{b/2}+1)^2=0$  which reflects the fact that  flat-connections on $D_2 \times S^1$ have trivial holonomy ($e^{b/2}=-1$) along the boundary cycle $B$. 
A solution for the difference equations is
\begin{align}
\langle m, e| D_2 \times S^1\rangle = \frac{1}2(-1)^{m} \bigg{(}\delta_{e,0}(x^{\frac{m}2}+x^{- \frac{m}2})- \delta_{e,2}- \delta_{e,-2} \bigg{)}
\end{align}
Then, the CS ptn for $k=0$ on $N_{pA+ q B}$ is given by 
\begin{align}
\begin{split}
& N_{pA+qB} =\big{(} (D_2 \times S^1) \cup N \big{)}/\sim \;,\quad 
\left(\begin{array}{c}A \in H_1 (\partial ({D_2 \times S^1}),\mathbb{Z}) \\ B \in H_1 (\partial ({D_2 \times S^1}),\mathbb{Z}) \end{array}\right) \sim \varphi \left(\begin{array}{c} A \in H_1 (\partial N, \mathbb{Z}) \\ B \in H_1 (\partial N, \mathbb{Z})  \end{array}\right)\;,
\\
&\Rightarrow \; \cI_{N_{pA+qB}}(x) = \langle D_2 \times S^1 | \hat{\varphi}|N \rangle\;, \quad \varphi =\left(\begin{array}{cc}r & s \\ p & q\end{array}\right) \in SL(2,\mathbb{Z})\;. \label{DvarphiN}
\end{split}
\end{align}
The operator $\hat{\varphi}$ satisfies
\begin{align}
 \hat{\varphi}^{-1}\hat{a} \hat{\varphi} = r \hat{a}+ s \hat{b}\;,\quad  \hat{\varphi}^{-1}\hat{b}\hat{\varphi} = p \hat{a}+ q \hat{b}\;.
\end{align}
The matrix element of the operator is given by 
\begin{align}
&\langle m, e | \hat{\varphi} | m', e'\rangle =\delta_{r m' +s e' , m}\delta_{p m' + q e' , e}\;.
\end{align}
Plugging the matrix element into \eqref{DvarphiN} with the completeness relation in \eqref{Basis(m,e)},
\begin{align}
\begin{split}
&\cI_{N_{pA+qB}}(x) 
\\
&= \sum_{(m, e, m', e')\in \mathbb{Z}^4}\langle D_2 \times S^1 | m, e \rangle \langle m, e |\hat{\varphi} | m',e' \rangle \langle m', e'|N \rangle
\\
&= \sum_{(m, e)\in \mathbb{Z}^2} \langle D_2\times S^1 | r m+ s e, p m+q e \rangle \langle m, e |N \rangle
\\
&= \sum_{(m, e)\in \mathbb{Z}^2} \frac{1}2 (-1)^{r m+ s e} \bigg{(} \delta_{p m+q e,0} (x^{\frac{r m+s e}2}+ x^{-\frac{r m+s e}2}) - \delta_{p m+qe,-2}-\delta_{p m+q e,2} \bigg{)}   \langle m, e |N\rangle\;.
\end{split}
\end{align}
Note that the final expression is dependent on the choice of $(r,s)$ since the expression is invariant under 
\begin{align}
(r,s) \rightarrow (r,s)+ \mathbb{Z}(p,q)
\end{align}
which is expected since $N_{pA+qB}$ depends only on $(p,q)$.
Combining with eq.~\eqref{3d index as overlap}, we finally have
\begin{align}
\begin{split}
&\textrm{For $SU(2)$ type $A$,}
\\
&\cI_{N_{pA+qB}}(x)  
\\
&= \sum_{(m, e)\in \mathbb{Z}^2} \frac{1}2 (-1)^{2r m+ s e} \bigg{(} \delta_{2p m+q e,0} (x^{\frac{2r m+s e}2}+ x^{-\frac{2r m+s e}2}) - \delta_{2p m+qe,-2}-\delta_{2p m+q e,2} \bigg{)}   \cI^{(A,B)}_{N}(m,e;x)\;,
\\
&:=\sum_{(m,e)\in \mathbb{Z}^2} \mathcal{K}^{SU(2)}(m,e;p,q;x)\mathcal{I}^{(A,B)}_N(m,e;x)\;,
\\
&\textrm{For $SO(3)$ type $A$,}
\\
&\cI_{N_{pA+qB}}(x)  
\\
&= \sum_{(m, e)\in \mathbb{Z}^2} \frac{1}2 (-1)^{r m+ 2s e} \bigg{(} \delta_{p m+2q e,0} (x^{\frac{r m+2s e}2}+ x^{-\frac{r m+2s e}2}) - \delta_{p m+2qe,-2}-\delta_{p m+2q e,2} \bigg{)}   \cI^{(A,B)}_{N}(m,e;x)\;,
\\
&:=\sum_{(m,e)\in \mathbb{Z}^2} \mathcal{K}^{SO(3)}(m,e;p,q;x)\mathcal{I}^{(A,B)}_N(m,e;x)\;.
 \label{3d index-2}
\end{split}
\end{align}
The  formulae in eq.~\eqref{3d index-1} and \eqref{3d index-2} can be straightforwardly extended to the case when $N$ is a link complement with several components and the case when performing dehn filling along several components. 
\begin{align}
\textbf{Conjecture : the 3d index is topological invariant} \label{conjecture on top invariance of index}
\end{align}
Different Dehn surgery representation of a 3-manifold gives different expressions for the 3d index and the conjecture says that they are  all equivalent.  Let us give some non-trivial evidence for the conjecture
\paragraph{Example : $(S^3\backslash \mathbf{4_1})_{-5\mu+\lambda} = (S^3\backslash \mathbf{5_2} )_{5\mu+\lambda}$} The indices for two knot complements are
\begin{align}
\begin{split}
&\cI_{S^3\backslash \mathbf{4_1}}^{(\mu, \lambda)} (m,e;x) =  \sum_{e_1 \in \mathbb{Z}}\cI_\Delta (m-e_1, m+e-e_1;x) \cI_{\Delta}(e-e_1,-e_1;x)\;,
\\
&\cI_{S^3\backslash \mathbf{5_2}}^{(\mu, \lambda)} (m,e;x) =  \sum_{e_1,e_2 \in \mathbb{Z}}(-x^{\frac{1}2})^{-(e_1+m)} \cI_{\Delta}(e_1,e_2;x) \cI_\Delta(e_1+m,-e-2 e_2-e_1-2 m;x) 
\\
&\qquad \qquad \qquad \qquad \qquad  \times \cI_\Delta(e_1+2 m,e+e_2+m;x)\;.
\end{split}
\end{align}
Then, from series expansion in $x$, we can check that
\begin{align}
\begin{split}
&\sum_{(m,e)\in \mathbb{Z}^2} \mathcal{K}^{SO(3)}(m,e;5,1;x) \cI^{(\mu,\lambda)}_{S^3 \backslash \mathbf{5_2}}(m,e;x) =  \sum_{(m,e)\in \mathbb{Z}^2} \mathcal{K}^{SO(3)}(m,e;-5,1;x) \cI^{(\mu,\lambda)}_{S^3 \backslash \mathbf{4_1}}(m,e;x) 
\\
&=1-x-2x^2-x^3-x^4+x^5+2 x^6+7 x^7+8 x^8+ 12 x^9+\ldots
\end{split}
\end{align}

\paragraph{Example : $(S^3\backslash \mathbf{5}^2_1)_{\mu_1  +k \lambda_1}  = (S^3\backslash {\rm K}_k)$} 
$\mathbf{5}^2_1$ denotes Whitehead link depicted in Figure~\ref{fig:Whitehead}. The corresponding link complement can be triangulation can be 4 tetrahedra. The gluing datum are (from SnapPy)
\begin{align}
\begin{split}
&C_1 = Z_1+Z_2+Z_3+Z_4 \;, \quad C_2 = 2Z_1'+Z_1''+2Z_2'+Z_2''+Z_3''+Z_4''\;, 
\\
&C_3 = Z_1''+Z_2''+2Z_3'+Z_3''+2Z_4'+Z_4''\;, \quad C_4 = C_1\;,
\\
&a_{\mu_1} = Z_3'' +Z_1'-Z_2\;, \quad b_{\lambda_1} = 2 (Z_2'+Z_2''-Z_3)\;, 
\\
&a_{\mu_2} = Z_1 - Z_2'-Z_3''\;, \quad b_{\lambda_2} = 2 (Z_1 -Z_3'-Z_3'')\;. \label{Gluing eqns for Whitehead}
\end{split}
\end{align}
Note that only two internal edges are linearly independent and both of $(a_{\mu_1}, b_{\lambda_1})$ and  $(a_{\mu_2}, b_{\lambda_2})$ are  $SO(3)$ type basis.
Corresponding index is 
\begin{align}
\begin{split}
&\cI_{S^3\backslash \mathbf{5}^2_1}^{(\mu_1, \mu_2 ; \lambda_1, \lambda_2)}(m_1, m_2, e_1, e_2;x) = \cI_{\mathbf{5}^2_1} (m_1, m_2,e_1,e_2;x)
\\
&:=\sum_{(n_1,n_2) \in \mathbb{Z}^2} (-x^{1/2})^{-e_1-e_2+m_1+m_2+2 n_1+2 n_2}
\cI_\Delta(n_1,n_2)  \cI_\Delta(-e_1-e_2+n_1,m_1+m_2+n_2)
\\
&\quad \quad \;  \times \cI_\Delta(e_2-n_1,-e_1-e_2+m_1+2 n_1+n_2)  \cI_\Delta(e_1-n_1,-e_1-e_2+m_2+2 n_1+n_2) \;.  \label{Whitehead index}
\end{split}
\end{align}
The Whitehead index $\cI_{\mathbf{5}^2_1}(m_1,m_2,e_1,e_2)$ enjoys following $\mathbb{D}_8$ symmetry in addition to the Weyl-symmetry  $\mathbb{Z}_2 : (m_i, e_i )\rightarrow (-m_i, -e_i)$: 
\begin{align}
\begin{split}
&\mathbb{D}_8=\big{\langle} S_1, S_2 : S_1^2 = S_2^2=1, (S_1 S_2)^8 =1 \big{\rangle}\;,
\\
&S_1 \;:\quad \left(\begin{array}{c}m_1 \\ m_2 \\ e_1 \\ e_2\end{array}\right) \rightarrow \left(
\begin{array}{cccc}
 0 & 1 & 0 & 0 \\
 1 & 0 & 0 & 0 \\
 0 & 0 & 0 & 1 \\
 0 & 0 & 1 & 0 \\
\end{array}
\right)  \left(\begin{array}{c}m_1 \\ m_2 \\ e_1 \\ e_2\end{array}\right) \;,
\\
&S_2 \;:\quad \left(\begin{array}{c}m_1 \\ m_2 \\ e_1 \\ e_2\end{array}\right) \rightarrow \left(
\begin{array}{cccc}
 -1 & 0 & 0 & 0 \\
 0 & 1 & 0 & -1 \\
 -2 & 0 & 1 & 0 \\
 0 & 0 & 0 & -1 \\
\end{array}
\right)  \left(\begin{array}{c}m_1 \\ m_2 \\ e_1 \\ e_2\end{array}\right) \;.
\end{split} \label{D_8 of Whitehead}
\end{align}
Recall that ${\rm K}_k$ denotes $k$-twist knot. There are two ways of computing the index for the knot complement $S^3\backslash {\rm K}_k$. First one is using the index for Whitehead link \eqref{Whitehead index} and applying the Dehn filling prescription in eq.~\eqref{3d index-2}. The other is using an ideal triangulation for the twist knot complement and apply the 3d index formula in eq.~\eqref{3d index-1}. We checked they  give the same index in $x$-expansion for several examples.

\section{$T[SU(2)]$ and $SU(2)/SO(3)$ types}\label{app:susotypes}

Here we discuss some properties of the Gaiotto-Witten $T[SU(2)]$ theory.
This theory plays two different roles:
\begin{enumerate}
\item In 3d SCFTs, the $SL(2, \BZ) $ transformations of $su(2)$ type uses duality wall theories, and $T[SU(2)]$ corresponds to the operation of $S \in SL(2,\BZ)$.
\item In $S^1$ compactification of 6d $\CN=(2,0)$ theory, the codimension-2 defect becomes $T[SU(2)]$ coupled to the 5d gauge field, 
which gives properties of knots in complex Chern-Simons theory on 3-manifolds.
\end{enumerate} 
Therefore, the properties of this theory are important in both sides of 3d/3d correspondence.

\subsection{Brief review of $T[SU(2)]$}\label{app:brief}
Let us first review the $T[SU(2)]$ theory. It is a 3d $\mathcal{N}=4$ supersymmetric field theory obtained by
a $U(1)$ gauge multiplet with two hypermultiplets of charge $\pm 1$. In terms of 3d $\mathcal{N}=2$ supersymmetry,
there are one $U(1)$ vector multiplet $V$, one neutral chiral multiplet $\phi$, and two pairs of chiral multiplets $(E^i, \tilde{E}_i)~(i=1,2)$ where $E^i$ has $U(1)$ charge $+1$
and $\tilde{E}_i$ has charge $-1$.
The superpotential is given by
\beq
W = \phi \tilde{E}_i E^i.
\eeq
At the level of Lie algebra, this theory has global symmetry $su(2)_H \times su(2)_C$.
The $su(2)_H$ acts on the index $i$ of $(E^i, \tilde{E}_i)~(i=1,2)$. On the other hand, the $su(2)_C$ arises at the quantum level.
The $u(1)_C \subset su(2)_C$ comes from the topological symmetry of the gauge $U(1)$ symmetry whose current is $j = \frac{1}{2\pi}f$, where $f=da$ is the field strength of the gauge field $a$.
This topological symmetry is enhanced to $su(2)_C$ at the quantum level. 

Because of the $\CN=4$ supersymmetry, the $\CN=4$ conserved current supermultiplets contain $\CN=2$ chiral operators which are in the adjoint representation of the symmetry.
They are called (holomorphic) moment map operators because they are associated to the moment maps of hyperkahler moduli spaces of the Higgs and Coulomb branch,
and their scaling dimensions are protected to be 2. In this paper we abuse the terminology 
and call these operators as moment map operators even if there is only $\CN=2$ supersymmetry.

For the $su(2)_H$, the holomorphic moment map operator is given by
\beq
(\mu_H)^i_j = E^i \tilde{E}_j - \frac{1}{2}(\tilde{E}_k E^k) \delta^i_j.
\eeq
For the $su(2)_C$, the holomorphic moment map operator is given by $\phi$ and monopole operators $v_\pm$,
\beq
\mu_C = (\phi, v_\pm).
\eeq
The $\phi$ corresponds to the Cartan of $su(2)_C$, while $v_\pm$ are off-diagonal components of the $su(2)_C$.
These $v_\pm$ have charge $\pm 1$ under the topological $u(1)_C$ symmetry with the current $ j =  \frac{1}{2\pi }f$.

There is a mirror symmetry which exchanges the Higgs branch and Coulomb branch of the theory.
Under the mirror symmetry, the symmetries and operators are exchanged as
\beq
\text{mirror}:~~~\left( su(2)_H, \mu_H \right) \longleftrightarrow \left(su(2)_C, \mu_C \right).
\eeq
This mirror symmetry also guarantees that the topological symmetry $u(1)_C$ is enhanced to $su(2)_C$.

\subsection{The global structure of the symmetries and 't~Hooft anomaly}
Now we study the global structure of the symmetries $su(2)_H$ and $su(2)_C$.
We claim that both of them are $SO(3)$ type in the sense that all gauge invariant operators (in the absence of background fields) 
are in representations of $SO(3)$ (i.e, integer spin representations of $su(2)$).
We denote them as  $SO(3)_H$ and $SO(3)_C$, respectively. However, we will also show that there is a mixed 't~Hooft anomaly between these groups $SO(3)_H$ and $SO(3)_C$
which forbid gauging both of them as $SO(3)$ groups. In other words, this anomaly implies that if we gauge one of them as $SO(3)$ gauge group, then the other symmetry becomes $SU(2)$.

It is easy to see that $su(2)_H$ is of $SO(3)$ type. The fields $(E^i, \tilde{E}_i)$ transform under $su(2)_H$. Now, the center $-1 \in SU(2)_H$ multiplies
the $(E^i, \tilde{E}_i)$ by $(-1)$, but this can be cancelled by a $U(1)$ gauge transformation. Therefore, the action of the center $-1 \in SU(2)_H$ to all gauge invariant operators is trivial.
This shows that the symmetry which acts faithfully to gauge invariant operators is $SO(3)_H$.

By mirror symmetry, it is obvious that the symmetry $su(2)_C$ must also be of $SO(3)$ type. 
More direct way to see this is to notice that all monopole operators have integer charges under $u(1)_C \subset su(2)_C$. Thus, all operators are in integer spin representations of $su(2)_C$,
meaning that it is $SO(3)_C$.

However, there is a subtle mixed anomaly between $SO(3)_H$ and $SO(3)_C$ as we now see. Including the gauge group $U(1)$, the operators are in representations of
\beq
\frac{U(1) \times SU(2)_H}{\BZ_2} = U(2).
\eeq
Then, gauge invariant operators are in representations of $SU(2)_H/\BZ_2 = SO(3)_H$. Now let us consider a monopole background of the above $U(2)$ on $S^2$
which is obtained by embedding a $U(1)$ magnetic flux into $U(2)$ as $\diag(+1,0)$.
This is separated into gauge and flavor parts as
\beq
\diag(+1,0) = \diag(+1/2, +1/2) + \diag(+1/2, -1/2) .
\eeq
So, the gauge $U(1)$ has magnetic flux $+1/2$ on $S^2$. The $SO(3)_H$ also has nontrivial magnetic flux $ \diag(+1/2, -1/2) $ which is measured by a nontrivial value
of the second Stiefel-Whitney class $w_2 \in H^2(X, \BZ_2)$ of the $SO(3)_H$ bundle, where $X$ is the spacetime on which the theory is placed. 
Roughly speaking, the Stiefel-Whitney class $w_2$ is defined such that half of it, $\frac{1}{2}w_2 \mod 1$, is
the fractional part of the magnetic flux of $SO(3)$. 
The integer part is not topological invariant in non-abelian $SO(3)$ group. The topologically invariant magnetic fluxes are classified by $\pi_1(SO(3))=\BZ_2$.

The above argument implies the following. Suppose we gauge the group $SO(3)_H$ as an $SO(3)$ gauge group.
Then it is possible to consider a monopole operator of $SO(3)_H$ which has nontrivial Stiefel-Whitney class. However, for this monopole operator to make sense,
we also have to turn on a half-integral magnetic flux of the gauge $U(1)$. Then, these monopole operators have half-integral charges under the topological $u(1)_C$ 
and hence they are in half-integer spin representations of the Coulomb branch symmetry $su(2)_C$. Therefore, by gauging $SO(3)_H$, the symmetry $su(2)_C$ becomes $SU(2)_C$.
This fact forbids to gauge both of the $su(2)_H$ and $su(2)_C$ symmetries as $SO(3)$ type symmetries, and this means there is an anomaly.

More formally, the anomaly is shown as follows~\cite{Benini:2017dus}. 
(See also \cite{Komargodski:2017dmc, Komargodski:2017smk, Shimizu:2017asf,Gaiotto:2017tne,Tanizaki:2017qhf,Tanizaki:2017mtm,Sulejmanpasic:2016uwq,Sulejmanpasic:2018upi} 
where nontrivial mixture of the center of gauge and flavor symmetries lead to 't~Hooft anomalies.)
Let $f=da$ be the field strength of the $U(1)$ gauge field, $F=dA+A^2$ be the field strength of the mixed gauge-flavor symmetry $U(2) = [U(1) \times SU(2)_H]/\BZ_2$,
and $B$ be the gauge field of $u(1)_C$. Then, the coupling of the background $B$ and $F$ in the Lagrangian is given by
\beq
\int_X \frac{i}{2\pi } B \wedge f = \frac{i}{4\pi } \int_X B \wedge  \tr F. 
\eeq
where $X$ is a 3-manifold in which our $T[SU(2)]$ theory lives. To make the definition manifestly gauge invariant,
we consider a 4-manifold $Y$ whose boundary is $X$, $\partial Y = X$, and define the above coupling as
\beq
\frac{i}{4\pi } \int_Y G \wedge  \tr F 
\eeq
where $G=dB$. However, this depends on the extension of the manifold and the gauge field from $X$ to $Y$.
Choose another extension $Y'$. Then glue $Y$ and $Y'$ together along their common boundary to make a closed manifold $Z$.
Then we get
\beq
\frac{i}{4\pi } \int_Y G \wedge  \tr F  - \frac{i}{4\pi } \int_{Y'} G \wedge  \tr F = \frac{i}{4\pi } \int_Z G \wedge  \tr F .
\eeq
This shows the dependence on the extension. 
Now, let $w_2(SO(3)_H)$ and $w_2(SO(3)_C)$ be the Stiefel-Whitney classes of $SO(3)_H$ and $SO(3)_C$, respectively. We have
\beq
\frac{1}{2} G  &=\pi w_2(SO(3)_C) \mod 2\pi, \\
\frac{1}{2}\tr F &= \pi w_2(SO(3)_H) \mod 2\pi
\eeq
and hence
\beq
\frac{i}{4\pi } \int_Z G \wedge  \tr F = i\pi \int_Z w_2(SO(3)_C) w_2(SO(3)_H)  \mod 2\pi i.
\eeq
This represents the anomaly. The anomaly polynomial is $\frac{1}{2}w_2(SO(3)_C) w_2(SO(3)_H)$.

\subsection{$S$-transformation of 3d SCFT }
Now it is clear why the $SU(2)$ and $SO(3)$ types are exchanged under the Gaiotto-Witten's $S \in SL(2,\BZ)$ transformation of 3d SCFT.
We start from some 3d theory $\CT$. Then, the $S$ transformation is performed as follows.

If the $\CT$ has symmetry $SU(2)$, then we add $T[SU(2)]$ and gauge the diagonal $SU(2)$ subgroup of the $SU(2)$ of $\CT$ and the $SU(2)_H$ of $T[SU(2)]$.
Then, $w_2(SO(3)_H)=0$, and hence the anomaly vanishes. The $su(2)_C$ symmetry is $SO(3)_C$.
On the other hand, if the $\CT$ has the symmetry $SO(3)$, then we can (and choose to do) gauge the diagonal $SO(3)$ subgroup of the $SO(3)$ of $\CT$ and the $SO(3)_H$ of $T[SU(2)]$.
Because of the anomaly, or more explicitly by the consideration of the monopole operators discussed above, the $su(2)_C$ becomes $SU(2)_C$.

In summary, if the original $\CT$ has $SO(3)$ symmetry, the $S$-transformed theory $S \cdot \CT$ has $SU(2)$ symmetry and vice versa.
In this way, $SU(2)$ and $SO(3)$ are exchanged under the $S$-transformation.

\subsection{$SU(2)/SO(3)$ symmetry types of knots from six dimensions}
From the point of view of 6d $\CN=(2,0)$ theory, the symmetry type is determined as follows.
First we have to recall some of the properties of this theory~\cite{Witten:1998wy}. For simplicity we focus on the case of the $A_1$ theory corresponding to $su(2)$. 

Let $X_6$ be a six manifold. To determine the partition function on $X_6$, we have to give some additional data.
Let $H_3(X_6, \BZ_2)$ be the third homology with $\BZ_2$ coefficients. By Poincare duality, it is possible to split this homology as
\beq
H_3 (X_6, \BZ_2) = A \oplus B. \label{eq:polarization}
\eeq
This is chosen such that any two elements $a, a' \in A$ have zero intersection $ \langle a , a' \rangle=0$, and similarly for $B$,
and the pairings between $A$ and $B$ are non-degenerate.
The splitting of $H_3 (X_6, \BZ_2) $ into $A$ and $B$ is not unique, but we have to choose one to define partition functions of the 6d theory.
We call this splitting as polarization, and call $A$ and $B$ as A-cycles and B-cycles, respectively.
The partition function of the theory not only depends on the manifold $X_6$, but also on the polarization.

Let us compactify the 6d theory on $S^1$ and consider $X_6 = S^1 \times X_5$. Then we get 5d SYM theory with gauge algebra $su(2)$.
Then, the above splitting determines the $SU(2)/SO(3)$ types of the 5d gauge theory. 
First, notice that the cohomology is given as
\beq
H_3( S^1 \times X_5, \BZ_2) \cong  H_2(X_5, \BZ_2) \oplus H_3(  X_5, \BZ_2).
\eeq
Under this isomorphism, we define
\beq
\hat{A} = A \cap H_2(X_5, \BZ_2),~~~~~\hat{B}= B \cap H_2(X_5, \BZ_2).
\eeq
Then, the 5d $su(2)$ theory has the following properties. 
Roughly speaking, the theory is $SU(2)$ type for A-cycles and $SO(3)$ type for B-cycles, respectively.
More precisely, let $w_2$ be the second Stiefel-Whitney class of the $su(2)$ bundle on $X_5$. 
For a 2-cycle $\hat{a} \in \hat{A}$, we require that $\int_{\hat{a}} w_2=0$.\footnote{In the presence of background fields for 1-form center symmetry, it can take nonzero but fixed values determined
by the background field.} On the other hand, for $\hat{b} \in \hat{B}$, we sum over all gauge configurations with different values of $\int_{\hat{b}} w_2$ in the path integeral.

Applications of the above framework to 4d class S theories were studied in \cite{Tachikawa:2013hya}.
Here we want to do it for 3d/3d correspondence.
More specifically, we want to determine the $SU(2)/SO(3)$ types of the flavor symmetry associated to a knot.

Take the 6d manifold as $X_6= X_3 \times M_3$, where $X_3$ is ``space-time'' and $M_3$ is ``internal space'' which is closed. The holomogy is given by
\beq
 &H_3 (X_3 \times M_3, \BZ_2)  \nonumber \\
=& H_3(X_3,\BZ_2) \oplus [H_2(X_3,\BZ_2) \otimes H_1(M_3,\BZ_2)] \oplus [H_1(X_3,\BZ_2) \otimes H_2(M_3,\BZ_2)] \oplus H_3(M_3,\BZ_2)
\eeq
First we have to choose a polarization \eqref{eq:polarization}. There is no unique way to do it.
However, there are only a few choices which preserve the diffeomorphism invariance of $X_3$ and $M_3$,\footnote{There is no way to preserve the diffeomorphism invariance of the full
6d space $X_6$ because the splitting \eqref{eq:polarization} breaks it. } and we assume that one of those choices is realized in 3d/3d correspondence.

One possible choice is to take
\beq
A& = H_3(X_3,\BZ_2)  \oplus [H_1(X_3,\BZ_2) \otimes H_2(M_3,\BZ_2)]  , \nonumber \\
B&=  H_3(M_3,\BZ_2) \oplus [H_2(X_3,\BZ_2) \otimes H_1(M_3,\BZ_2)]. \label{eq:onepolarication}
\eeq
In this case, by taking $X_3 = S^1 \times X_2$, we get
\beq
\hat{A}& = H_2(X_2,\BZ_2)  \oplus   H_2(M_3,\BZ_2)    , \\
\hat{B}&=   H_1(X_2,\BZ_2) \otimes H_1(M_3,\BZ_2).
\eeq
One of the consequences of this choice is as follows. Let $K \in M_3$ be a knot, and take a codimension-2 defect along $X_3 \times K \subset X_6$.
After the reduction on the $S^1$ of the $X_3 = S^1 \times X_2$, 
the defect becomes the $T[SU(2)]$ theory coupled to the 5d gauge theory along $X_2 \times K$.
Then, there are two cases, depending on whether the homology class of $K$, which we denote $[K]$, is nontrivial or not in $H_1(M_3,\BZ_2)$.
If $[K]$ is nonzero, then the $X_2 \times K$ can contain nontrivial elements of $ H_1(X_2,\BZ_2) \otimes H_1(M_3,\BZ_2)$ which have nonzero values of the Stiefel-Whitney class $w_2$ on them.
Then, by coupling the symmetry $su(2)_H$ of $T[SU(2)]$ to the 5d gauge group, the $su(2)_C$ becomes $SU(2)_C$ by the anomaly explained in the previous subsections.
On the other hand, if $[K]$ is zero in $ H_1(M_3,\BZ_2)$, then the $H_1(X_2,\BZ_2) \otimes H_1(M_3,\BZ_2)$ restricted to the $X_2 \times K$ is trivial.
Hence, the $su(2)_C$ is $SO(3)_C$ type.

Let us summarized the above result. Under the choice of polarization \eqref{eq:onepolarication};
\begin{itemize}
\item If the knot $K$ has a nontrivial homology in $H_1(M_3, \BZ_2)$, then the type of the symmetry associated to the knot is $SU(2)$.
\item If the knot $K$ has a trivial homology in $H_1(M_3, \BZ_2)$, then the type of the symmetry associated to the knot is $SO(3)$.
\end{itemize}
Let us compare this result with the criterion of $SU(2)/SO(3)$ given in Sec.~\ref{sec:T[M]}.
There, we consider the knot complement $N_3:=M_3 \backslash K$. The boundary of $N_3$ is
a torus, $\partial N_3 \cong T^2$, and let $A$ be the A-cycle of the torus which is contractible on the ambient manifold $M_3$.
Let $[A] \in H_1(N_3, \BZ_2)$ be the image of $A$ in the $\BZ_2$ homology of the knot complement $N_3$.
The proposal in Sec.~\ref{sec:T[M]} is that if $[A]$ is trivial in $H_1(N_3, \BZ_2)$, 
then the knot is of $SU(2)$ type, and if $[A]$ is nontrivial, it is of $SO(3)$-type.

Suppose that $[A]$ is nonzero in $H_1(N_3, \BZ_2)$. Poincare-Lefschetz duality implies that
there is a dual cycle $\CB \in H_2(N_3,  \partial N_3, \BZ_2)$ in the relative homology group  $H_2(N_3, , \partial N_3, \BZ_2)$ such
that $[A]$ and $\CB$ has intersection number $1 \mod 2$. By regarding $\CB$ as a chain of $N_3$,
we can take 
the boundary $\partial \CB \in H_1(\partial N_3, \BZ_2)$ (or more precisely, this map is a connection homomorphism in the long exact sequence
of $H_*(N_3, \BZ_2)$, $H_*(N_3,\partial N_3, \BZ_2)$ and $H_*(\partial N_3, \BZ_2)$).
On $\partial N_3 \cong T^2$, the $[A]$ and $\partial \CB$ regarded as elements of $H_1(\partial N_3, \BZ_2)$ 
has intersection number $1 \mod 2$ because, roughly speaking, the intersection of $[A]$ and $\CB$ must happen on the boundary $\partial N_3$. 
Therefore, $\partial \CB$ is of the form $[B+p A]$, where $B$ is the B-cycle on $\partial N_3$,
and $p $ is an integer. If we embed $\CB$ into the ambient manifold $M_3$,
then we get $\partial \CB = [B+pA]=[K]$ in $H_1(M_3, \BZ_2)$ because $A$ is contractible in $M_3$ and $B$ is homotopic to $K$. 
Therefore, $[K]$ is trivial. Conversely, it is also true by Poincare-Lefschetz duality that if $[K]$ is trivial, then $[A]$ is nonzero in $H_1( N_3, \BZ_2)$.
Thus we get
\beq
\text{$[A] \in H_1(N_3, \BZ_2)$ is nonzero} \Longleftrightarrow  \text{$[K] \in H_1(M_3, \BZ_2)$ is zero.} 
\eeq
This means that the result for $SU(2)/SO(3)$ types obtained in this appendix is the same as the proposal in Sec.~\ref{sec:T[M]}.

Remember that there are only a few polarization choices which preserve the diffeomorphism invariance of $M_3$ and $X_3$.
One of them \eqref{eq:onepolarication} reproduces the rules for the $SU(2)/SO(3)$ symmetry types of knots.
This is a nontrivial check of our proposal.
Thus we assume that the choice \eqref{eq:onepolarication} is realized in 3d/3d correspondence.

There are other consequences of the above choice of polarization.
The fact that $H_2(M_3,\BZ_2) $ is in $\hat{A}$ means that the 5d gauge bundle does not have a nontrivial Stiefel-Whitney class on $M_3$.
This means that the complex Chern-Simons theory on the 3-manifold has the $SU(2)_\BC=SL(2,\BC)$ bundles instead of $SO(3)_\BC=PSL(2,\BC)$,
as far as the Stiefel-Whitney class (i.e., discrete magnetic flux) on $M_3$ is concerned. However, this does not mean that the periodicity of holonomy is of $SL(2,\BC)$ type.
Indeed, because $H_1(X_2,\BZ_2) \otimes H_1(M_3,\BZ_2)$ is in $\hat{B}$, the holonomies have periodicities of $PSL(2,\BC)$ type.
This fact can be seen as follows. There can be a nontrivial magnetic flux $w_2$ on $S^1_X \times S^1_M$ where $S^1_X \subset X_2$ and $S^1_M \subset M_3$.
This is possible only if $[S^1_M] \in H_1(M_3,\BZ_2)$ is nonzero.
Now consider holonomies in the spin $J$ representation of $su(2)$ gauge algebra around the cycle $p \times S^1_M$, where $p$ is a point on $S^1_X$.
If we let the point $p$ go around $S^1_X$ and return to the same point, the value of the holonomy changes as 
\beq
\exp ( 2J\pi i \int_{S^1_X \times S^1_M} w_2 ),
\eeq
where we have assumed that the holonomy is taken in the spin $J$ representation of $su(2)$. 
Thus, the well-definedness of the holonomy requires that we only consider representations with integer spin $J \in \BZ$. This means that only the representations of $SO(3)$ type are consistent.
This is a version of the Dirac quantization condition argument.
On the other hand, if $[S^1_M] \in H_1(M_3,\BZ_2) $ is zero, then holonomy may be well-defined for half-integer representations of $su(2)$.

Thus, in complex Chern-Simons theory on $M_3$, we get the following conditions;
\begin{itemize}
\item We only consider gauge bundles of zero Stiefel-Whitney class $w_2=0$ on $M_3$.
\item Holonomies around cycles $S^1_M \in M_3$ are defined only for $PSL(2,\BC)$ representations if $[S^1_M] \in H_1(M_3,\BZ_2)$ is nonzero,
while they may be defined for $SL(2,\BC)$ representations if $[S^1_M]$ is zero.

\end{itemize}
The first condition is consistent with the connection coming from hyperbolic metric, because the tangent bundle of any orientable 3-manifold has $w_2(\text{tangent})=0$.
The second condition is consistent with constructing $PSL(2,\BC)$ connections by ideal triangulations. 
If we try to uplift the holonomy to $SL(2,\BC)$, then there may be 
$\pm$ ambiguity. For example, the holonomies \eqref{Hol-m004-1} contain square roots $\sqrt{z_1},~\sqrt{z_2}$, and so on, which represent this ambiguity.




\bibliographystyle{JHEP}
\bibliography{ref}

\end{document}